\documentclass[useAMS,usenatbib,usegraphicx]{mn2e}

\usepackage{amssymb}
\usepackage{epsfig}
\usepackage{lscape}
\usepackage{graphicx}
\usepackage{rotating}
\usepackage{times}
\bibpunct{(}{)}{;}{a}{}{,}

\def\l{\ensuremath{\lambda}}
\newcommand{\rsi}{$\mathcal{R}({\rm Si})$}
\newcommand{\snia}{SN~Ia}
\newcommand{\sneia}{SNe~Ia}

\let\ts=\thinspace
\newcommand{\one}{\ts {\,\sc i}}
\newcommand{\two}{\ts {\,\sc ii}}
\newcommand{\three}{\ts {\,\sc iii}}

\newcommand{\nifs}{\ensuremath{^{56}\rm{Ni}}}

\newcommand{\msun}{\ensuremath{\rm{M}_{\odot}}}
\newcommand{\kms}{\ensuremath{\rm{km\,s}^{-1}}}

\newcommand{\dmft}{\ensuremath{\Delta m_{15}(B)}}

\title[2D \snia\ model comparison]{Confronting 2D
  delayed-detonation models with light curves and
  spectra of Type Ia supernovae}
\author[S. Blondin et al.]
{
St\'ephane Blondin,$^{1}$\thanks{E-mail: blondin@cppm.in2p3.fr}
Daniel Kasen,$^{2,3}$ Friedrich K. R\"opke,$^{4,5}$ Robert P. Kirshner$^{6}$
\newauthor and Kaisey S. Mandel$^{6}$\\
$^{1}$Centre de Physique des Particules de Marseille (CPPM),
  Universit\'e Aix-Marseille, CNRS/IN2P3, 163 avenue de Luminy, 13288
  Marseille Cedex 9, France\\
$^{2}$Department of Physics, University of California at Berkeley, 366
  LeConte, Berkeley, CA 94720, USA\\
$^{3}$Nuclear Science Division, Lawrence Berkeley National Laboratory,
  Berkeley, CA 94720, USA\\
$^{4}$Universit{\"a}t W{\"u}rzburg, Am Hubland, D-97074 W{\"u}rzburg,
  Germany\\
$^{5}$Max-Planck-Institut f\"ur Astrophysics, Karl-Schwarzschild-Strasse
  1, D-85741 Garching, Germany\\
$^{6}$Harvard-Smithsonian Center for Astrophysics, 60 Garden Street,
  Cambridge, MA 02138, USA
}

\begin{document}

\date{Accepted 2011 June 28. Received 2011 June 27; in original form 2011 April 18}

\pagerange{\pageref{firstpage}--\pageref{lastpage}} \pubyear{2011}

\maketitle

\label{firstpage}


\begin{abstract}
We compare models for Type Ia supernova (\snia) light curves and
spectra with an extensive set of observations. The models come from
a recent survey of 44 two-dimensional delayed-detonation models
computed by \cite{KRW09}, each viewed from multiple directions. The
data include optical light curves of 251 \sneia, some of which
have near-infrared observations, and 2231 low-dispersion spectra from
the Center for Astrophysics, plus data from the literature. These
allow us to compare a wide range of \snia\ models with observations
for a wide range of luminosities and decline rates. The analysis uses
standard techniques employed by observers, including MLCS2k2, SALT2,
and SNooPy for light-curve analysis, and the Supernova Identification
(SNID) code of Blondin \& Tonry for spectroscopic comparisons to assess how
well the models match the data.   
The ability to use the tools developed for observational
data directly on the models marks a significant step forward in the
realism of the models.
We show that the models that match observed spectra best
 lie systematically on the observed width-luminosity
relation. Conversely, we 
reject six models with highly asymmetric ignition conditions
and a large
amount ($\gtrsim1$\,M$_{\sun}$) of synthesized \nifs\ that yield poor
matches to observed \snia\ spectra. More subtle
features of the comparison include the general difficulty of the
models to match the $U$-band flux at early times, caused by
a hot ionized ejecta that affect the subsequent redistribution
of flux at longer wavelengths. The models have systematically higher
velocities than the observed spectra at maximum light, as inferred
from the Si\two\,\l6355 line. We examine ways in which the asymptotic
kinetic energy of the explosion affects both the predicted velocity
and velocity gradient in the Si\two\ and Ca\two\ lines. 
Models with an asymmetric distribution of \nifs\ are found to
result in a larger variation of photometric and spectroscopic
properties with viewing angle, regardless of the initial ignition setup.
We discuss more generally whether
highly anisotropic ignition conditions are ruled out by
observations, and how detailed 
comparisons between models and observations involving both light
curves and spectra can lead to a better understanding of
\snia\ explosion mechanisms.
\end{abstract}



\section{Introduction}\label{sect:intro}

Type Ia supernovae (\sneia) play a major role in many astrophysical
phenomena. They produce a large fraction of iron in the universe
(e.g., \citealt{Truran/Cameron:1971}), heat the interstellar medium
(e.g., \citealt{Ciotti/etal:1991}), and form an endpoint of
binary star evolution (e.g., \citealt{Iben/Tutukov:1984}). 
\sneia\ provide the most reliable and precise cosmological distances
to establish the acceleration of cosmic expansion \citep{R98,P99}.

Despite their astrophysical importance, however, they remain enigmatic
objects. There is a general consensus that they result from the
thermonuclear disruption of a carbon-oxygen white dwarf (WD) star
\citep{Hoyle/Fowler:1960} approaching the Chandrasekhar
mass ($M_{\rm Ch}\approx1.4$\,M$_{\sun}$), either through
accretion from a non-degenerate binary companion (the
``single-degenerate'' scenario), or through merger with
another WD (the ``double-degenerate'' scenario;
\citealt{Iben/Tutukov:1984,Webbink:1984}). Which of these two
possibilities constitutes the dominant (or sole) progenitor channel for
\sneia\ is still debated (see \citealt{Howell:2010} for a recent
review).

The explosion mechanism itself is also largely unknown (see
\citealt{Hillebrandt/Niemeyer:2000} for a review). In the preferred 
``delayed-detonation'' model \citep{Khokhlov:1991}, the
burning starts as a turbulent subsonic deflagration near the WD center
and transitions to a supersonic detonation near its surface. The
deflagration pre-expands the WD so that the subsequent
detonation does not burn the entire star to nuclear statistical
equilibrium (NSE) material (including \nifs\ to power the light
curve), but instead 
synthesizes appropriate fractions of high-velocity ($\sim10000$\,\kms)
intermediate-mass elements (IME; 
such as Mg, Si, S, Ca, etc.) needed to reproduce the observed spectra. 
The first simulations were carried out in 1D, but recent studies
show that multi-dimensional simulations are needed to capture 
hydrodynamical instabilities (e.g.,
\citealt{Gamezo/Khokhlov/Oran:2005,Roepke/Niemeyer:2007}) and to 
provide a physical basis for the transition from deflagration to
detonation. (e.g., \citealt{Woosley:2007,Roepke:2007,Woosley/etal:2009}).

The empirical relation between the peak luminosity and the width of
the light-curve (the so-called width-luminosity relation, or WLR;
\citealt{Pskovskii:1977,Phillips:1993}), instrumental to the use of
\sneia\ as distance indicators, can be physically interpreted in
terms of (1) varying opacity with the amount of synthesized \nifs\
\citep{Hoeflich/Khokhlov:1996}, (2) varying mass of the progenitor
WD \citep{Pinto/Eastman:2000a}, or (3) the {\sc
  iii}$\rightarrow${\sc ii} (i.e. doubly- to singly-ionized)
recombination timescale of iron-group 
elements in the SN atmosphere \citep{Kasen/Woosley:2007}. More
detailed calculations are also needed to explain the observed scatter
in the WLR.

In a recent paper, \cite{KRW09} (hereafter KRW09) conducted a 2D survey of
delayed-detonation models, in which they varied the 
radial/angular distribution and number of
ignition points in a Chandrasekhar-mass C+O WD star, as well as the
criterion for deflagration-to-detonation transition. By synthesizing
light curves and spectra for different viewing angles, they were able
to broadly reproduce both the observed width-luminosity relation (for
all but subluminous \sneia) and its scatter, illustrating the
importance of multi-dimensional computations to reproduce observed trends in
\snia\ properties. Furthermore, they showed that variations
in the metallicity of the progenitor WD affect both the slope
  and normalization of the WLR, and that ignoring these effects
could potentially lead to systematic overestimates of $\sim2$\%\ on
distance determinations to \sneia.

To study such subtle effects, one must ensure the models reproduce
all observed trends in some detail, and not just the
width-luminosity relation. This is precisely what we set out to do in
the present study, where we conduct an extensive analysis of the
delayed-detonation models of KRW09 through a detailed and direct
comparison with observations of \sneia. We present the models
and data in \S~\ref{sect:modeldata}, and our methods for
quantitatively evaluating each model in \S~\ref{sect:method}. 
We then proceed to a detailed comparison of their photometric
(\S~\ref{sect:phot}) and spectroscopic (\S~\ref{sect:spec}) properties
with actual data. We discuss whether explosion models with
asymmetric ignition conditions
are ruled out by observations in \S~\ref{sect:disc}, and conclude in
\S~\ref{sect:ccl}.


\section{Models and data}\label{sect:modeldata}

\subsection{2D delayed-detonation models from KRW09}\label{sect:model}

\cite{KRW09} simulated 44 axisymmetric two-dimensional
delayed-detonation explosions in a non-rotating spherical
Chandrasekhar-mass WD star composed of equal mass 
fractions of carbon and oxygen. They varied both the radial/angular
distributions and number of ignition points used to trigger the
deflagration (which control the level of asymmetry of the explosion),
as well as the criterion for the transition to a detonation (termed
``dc'', and parameterized via the critical Karlovitz number, which
controls the level of turbulence in the combustion). We refer the
reader to KRW09 for more information on the initial conditions for
each model, but note that these were chosen {\it a priori}, and
  not tuned to match the range of observed \snia\ properties.

Table~\ref{tab:modelinfo} gives the asymptotic kinetic energy ($E_{\rm
kin}$), abundances, peak bolometric luminosities ($L_{\rm bol,peak}$)
and \dmft\footnote{the difference in $B$-band magnitude between
maximum light and 15\,d after maximum; \citealt{Phillips:1993})}
decline-rate ranges for all the models of KRW09.  
Models with an isotropic (anisotropic) distribution of ignition
points are labeled DD2D\_iso (DD2D\_asym). One exception concerns the
DD2D\_asym\_01 model series, which in fact has an isotropic
distribution of ignition points. The number that follows
(01--08) corresponds to a particular ignition setup, and the criterion
for deflagration-to-detonation transition (dc) is also included in the
model name (see Tables~1 \& 2 in KRW09).
The models span a large range in mass of synthesized
\nifs\ (0.29--1.10\,M$_{\sun}$) and decline rate parameters
\dmft\ (0.74--1.57\,mag). The models are unable to reproduce
rapidly-declining \sneia\ such as the subluminous SN~1991bg
(see \citealt{Taubenberger/etal:2008} for a discussion on this class of
  objects). Model 
DD2D\_iso\_01\_dc4 was accidently omitted in the radiative transfer
calculations, but this does not affect the results presented in this
paper. We do not study the impact of metallicity variations in the
progenitor WD star (between 1/3 to 3$\times$ solar; see KRW09).
We selected eight models for
detailed analysis, and rejected six models based on their spectra (see
\S~\ref{sect:rank}). These are marked as ``subset'' and ``rejected''
in Table~\ref{tab:modelinfo}.

\begin{figure*}
\centering
\includegraphics[width=8.5cm,angle=90]{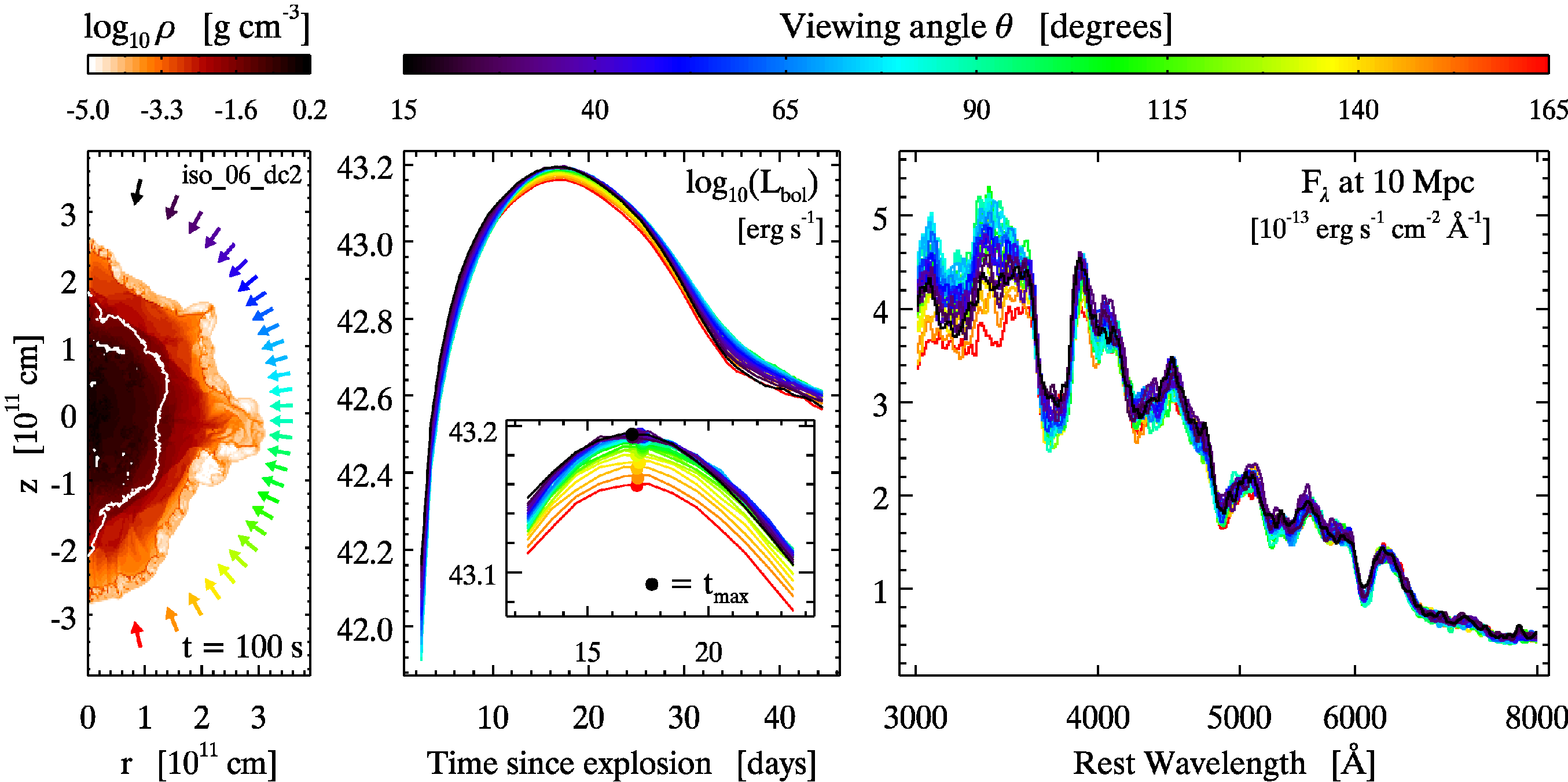}\vspace{.75cm}
\includegraphics[width=8.5cm,angle=90]{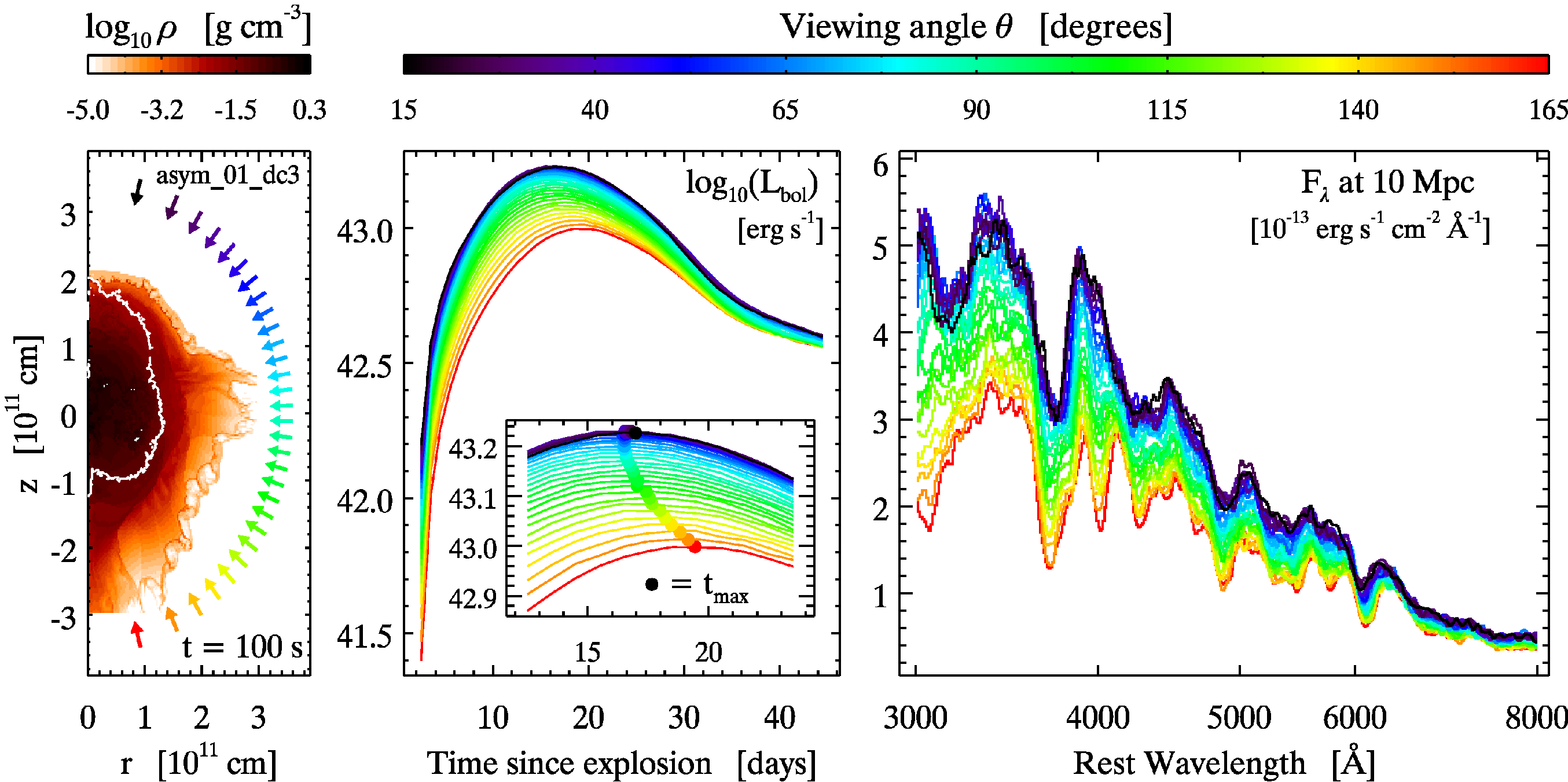}
\caption{\label{fig:dens}
{\it Top left:}
Total mass density distribution at $t=100$\,s for model
DD2D\_iso\_06\_dc2. The ejecta are in homologous expansion, and 
$10^{11}$\,cm in $(r,z)$ corresponds to $10^4$\,\kms\ in $(v_r,v_z)$ .
The white contours correspond to a \nifs\ mass fraction of
$10^{-3}$. The arrows denote the 30 different viewing angles.
{\it Top center:}
Bolometric light curves as a function of viewing angle. The inset
shows the maximum-light region, with filled circles corresponding to the time
of bolometric maximum.
{\it Top right:}
Optical spectra (3000--8000\,\AA) as a function of viewing angle.
{\it Bottom row:}
Same as above but for model DD2D\_asym\_01\_dc3.
}
\end{figure*}

The models were evolved hydrodynamically until $\sim100$\,s past
ignition, at which point the ejecta had reached a phase of homologous
expansion (velocity proportional to radius). Figure~\ref{fig:dens}
shows the total mass density distribution for models DD2D\_iso\_06\_dc2
({\it top left}) and DD2D\_asym\_01\_dc3 ({\it bottom
  left}). The white contours correspond to a \nifs\ mass fraction
of $10^{-3}$. The \nifs\ distribution is roughly symmetric about
$(z=0)$ in model DD2D\_iso\_06\_dc2, while it is strongly skewed toward
$z>0$ in model DD2D\_asym\_01\_dc3, this despite the isotropic
distribution of ignition points in both models.

A crude nuclear network was used to determine the distributions of
radioactive $^{56}$Ni, stable iron-group elements (IGE), and
intermediate-mass elements (IME; such as Mg, Si, S, and Ca), but
detailed abundances of all important chemical species were determined
based on a more elaborate nuclear network applied to a representative model
(DD2D\_iso\_06\_dc2\footnote{and not DD2D\_iso\_06\_dc1, as stated by
  KRW09.}) using
  12800 tracer particles \citep[see][]{Travaglio/etal:2004} and
interpolated accordingly for the other models. The
nucleosynthetic yields are thus subject to some uncertainty which
can impact the derived observables.

The abundances were then remapped on a $64\times128$ cylindrical
grid with a velocity cutoff of $\sim25000$\,\kms\ for the radiative
transfer calculations. These were done with the time-dependent Monte
Carlo radiative transfer code SEDONA \citep{SEDONA} using $10^8$
photon packets and 13 chemical species (ionization stages
  I--V): C, O, Na, Mg, Si, S, Ar, Ca, Ti, Cr, Fe, Co, and Ni.
Local thermodynamic equilibrium (LTE) was assumed for 
the atomic level populations, although non-LTE effects in the
radiation field were treated approximately through an equivalent
two-level atom formalism. The photon packets were collected into 30
separate viewing angle bins ($15^{\circ}\lesssim \theta \lesssim
165^{\circ}$), offering a 2D view of the explosion. 
The bins are equally spaced in $\cos \theta$, so that each bin subtends
the same solid angle and has an equal probability of being
observed. The arrows in the left panels of Fig.~\ref{fig:dens}
  denote the viewing angles considered by KRW09. 
The limited number of photon packets and the large number of
viewing angles results in moderate S/N in
individual angle-dependent light curves and spectra (typically
  S/N$\approx$15--20 per 10\,\AA\ in $B$ and $V$ at maximum light). 
The velocity
cutoff of $\sim25000$\,\kms\ also limits the formation of
high-velocity absorption features, which are common in early-time
\snia\ spectra \citep[e.g.][]{Mazzali/etal:2005}.

Spectra were computed between 2.5\,d and 44.5\,d after explosion
(i.e. between $-17(-15)$\,d and +25(+27)\,d from $B$-band maximum for
a 19.5(17.5)\,d rise time) in 1\,d steps. At later times, non-LTE
effects and 
non-thermal excitation by fast electrons become increasingly important
\citep[e.g.,][]{Kozma/Fransson:1992}
and the radiative transfer calculations are less reliable\footnote{We
  do note, however, that KRW09 present light curves for model
  DD2D\_iso\_06\_dc2 until +60\,d past $B$-and maximum in their
  Fig.~2.}. 
The mass fraction of (unburnt) carbon dominates over all other
intermediate-mass elements at $v\gtrsim$16000--21000\,\kms,
but we are unable to study the presence of corresponding spectral lines
(e.g. C\two\,\l6580) in the early-time spectra due to limited signal
in the Monte Carlo spectra. The total mass of (unburnt) carbon is
$\lesssim 0.04$\,M$_{\sun}$ for all models.

Figure~\ref{fig:dens} shows the variation of bolometric light
curves ({\it middle panels}) and optical spectra ({\it right
  panels}) with viewing angle for models DD2D\_iso\_06\_dc2 ({\it top}) and
DD2D\_asym\_01\_dc3 ({\it bottom}). In model DD2D\_iso\_06\_dc2, the
peak bolometric 
luminosity varies by $<10$\%\ with viewing angle, and the spectra
only change significantly blueward of $\sim3500$\,\AA. In model
DD2D\_asym\_01\_dc3, however, the peak bolometric luminosity varies by
$\sim70$\%\ with viewing angle, and the spectra are affected over a
large fraction of the optical range (out to $\sim6000$\,\AA).

These variations are not related to the total mass of
\nifs\ synthesized during the explosion
($\sim$0.70\,M$_{\sun}$ and $\sim$0.64\,M$_{\sun}$ for DD2D\_iso\_06\_dc2 and
DD2D\_asym\_01\_dc3, respectively); rather, they are a consequence
of the {\it distribution} of \nifs\ in the 
ejecta. When the model is viewed from directions where \nifs\
extends to larger radii, the bolometric light curves peak earlier at
a higher luminosity and are  broader than for viewing angles
where \nifs\ is confined to the deeper ejecta regions \citep[see
  also][their Fig.~4]{Pinto/Eastman:2000a}. 
The impact on the spectra
is dramatic, owing to the higher ejecta temperatures (resulting
in a bluer SED) and higher ionization (affecting the relative shapes
and strengths of spectral features).
For both models, the variation is largest at ultraviolet
wavelengths, and is related to differences in the abundance of
iron-group elements in the outer layers of the ejecta
\citep[see, e.g.,][]{Sauer/etal:2008}.

The variation in peak bolometric luminosity with viewing angle 
in a given model (see Table~\ref{tab:modelinfo}) can thus be
used to gauge the level of  
asymmetry in the \nifs\ distribution, but it does not constrain the
level of isotropy in the distribution of ignition points. Most
DD2D\_iso models display variations in $L_{\rm bol,peak}$ with viewing
angle well below the 20\%\ level, but others (DD2D\_iso\_04\_dc3 and
dc4) display factor-of-two variations. Likewise, some DD2D\_asym
models with the highest level of anisotropy in the distribution of
ignition points (e.g. DD2D\_asym\_07\_dc2 and dc3) display
$\lesssim10$\% variations in peak bolometric luminosity with viewing
angle, whereas the DD2D\_asym\_01 model series (which has an {\it
  isotropic} distribution of ignition points; see above) displays the
largest variations (65--70\%). This is not entirely surprising, since
the ignition points are randomly distributed within a given setup,
and individual ignition points that happen to placed farther out than
the bulk of the ignition sparks can easily dominate the flame
morphology (this is especially true in 2D, which tends to favour large
Rayleigh-Taylor-structures; see \citealt{Roepke/etal:2006}).

\subsection{Data}\label{sect:data}

We largely rely on our database of light curves and spectra from the
CfA Supernova Program to compare the models with observations. These
include optical light curves of 251
\sneia\ \citep{Riess/etal:1999a,Jha/etal:2006,Hicken/etal:2009a}, with
near-infrared (NIR) photometry for the brighter objects \citep{Wood-Vasey/etal:2008},
and 2231 optical spectra obtained
for the most part using the FAST spectrograph
\citep{Fabricant/etal:1998} mounted on the 1.5\,m Tillinghast
telescope at the Fred Lawrence Whipple Observatory (FLWO). The
spectra have a typical FWHM resolution of 6--7\,\AA\ with a
rest-frame wavelength range 3500--7500\,\AA.
Already 577 \snia\ spectra have been published in
several papers (the largest collection of 432 spectra of 32
\sneia\ has been published by \citealt{Matheson/etal:2008}) and are
publicly available via the CfA Supernova
Archive\footnote{http://www.cfa.harvard.edu/supernova/SNarchive.html}.
Additional spectra will be published in a forthcoming paper.

We complemented this data set with published data from the
literature, and reference our sources where appropriate. Our
study of rise times (\S~\ref{sect:trise}) makes use of measurements on
\snia\ light curves from the SDSS-II Supernova Survey by
\cite{Hayden/etal:2010a}, as well as pseudo-bolometric light curves
published by \cite{Stritzinger:2005}.

\begin{figure}
\centering
\resizebox{0.475\textwidth}{!}{\includegraphics{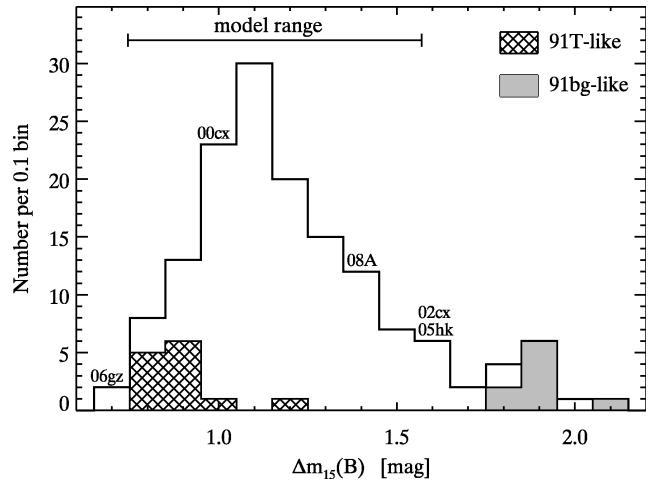}}
\caption{\label{fig:dm15hist}
Distribution of the \dmft\ decline rate parameter in our
\snia\ sample with a maximum-light spectrum. \sneia\ with
spectra resembling those of the luminous SN~1991T and the
faint SN~1991bg are highlighted. Names of peculiar \sneia\ are
given in their respective \dmft\ bin. The horizontal error bar shows
the \dmft\ range in the 2D models of KRW09.
}
\end{figure}

The observed \snia\ sample used in this paper offers a fair
representation of the true \snia\ population in the local
Universe. Figure~\ref{fig:dm15hist} shows the \dmft\ distribution of
\sneia\ for which we have a spectrum within three days from maximum
light. This sample spans a large range in \dmft\ ($\sim$0.7--2.1\,mag)
and includes fractions of luminous 1991T-like ($\sim9$\%), faint
1991bg-like ($\sim7$\%), and peculiar 2002cx-like ($\sim3$\%)
\sneia\ comparable to those found by \cite{Li/etal:2011a} for a
volume-limited sample (9\%, 15\% and 5\%, respectively). The fraction
of 1991bg-like \sneia\ is a factor of two smaller than that reported
by \cite{Li/etal:2011a} and reflects a possible magnitude bias (the 
fraction drops from 15\% to 3\% for an ideal magnitude-limited
sample; \citealt{Li/etal:2011a}, their Fig.~11). In any case, the
models of KRW09 do not extend beyond $\dmft\approx1.6$\,mag and none
of their spectra present the prominent Ti\two\ absorption band around
4000--4500\,\AA\ characteristic of 1991bg-like \sneia. The fraction of
2002cx-like \sneia\ is also smaller than for the volume-limited sample
of \cite{Li/etal:2011a}, but the reported rate has a large associated
error.

\section{Methods}\label{sect:method}

In this section we present the general procedure for fitting 
model light-curves with empirical templates and for inferring
intrinsic photometric properties from observed \snia\ light curves. We
also present the algorithm used to cross-correlate the synthetic
spectra with a large database of observed SN spectra, which we use as
a basis for quantitatively evaluating each model.


\subsection{Light-curve fits}\label{sect:lcfit}

\begin{figure*}
\centering
\resizebox{.8\textwidth}{!}{\includegraphics{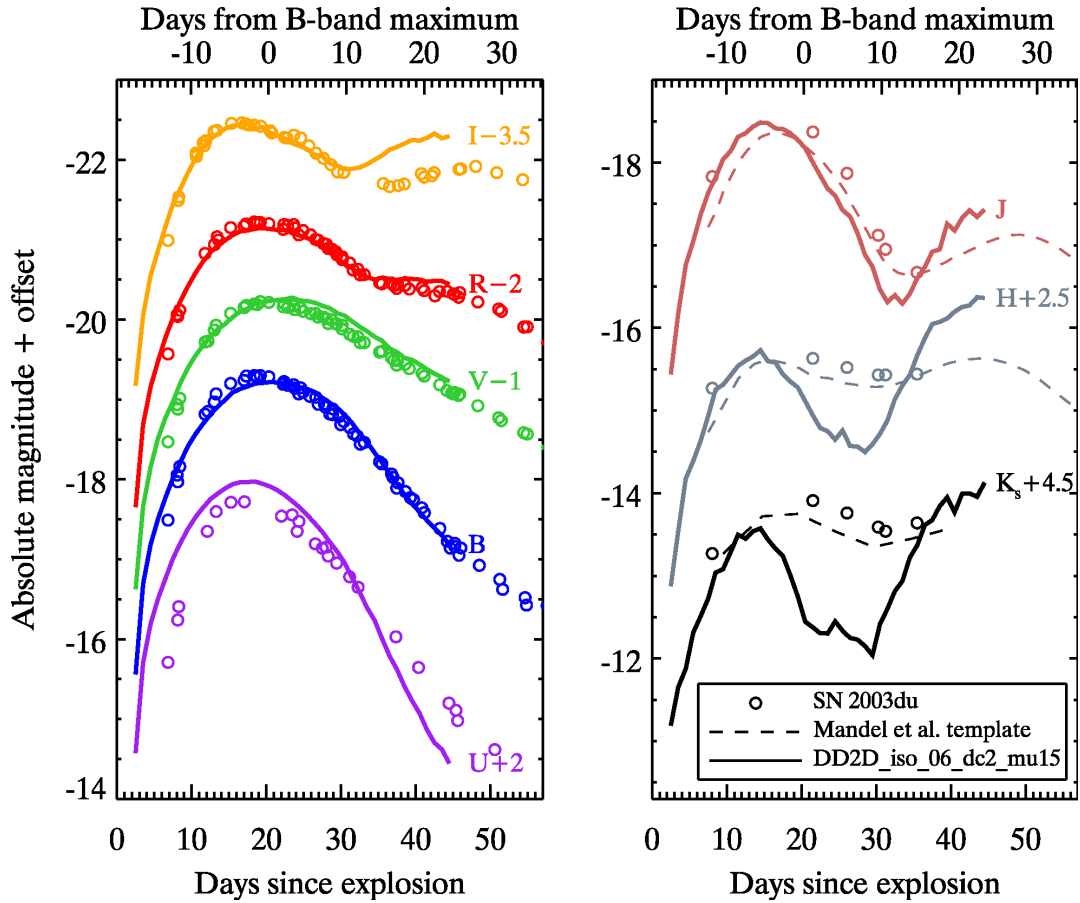}}
\caption{\label{fig:lceg}
$UBVRIJHK_s$ light curves for model DD2D\_iso\_06\_dc2 viewed along
$\theta=88^{\circ}$ ({\it solid lines}). The open circles correspond
to the light curves of SN~2003du
{\protect\citep{Anupama/Sahu/Jose:2005,Leonard/etal:2005,Stanishev/etal:2007}},
assuming a distance modulus $\mu=32.79$\,mag and no host-galaxy reddening
{\protect\citep{Stanishev/etal:2007}}. The dashed curves in the right
panel are the NIR templates of {\protect\cite{Mandel/etal:2009}}, with
updated templates for $J$ and $H$.
}
\end{figure*}

We derive synthetic $UBVRI$ \citep{Bessell:1990} and
$JHK_s$ (2MASS system; \citealt{Cohen/etal:2003}) magnitudes from the model 
spectra, with associated errors based on the number of photon packets
in a given frequency bin. Typical errors at maximum light are
$<0.005$\,mag in $UBVR$, $<0.01$\,mag in $I$, and progressively larger
errors (0.015--0.05\,mag) for $JHK_s$. We show the $UBVRIJHK_s$ light
curves for model DD2D\_iso\_06\_dc2 viewed along $\theta=88^\circ$ in
Fig.~\ref{fig:lceg}, compared to observations of SN~2003du
\citep{Anupama/Sahu/Jose:2005,Leonard/etal:2005,Stanishev/etal:2007}
and the NIR template light curves of \cite{Mandel/etal:2009}. The
models appear to match the data well in the optical (albeit with
significant differences in the $U$ and $I$ bands), but deviate  in
the near infrared.

\subsubsection{Direct polynomial fits}

We determine the time of maximum light and peak magnitude in the
optical bands ($UBVR$), as well as the corresponding $\Delta m_{15}$ decline
rate. Our final
estimates for the various quantities, along with 
their associated errors, are based on 1000 Monte Carlo realizations
using the magnitude errors associated with each light-curve
point. The resulting error on the fit parameters is below
0.1\% in all cases.

\subsubsection{Template-based fits}

We attempt to fit $BVRI$ light curves for each model viewed from
all 30 directions with three different light-curve ``fitters''
commonly used amongst SN observers: MLCS2k2 \citep{MLCS2k2}, SALT2
\citep{SALT2}, and SNooPy \citep{SNooPy}. To first order, these
algorithms all share the common approach of comparing an input light
curve with empirical templates based on observed data. Adding
the $U$-band light curves for the MLCS2k2 and SALT2 fits (there are no
standard $U$-band templates for SNooPy) has negligible impact on the
results. Since no extinction is applied to the model
synthetic magnitudes, we force the extinction to zero. This is not
possible with SALT2, as the effects of extinction and intrinsic colour
variations are described by a single parameter.

\begin{figure*}
\centering
\resizebox{\textwidth}{!}{
\includegraphics{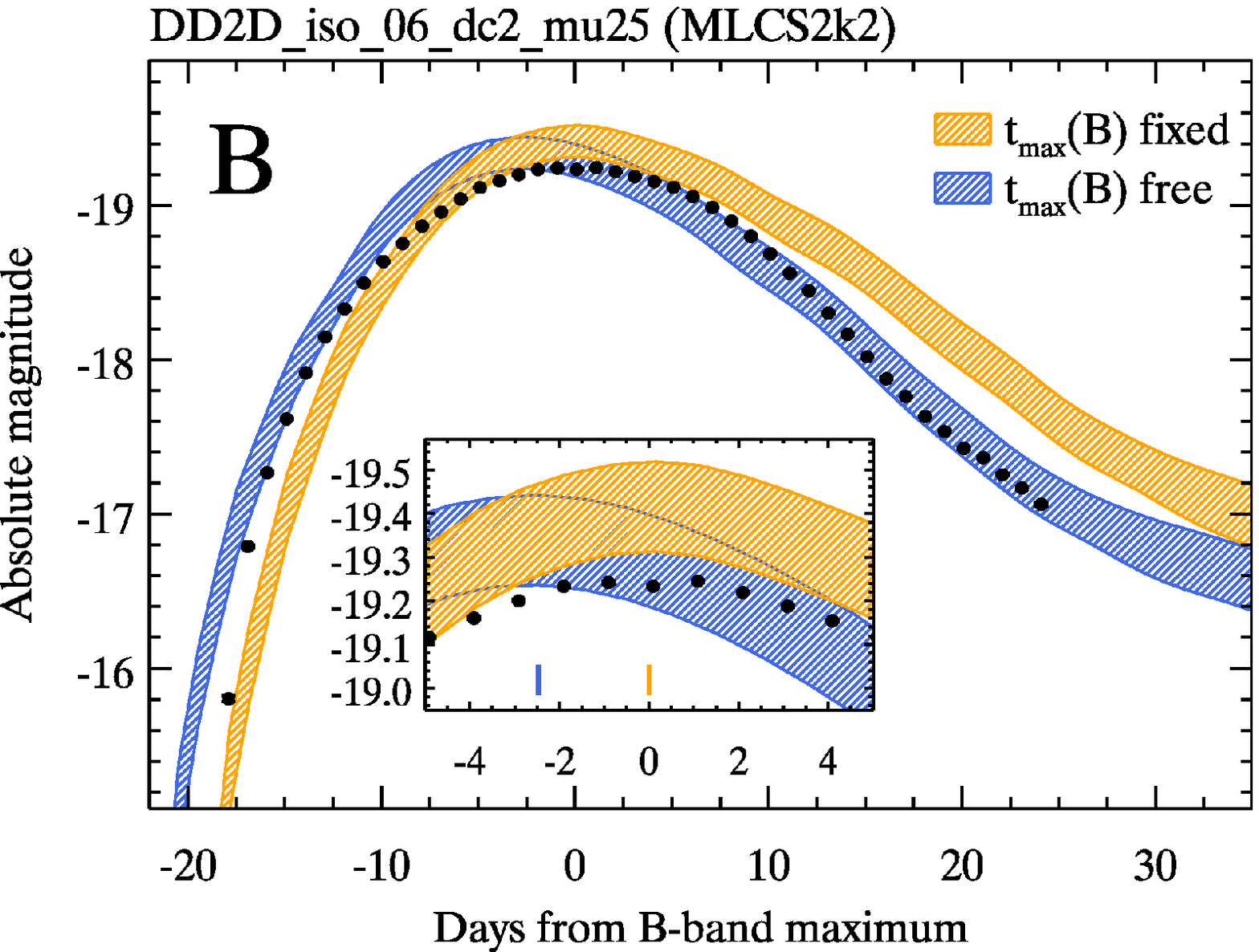}\hspace{1.5cm}
\includegraphics{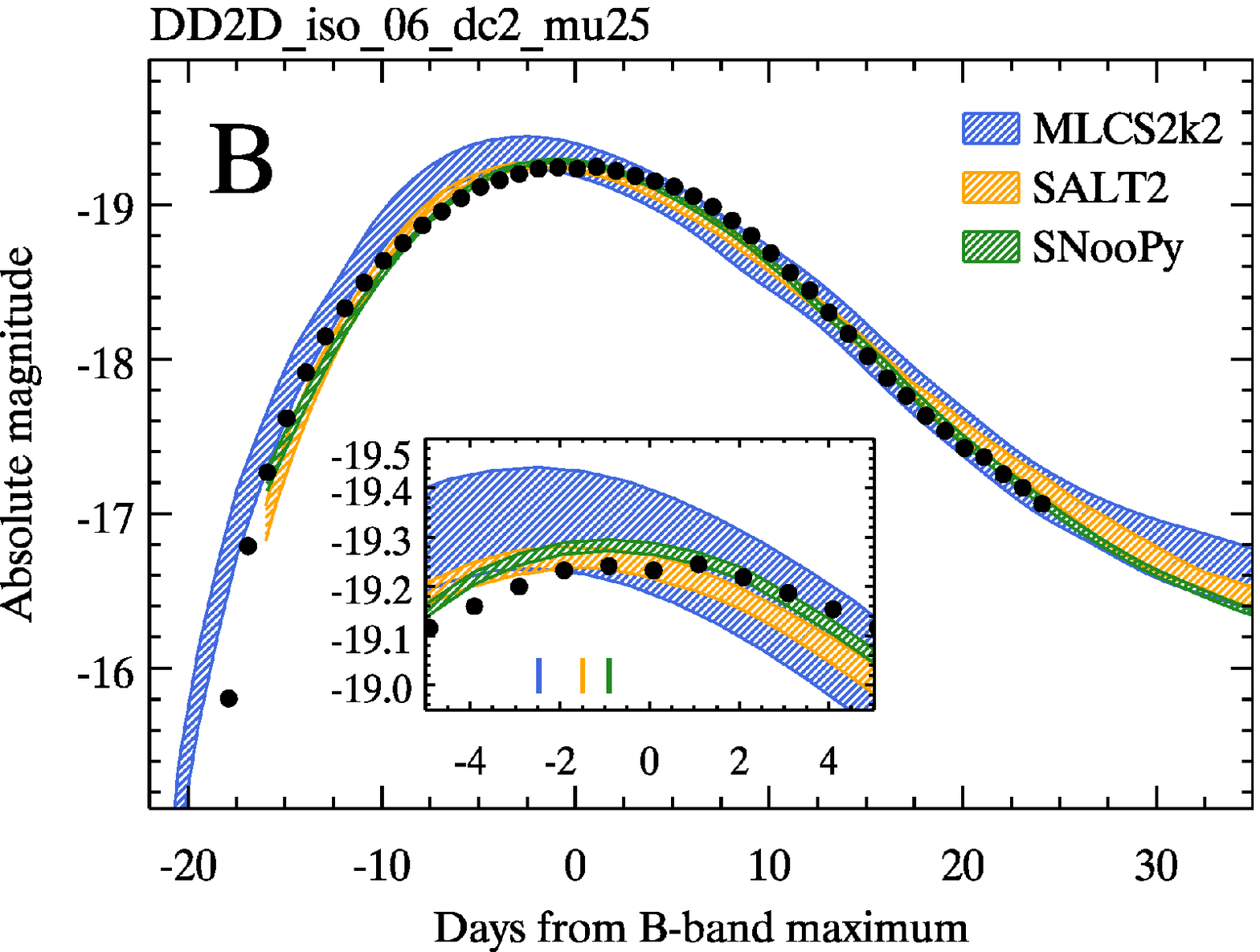}
}
\caption{\label{fig:lcfit}
{\it Left:}
Illustration of a light-curve fit using MLCS2k2
{\protect\citep{MLCS2k2}} on the $B$-band light curve for model
DD2D\_iso\_06\_dc2 viewed along $\theta=46^\circ$ ({\it black
  dots}). We show a fit 
fixing the time of $B$-band maximum ($t_{\rm max}(B)$) to its value
measured from polynomial fits ({\it orange}), and a fit leaving
$t_{\rm max}(B)$ as a free parameter ({\it blue}). The vertical tick marks
denote the time of $B$-band maximum for the MLCS2k2 templates.
Leaving $t_{\rm max}(B)$ free results in a formally better fit but the
time of $B$-band maximum is underestimated (here by $\sim3$ days).
{\it Right:} comparison of light-curve fits to the same model using
MLCS2k2 ({\it blue}), SALT2 ({\it orange}), and SNooPy ({\it green}).
The vertical tick marks denote the time of $B$-band maximum for
the template from the various fitters.
}
\end{figure*}

We show an example MLCS2k2 fit to the $B$-band light curve for model
DD2D\_iso\_06\_dc2 viewed along $\theta=46^\circ$ in the left panel of
Fig.~\ref{fig:lcfit} ({\it blue hatched area}). The fit is
formally acceptable overall, with a reduced
$\chi^2_\nu=0.79$. However, a closer inspection shows that the MLCS2k2
template reaches its peak magnitude $\sim2.5$\,d earlier than the
model predictions. This suggests that the
models rise more slowly to maximum light than actually observed. In 
MLCS2k2 we have the possibility of fixing the time of maximum in the
$B$ band, and we use the value found from direct polynomial
fits to the model light curves. The fits are severely degraded
($\chi^2_\nu=3.37$; {\it orange hatched area}), and the empirical
templates again appear to require a faster rise time than the models
suggest.

This tension between model and empirical light-curve shapes is not
specific to MLCS2k2, as can be seen from the comparison with SALT2 and
SNooPy on the $B$-band light curve for the same model/viewing angle
combination (Fig.~\ref{fig:lcfit}, {\it right panel}). In this
particular example, the time of $B$-band maximum is underestimated by
approximately 3\,d, 2\,d, and 1\,d for MLCS2k2, SALT2, and SNooPY,
respectively. The fits become formally worse following this same
sequence, but this is due to the progressively smaller errors
associated with the empirical templates (the mean absolute deviation
of the template from the synthetic light curve over the time interval
$-15\le t {\rm \ [d]} \le +20$ is $\sim0.7$\,mag for the three
fitters).

\begin{figure}
\centering
\resizebox{0.475\textwidth}{!}{\includegraphics{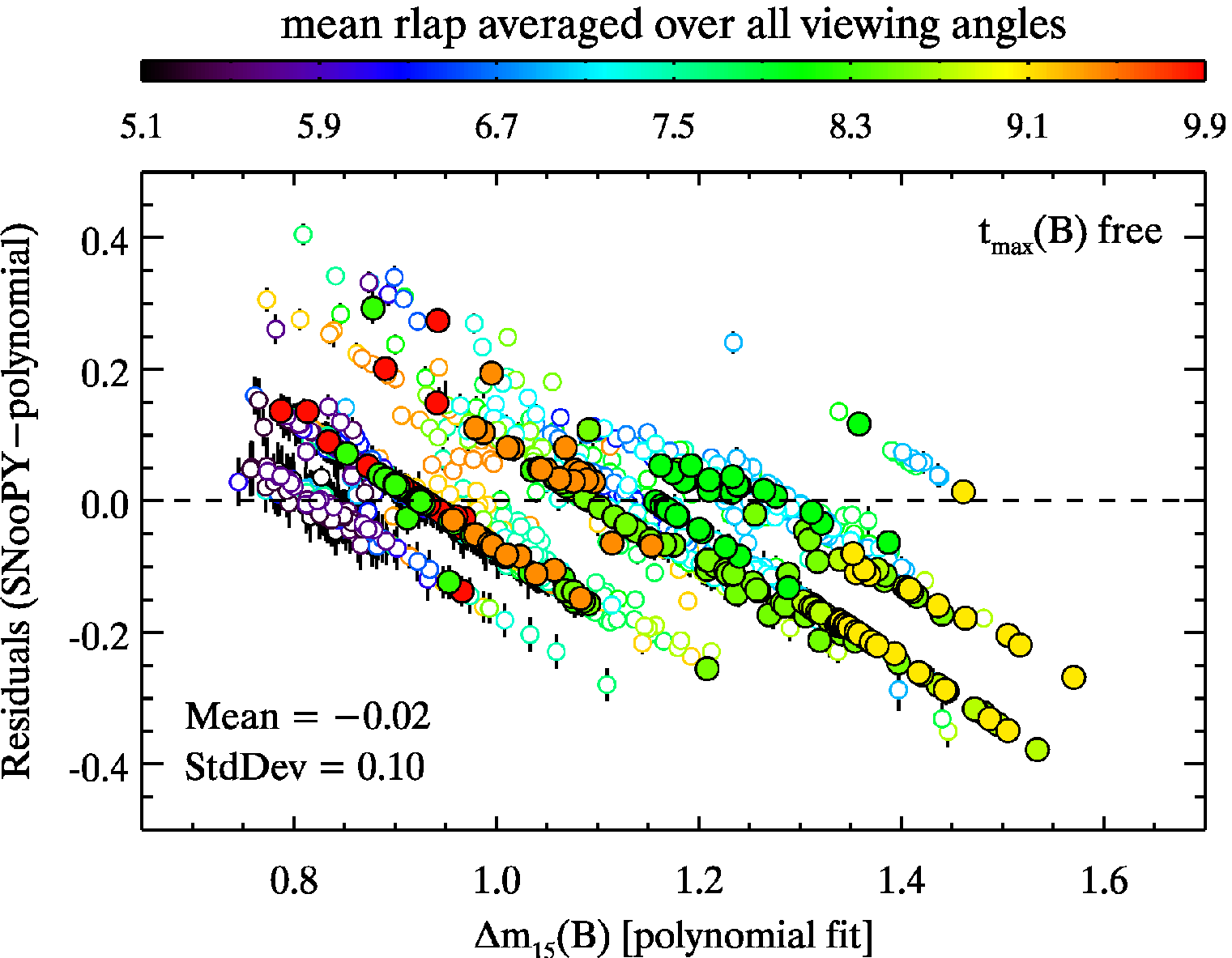}}
\caption{\label{fig:dm15diff}
Difference between \dmft\ inferred from SNooPy fits (where the time of
$B$-band maximum is a free parameter) and that actually
measured from polynomial fits.
The colour-coding is used to distinguish individual models, where
  each model has 30 associated data points (one per viewing angle;
{\it see the discussion of Fig.~\ref{fig:philrel} in \S~\ref{sect:rank}}).
}
\end{figure}

A comparison of light-curve fits for all model/viewing angle
combinations shows that, on average, MLCS2k2 underestimates $t_{\rm
    max}(B)$ by $\sim2$\,d, SALT2 by $\sim0.5$\,d, and SNooPy by
$\sim0.3$\,d.
One might then be tempted to assume that the models  have
an overall shape that is compatible with the SNooPy empirical
templates, but a comparison of \dmft\ values inferred from these fits
with those directly measured on the model light curves reveals a
strong systematic trend (despite the residuals having a mean
consistent with zero; see Fig.~\ref{fig:dm15diff}). The result is the
same when we fix $t_{\rm max}(B)$ to its actual value.

Given the difficulty for empirical templates to correctly reproduce
the overall shape of the model light curves, in what follows we use
the quantities derived from the direct polynomial fits (time of
maximum, peak magnitude, post-maximum decline rate) for the
models. The observed light curves are not always sufficiently well
sampled to enable us to determine the time of maximum and
\dmft\ directly from polynomial fits, in which case we use values
inferred from light-curve fitters. For \sneia\ for which we have both
direct polynomial and template-based fits, the difference in time of
maximum and decline rate from both methods is consistent with
zero.

The various observables derived from the models are intrinsic to the
supernova. Observations, however, are affected by extinction, both in
the Milky Way and the SN host galaxy. The former is usually determined
using the dust maps of \cite{SFD98}, but the latter requires
assumptions about the dust properties and the intrinsic colours of \sneia. 
In the next section we describe the BayeSN statistical model for
\snia\ light curves \citep{Mandel/etal:2011} used to infer the
host-galaxy extinction and hence the intrinsic \snia\ peak magnitudes
and colours.

\subsubsection{Inferring intrinsic \snia\ properties with BayeSN}\label{sect:bayesn}
 
\citet{Mandel/etal:2011} constructed hierarchical Bayesian models for \snia\
light curves spanning optical through near infrared ($BVRIJH$) data.
These statistical models describe the apparent distribution of light
curves as a convolution of intrinsic \snia\ variations and a dust
distribution.   They modeled the intrinsic covariance structure of the
full multi-band light curves, capturing population correlations
between the intrinsic absolute magnitudes, intrinsic colours and light
curve decline rates over multiple phases in the optical and NIR
filters, as well as the distribution of host galaxy dust and an
apparent correlation between the dust extinction $A_V$ and its
wavelength dependence, parameterized by $R_V$.   The models use the
optical and NIR time series data to fit individual \snia\
light curves, estimate their dust extinction and predict their
distances.   Probabilistic inference of the parameters of individual
\sneia\ and also those describing the intrinsic \snia\ and dust
populations are computed using a novel Markov Chain Monte Carlo code.
These models were trained on a nearby ($z < 0.07$) set of \sneia\ 
with optical (CfA3, \citealt{Hicken/etal:2009a}; Carnegie SN Program,
\citealt{Contreras/etal:2010}) and NIR (PAIRITEL,
\citealt{Wood-Vasey/etal:2008}) data, plus light curves from the
literature with joint optical and NIR observations.     The resulting
Markov chain is used to estimate the dust
extinction, apparent and absolute light curves, and intrinsic colours
for each \snia.  For this study, we employed the model that captures a
linear correlation between $R_V^{-1}$ and $A_V$, as this produced the
best cross-validated distance predictions in \citet{Mandel/etal:2011}.


\subsection{Spectral comparison using SNID}\label{sect:snid}

We use the SuperNova IDentification (SNID) code of \cite{SNID} to
cross-correlate the synthetic spectra with a large database of
observed spectra (referred to as ``templates''). SNID is commonly used
to determine the type, redshift, and age of a supernova spectrum 
(e.g. in IAU circulars). The database associated with the
public release of SNID includes 493 spectra of 48 \sneia\ in the age
range $-15 \le t {\rm \ [d]} \le +30$. We have augmented it with new
data from the CfA SN Program (see \S~\ref{sect:data}) and
recently published \snia\ data (SN~2005bl:
\citealt{Taubenberger/etal:2008}; SN~2005cf:
\citealt{Garavini/etal:2007b}, \citealt{WangX/etal:2009a}; SN~2005cg:
\citealt{Quimby/etal:2006}; SN~2005hj: \citealt{Quimby/etal:2007a};
SN~2007ax: \citealt{Kasliwal/etal:2008}). This revised SNID database
now includes 2046 spectra of 274 \sneia\ over the same age range. 
We refer the reader to \cite{SNID} for an extensive discussion of the
SNID algorithm. We point out here that both the input and
template spectra are ``flattened'' through division by a {\it pseudo}
continuum, such that the correlation relies on the {\it relative}
shape and strength of spectral features, and is therefore insensitive
to colour information, which can be intrinsic, or result from
extinction by dust and flux-calibration uncertainties.

We first run SNID on the model spectra closest to
$B$-band maximum light. We only consider \snia\ spectral templates,
and force the redshift (usually a free parameter in SNID) to $z=0$. We
restrict the rest-frame wavelength interval over which the correlation
is done to 3500--7500\,\AA\ to match the wavelength range of most of
the template spectra (see \S~\ref{sect:data}). Last, we restrict
the ages of the template spectra to be within 3\,d from $B$-band
maximum light (relaxing this age constraint does not
  significantly alter the results at maximum light). We are then left 
with 441 spectra of 175 \sneia\ in the SNID database. 

\begin{figure}
\centering
\resizebox{0.475\textwidth}{!}{\includegraphics{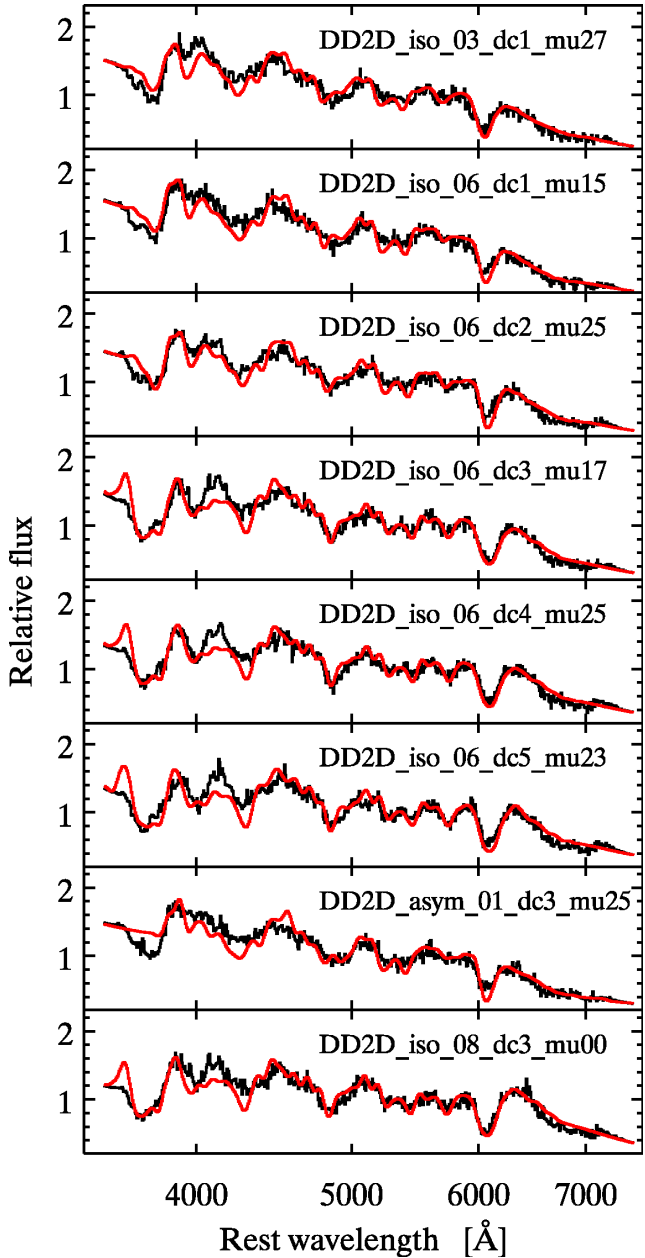}}
\caption{\label{fig:snidtmaxfits}
Results from SNID fits ({\it red}) to the $B$-band maximum-light
spectra ({\it black}) for a subset of eight models, viewed along
the direction which provides the best match to a spectrum in the SNID
database.
}
\end{figure}

The strength of a correlation between the input model spectrum and a
particular template is embodied in the $r$lap parameter, which is the
product of the height of the normalized cross-correlation peak (the
$r$-value of \citealt{TD79}) and the overlap in log-wavelength space
between the input and template spectra (here ${\rm lap}\approx
\ln(7500/3500)\approx0.7$). A correlation is considered good when
$r{\rm lap}\ge5$ \citep[see][]{SNID}. For each model/viewing angle
combination, we record the best-match template (the one with the
highest $r$lap), as well as the mean $r$lap value for the top five
matches, which we consider a more reliable estimate of how similar a
particular model spectrum is to
observations. Fig.~\ref{fig:snidtmaxfits} shows some example SNID fits
to our subset of eight models (see Table~\ref{tab:modelinfo}), viewed
along the direction which provides the best match to an observed
spectrum. The matches are good overall, with some discrepancies around
4000\,\AA\ for some models.

We have also run SNID on model spectra at ages other than maximum
light, namely $-10 \le t {\rm \ [d]} \le +20$, each time imposing a
$\pm3$\,d age constraint on the template spectra. Close to maximum light
($-5 \le t {\rm \ [d]} \le +5$), the fits are of similar quality as
at maximum, i.e. the mean $r$lap value for the cross-correlation
with observed spectra is on average the same. However, further away
from maximum light, the fits are  degraded, and in some cases
no good matches are found. Relaxing the age constraint for the
template spectra results in formally better matches (higher mean
$r$lap), but the mean age of the top five matches tends to over-
(under-) estimate the actual age for model spectra before (after)
maximum light. This again suggests some differences between the rate
at which the synthetic spectra evolve compared to observed spectra.

\subsection{Model selection and the width-luminosity relation}\label{sect:rank}

\begin{figure*}
\centering
\resizebox{\textwidth}{!}{\includegraphics{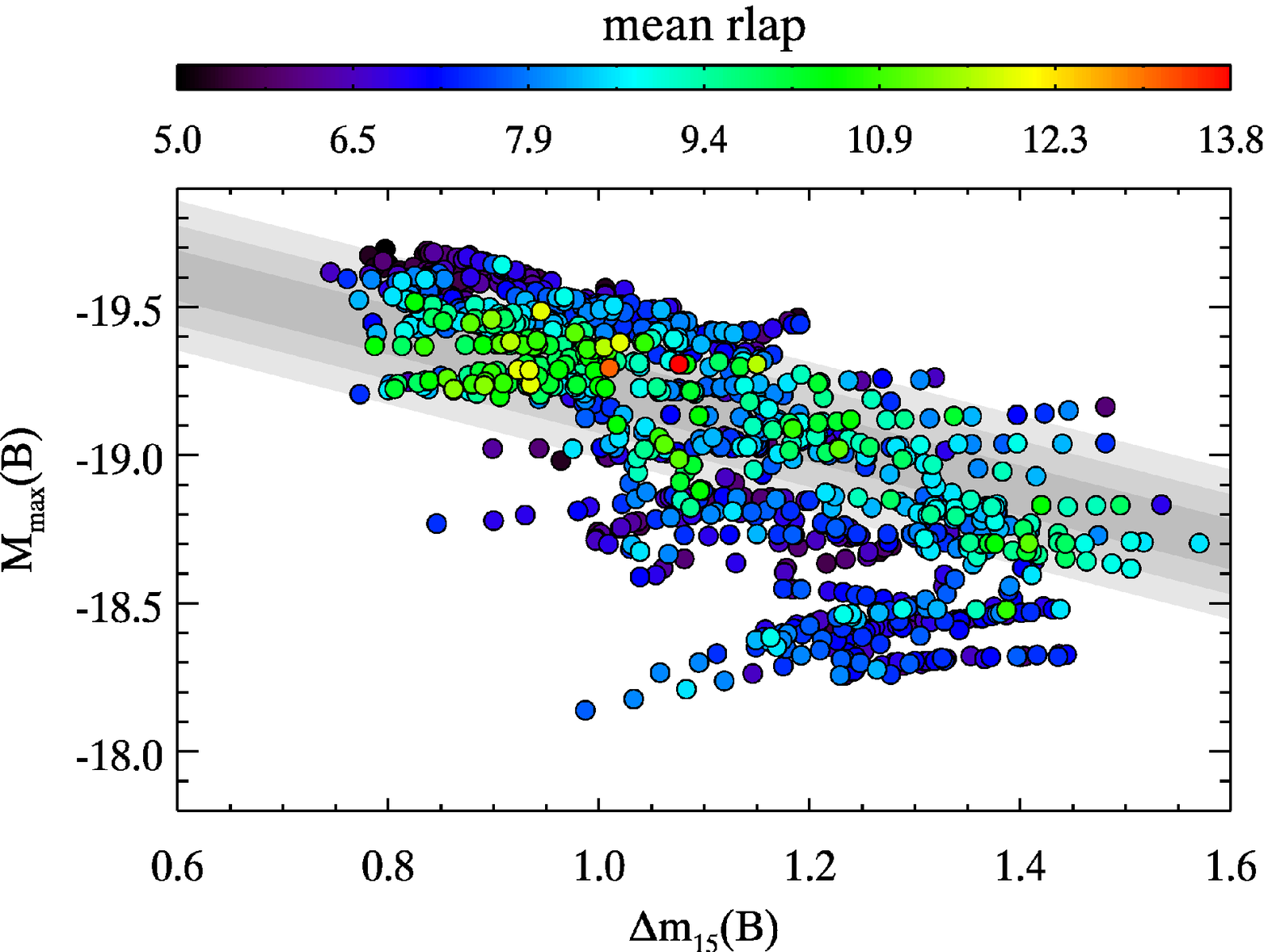}
\hspace{1.5cm}\includegraphics{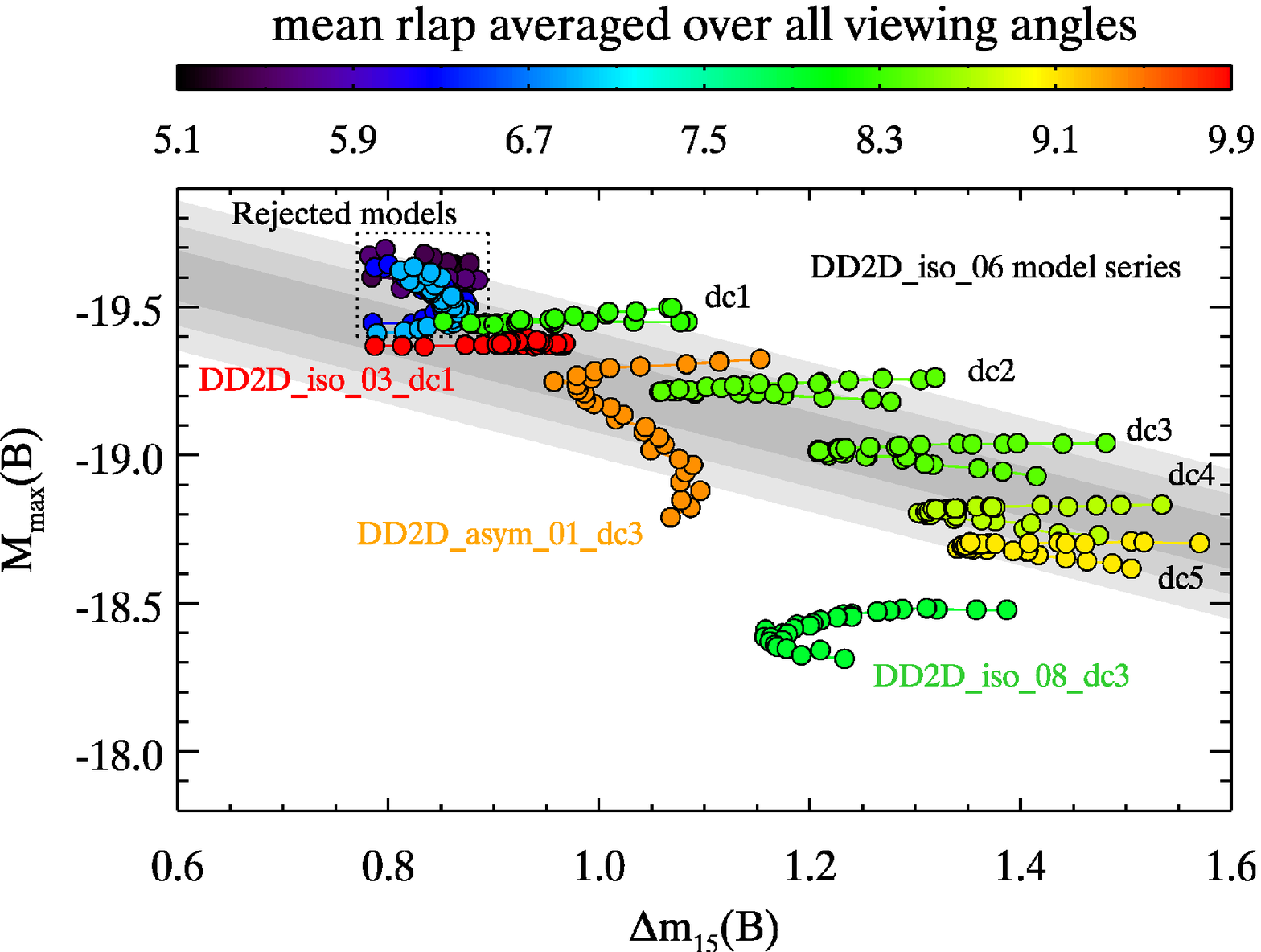}}
\caption{\label{fig:philrel}
{\it Left:} peak absolute $B$-band magnitude {\it vs.} \dmft, for
the 2D delayed-detonation models of KRW09. Each model has 30
associated data points, corresponding to a specific viewing
angle. Each point  is colour-coded 
according to the mean value of the SNID $r$lap parameter for the best
five matches to the maximum-light synthetic spectrum. The grey
contours show the 1--3$\sigma$ levels of the observed width-luminosity
relation of {\protect\cite{Folatelli/etal:2010}} with $\sigma=0.1$\,mag.
{\it Right:} Same as left panel but for a subset of models. The
colour-coding corresponds to the SNID $r$lap parameter averaged over all viewing
angles for any given model. For all models other than
DD2D\_iso\_03\_dc1 and DD2D\_iso\_08\_dc3, $\theta\approx
15^{\circ}$ corresponds to the most luminous point in each model
series.
Models the lie in the
dotted box are rejected based on the lack of any match to a
maximum-light \snia\ spectrum in the SNID database for several viewing
angles. These are: DD2D\_asym\_03 (dc2 and dc3), DD2D\_asym\_06
(dc2 and dc3), and DD2D\_asym\_08 (dc2 and dc3).
}
\end{figure*}

Based on the SNID fits to the maximum-light model spectra, we can
assign a ``grade'', namely the mean $r$lap value for the top five
matches, to each model/viewing angle
combination. Fig.~\ref{fig:philrel} ({\it left}) shows the
width-luminosity relation ($M_{\rm max}(B)$ {\it vs.} \dmft)
for the 2D delayed-detonation models of KRW09 (see their Fig.~3),
where each point has now been colour-coded according to the mean
$r$lap value for that particular model/viewing angle. Points with
higher $r$lap values tend to lie systematically on the observed
width-luminosity relation, i.e. the agreement between synthetic
and observed spectra appears to translate into an agreement in
photometric properties, thereby justifying our ranking scheme
based on maximum-light spectra only (but see \S~\ref{sect:colevol}).

Since each model has equal probability of being observed from one of
the 30 different viewing angles (see \S~\ref{sect:modeldata}), it can
only be validated if {\it all} viewing angles yield a good match to
an observed \snia\ spectrum. Conversely, a lack of a good match for a {\it
  single} 
viewing angle should in principle lead us to reject the model as a
whole, by which we mean it cannot be considered a valid
  approximation of observed \sneia. Six models fall in this category:
DD2D\_asym\_03 (dc2 and dc3), 
DD2D\_asym\_06 (dc2 and dc3), and DD2D\_asym\_08 (dc2 and dc3). As their
names suggest, all are models with an anisotropic
distribution of ignition 
points. Moreover, they all have slow-declining light curves, with
$0.76\le \dmft \le 0.89$. While the DD2D\_asym\_03 models (dc2 and
dc3) have 1-2 viewing angles with no good SNID matches, both
DD2D\_asym\_06 and DD2D\_asym\_08 each have no good matches for more
than half of the viewing angles. 

We have selected a subset of eight models for detailed studies
(Fig.~\ref{fig:philrel}, {\it right}): DD2D\_iso\_03\_dc1 ({\it red})
is the model with the highest $r$lap value overall;
DD2D\_asym\_01\_dc3 ({\it orange}) is the best model
with a highly asymmetric distribution of \nifs\ ($\sim70$\%
variation in peak bolometric luminosity; see
Table~\ref{tab:modelinfo}); the complete  DD2D\_iso\_06 model series
(dc1 through dc5) has consistently high $r$lap values and spans a
large range of \dmft\ (0.8--1.6\,mag), thus enabling us to isolate
the impact of the criterion for deflagration-to-detonation
transition.  It also includes the reference model
DD2D\_iso\_06\_dc2 used by KRW09 for detailed post-processing
nucleosynthesis calculations. These seven models all lie on the 
observed WLR and span a large range in 
synthesized \nifs\ mass (0.42--0.88\,M$_{\sun}$).
Last, DD2D\_iso\_08\_dc3 ({\it green; lower right}) is the best model that
does {\it not} lie on the width-luminosity relation. 
For all models other than DD2D\_iso\_03\_dc1 and
DD2D\_iso\_08\_dc3, $\theta\approx 15^{\circ}$ corresponds to the
direction of largest extent of \nifs\ distribution, and hence to the 
most luminous point in each model series. This is in part a
consequence of the symmetry imposed on the explosion by the 2D
axisymmetric setup in the models of KRW09. These models will
be highlighted and their colour-coding preserved in subsequent
  figures.

In what follows we compare photometric (\S~\ref{sect:phot}) and
spectroscopic (\S~\ref{sect:spec}) properties separately, as is
commonly done in observational studies of \sneia. This in part
reflects the way the observations are done (generally there are
separate follow-up programmes for photometric and spectroscopic
observations) and calibrated (individual photometric measurements
have a typical absolute calibration accurate to a few percent,
whereas individual spectra have a relative flux calibration that is
generally less accurate; see, e.g., \citealt{Matheson/etal:2008}). In
principle one would want to compare light curves and spectra
simultaneously, and in fact we will refer to conclusions drawn in
\S~\ref{sect:phot} when discussing spectroscopic properties in
\S~\ref{sect:spec}.

\section{Comparison of photometric properties}\label{sect:phot}

In this section we focus on multi-band light curves only, by comparing
the rise times, maximum-light colours and colour evolution in models
and data.


\subsection{Rise times}\label{sect:trise}

\begin{figure}
\centering
\resizebox{.475\textwidth}{!}{
\includegraphics{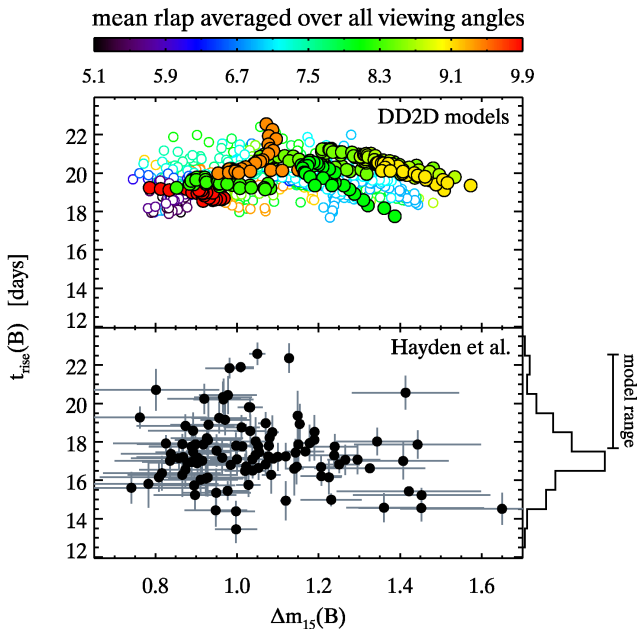}}
\caption{\label{fig:trise}
$B$-band rise time {\it vs.} \dmft, for the 2D
delayed-detonation models of KRW09 ({\it top}) and for \snia\ data from
{\protect\cite{Hayden/etal:2010a}} ({\it bottom}). The colour-coding is
the same as in the right panel of Fig.~\ref{fig:philrel} for the
subset of selected models ({\it filled circles}). The open circles
correspond to all other models. 
The rotated histogram in the lower panel shows the distribution
of the Hayden et al. measurements, while the vertical error bar
corresponds to the range of model rise times.
}
\end{figure}

As noted earlier (\S~\ref{sect:lcfit}), the models 
rise too slowly to maximum light in the $B$ band. We show the
relation between the $B$-band rise time
($t_{\rm rise}(B)$) and \dmft\ in Fig.~\ref{fig:trise}, for the 2D
models of KRW09 ({\it upper panel}) and as measured on
\snia\ light-curves from the SDSS-II Supernova Survey by
\cite{Hayden/etal:2010a} (where we exclude 
\sneia\ with $\dmft\gtrsim 1.6$ to match the \dmft\ range of the
models). In both cases $t_{\rm rise}(B)$ appears to be largely
independent of \dmft.
The range of model rise times (18--23\,d) overlaps with the
observed distribution (13--23\,d), although the bulk of the data
display rise times smaller than 18\,d.
Within a given model, the
rise time changes with viewing angle according to the
distribution of \nifs\ along that particular inclination, anywhere
between $\sim1$\,d and $\sim3$\,d.
For model DD2D\_iso\_06\_dc2, the $B$-band rise time is $\sim20$\,d
regardless of viewing angle, while for model DD2D\_asym\_01\_dc3,
it varies between $\sim20$\,d ($\theta<90^{\circ}$) and $\sim23$\,d
($\theta=165^{\circ}$).

The model $B$-band rise times are more consistent with earlier
measurements (e.g., $19.5\pm0.2$\,d found by 
\citealt{Riess/etal:1999c}), but the latter were based on empirical
templates that were ``stretched'' by equal amounts in the rising
(pre-maximum) and falling (post-maximum) portions of the light
curve. \cite{Strovink:2007} considered asymmetric templates and
found a mean rise time of $17.4\pm0.4$\,d, subsequently confirmed by
\cite{Hayden/etal:2010a} using a larger data set.

\begin{figure}
\centering
\resizebox{.475\textwidth}{!}{
\includegraphics{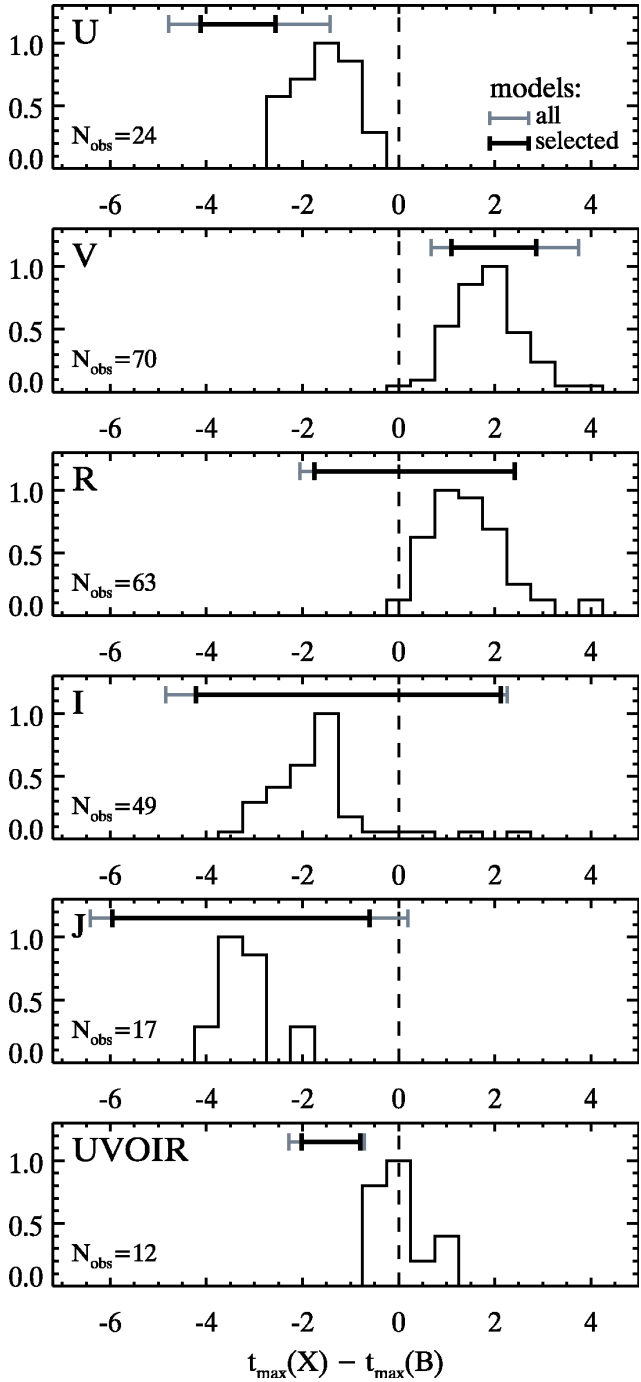}}
\caption{\label{fig:trdiff}
Difference between the time of maximum light in $UVRIJ$ (and in
UVOIR flux) and the time of $B$-band maximum as observed in
actual data ({\it open histograms}), and ranges of the same
quantity for all models ({\it gray error bars}) and our subset of
selected models ({\it black error bars}).
}
\end{figure}

The variation in rise times between various bands is linked to
the redistribution of flux in frequency in the supernova
ejecta. Fig.~\ref{fig:trdiff} shows the difference in time 
of maximum light with respect to the $B$ band, for the $UVRIJ$
photometric bands as well as for the integrated $UBVRI$ (UVOIR)
flux, where we have 
used polynomial fits to observed \snia\ $UBVRIJ$ light curves (again
excluding \sneia\ with $\dmft>1.6$) from the literature for the data
distributions ({\it open histograms}). The $I$-band and NIR light
curves of \sneia\ display characteristic secondary maxima; here we
report the time corresponding to the {\it first} maximum. The times of
UVOIR maximum were derived from light curves published
by \cite{Stritzinger:2005}, where we used only \sneia\ with $UBVRI$
light curves so as to minimize the uncertainty introduced by  
bolometric corrections which account for missing $U$-band data. These
distributions are broadly consistent with those found on a more
limited data set by \cite{Contardo/Leibundgut/Vacca:2000} (their
  Fig.~4). 

The data indicate that \sneia\ peak in the $U$ band typically
$\sim2$\,d earlier than in $B$ (consistent with \cite{Jha/etal:2006}, who
found $t_{\rm max}(U)-t_{\rm max}(B)=-2.3\pm0.4$\,d), and
$\sim1$--2\,d later in $V$ and $R$. The $I$-band has a large
spread in times of (first) maximum, but most \sneia\ peak $\sim2$\,d {\it earlier}
than in $B$. The redder NIR bands peak even earlier (only
$J$ is shown in Fig.~\ref{fig:trdiff}), as previously noted 
by several authors \cite[e.g.,][]{Meikle/Hernandez:2000}.
Where a simplistic model would 
predict the reddest bands to peak the latest as the temperature of the
ejecta decreases (and this  explains the $t_{\rm max}(U)<t_{\rm
  max}(B)<t_{\rm max}(V)$ sequence), the fact that the $IJHK_s$ bands
peak before the $B$ band shows that flux is efficiently redistributed
from the ultraviolet to the near infrared
\citep[e.g.,][]{Pinto/Eastman:2000b}. The magnitude of this effect is
strongly dependent on the ionization stage of iron-group elements in
the \snia\ core, and is responsible for the secondary maxima observed
in $IJHK_s$ \snia\ light curves \citep[see][]{Kasen:2006}.

This trend is broadly reproduced by the
models, with a large overlap with the observations in $VRIJ$, and
a more marginal overlap in $U$ (especially when considering our
subset of selected models). The distribution of the difference between
UVOIR and $B$-band maxima in the models, however, has no
overlap with the observed distribution. The data seem to indicate
that the times of UVOIR and $B$-band maxima coincide \citep[see
  also][]{Contardo/Leibundgut/Vacca:2000}, whereas the
model bolometric light curves peak 1--2\,d earlier than the $B$-band. The
$U$ band contributes most to this (modest) discrepancy, as it is the brightest
band at maximum light in these models.


\subsection{Maximum-light colours}\label{sect:col}

\begin{figure}
\centering
\resizebox{0.475\textwidth}{!}{\includegraphics{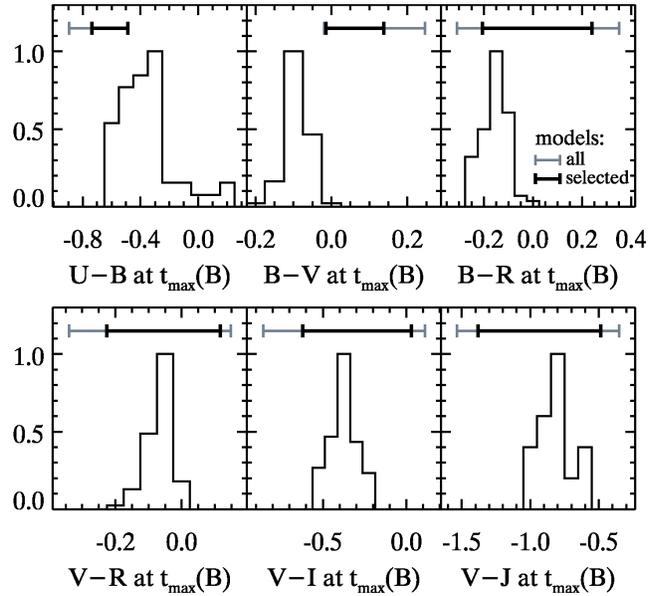}}
\caption{\label{fig:colorcomp}
Comparison of intrinsic rest-frame colours at $B$-band maximum as
inferred from BayeSN fits to actual data ({\it open histograms}) with
those for all models ({\it gray error bars}) and our subset of
selected models ({\it black error bars}).
The $U-B$ colour distribution is derived from MLCS2k2 fits to \sneia\
with low visual extinction ($A_V<0.1$\,mag).
}
\end{figure}

Fig.~\ref{fig:colorcomp} shows
the normalized distributions of intrinsic $U-B$, $B-V$, $B-R$, $V-R$,
$V-I$, and $V-J$ colours at $B$-band maximum as 
inferred from BayeSN fits to actual data ({\it
 open histograms}) compared with the ranges for the 2D models of
KRW09 ({\it gray}: all models; {\it black}: subset of selected
models). The data indicate that \sneia\ have an intrinsic
$B-V\approx -0.1$\,mag at $B$-band maximum, whereas the models
have a systematically redder $B-V$ colour (0.0--0.2\,mag). This
was already noted by 
KRW09 and attributed to the approximate treatment of non-LTE
effects. The distributions for $B-R$, $V-R$, and $V-I$ have a large
overlap between the models and data, but contrary to the $B-V$ colour
the former have a much larger spread (this is also true when only
considering the subset of selected models). The $V-J$ colour displays
differences of up to $\sim0.5$\,mag
between the models and the data, albeit with a large overlap with the
observed distribution. 

The models have a systematically redder $B-V$ colour at $B$-band
maximum, but they are systematically {\it bluer} in $U-B$ colour
(Fig.~\ref{fig:colorcomp}). \cite{Mandel/etal:2011} did not include
$U$-band data in their analysis to avoid calibration uncertainties, so we
instead infer the intrinsic $U$-band magnitude based on MLCS2k2 fits
to \sneia\ with low visual extinction
($A_V<0.1$\,mag). The models show some overlap in $U-B$ colour
with the data, but extend $\gtrsim0.2$\,mag further in the blue.
This excess of $U$-band flux can be explained either by
a hotter ejecta temperature affecting the overall SED or an incomplete
description of line-blanketing of UV photons and subsequent
redistribution of flux to redder wavelengths, or a combination of
both. Since the impact of such effects is most easily seen on the
spectra, we will discuss them in \S~\ref{sect:spec}.

\begin{figure*}
\centering
\includegraphics[width=7cm]{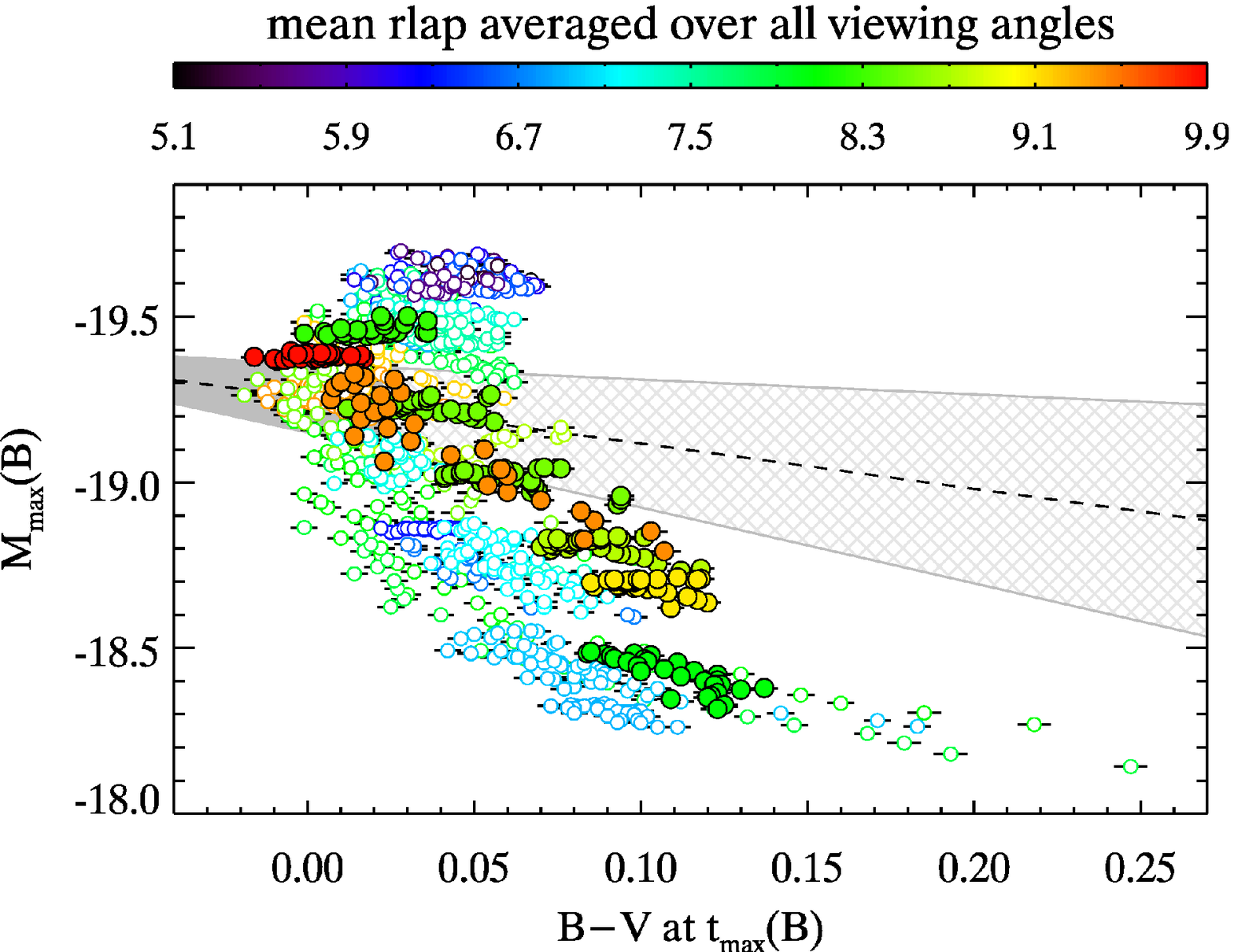}\hspace{.5cm}
\includegraphics[width=7cm]{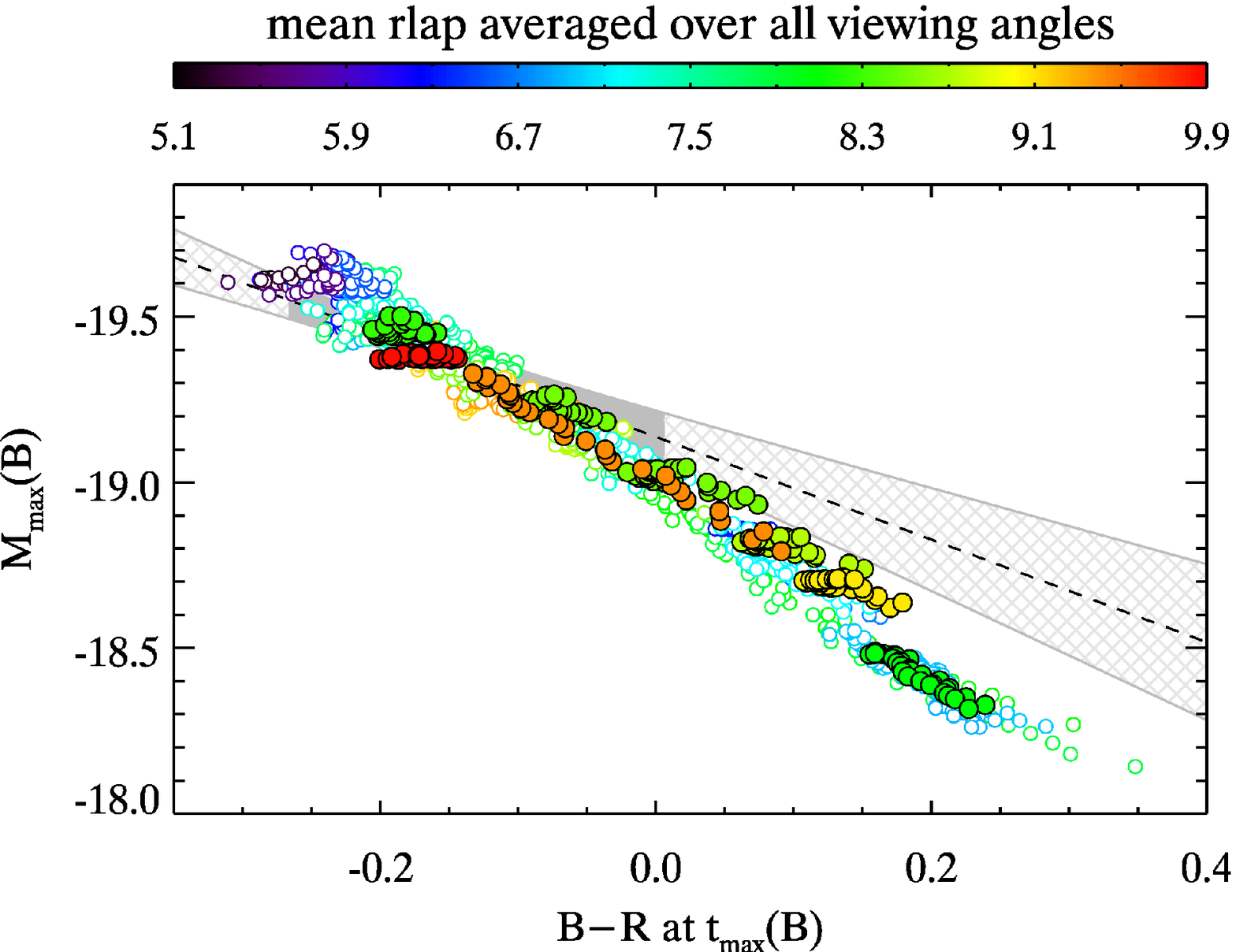}\vspace{.5cm}
\includegraphics[width=7cm]{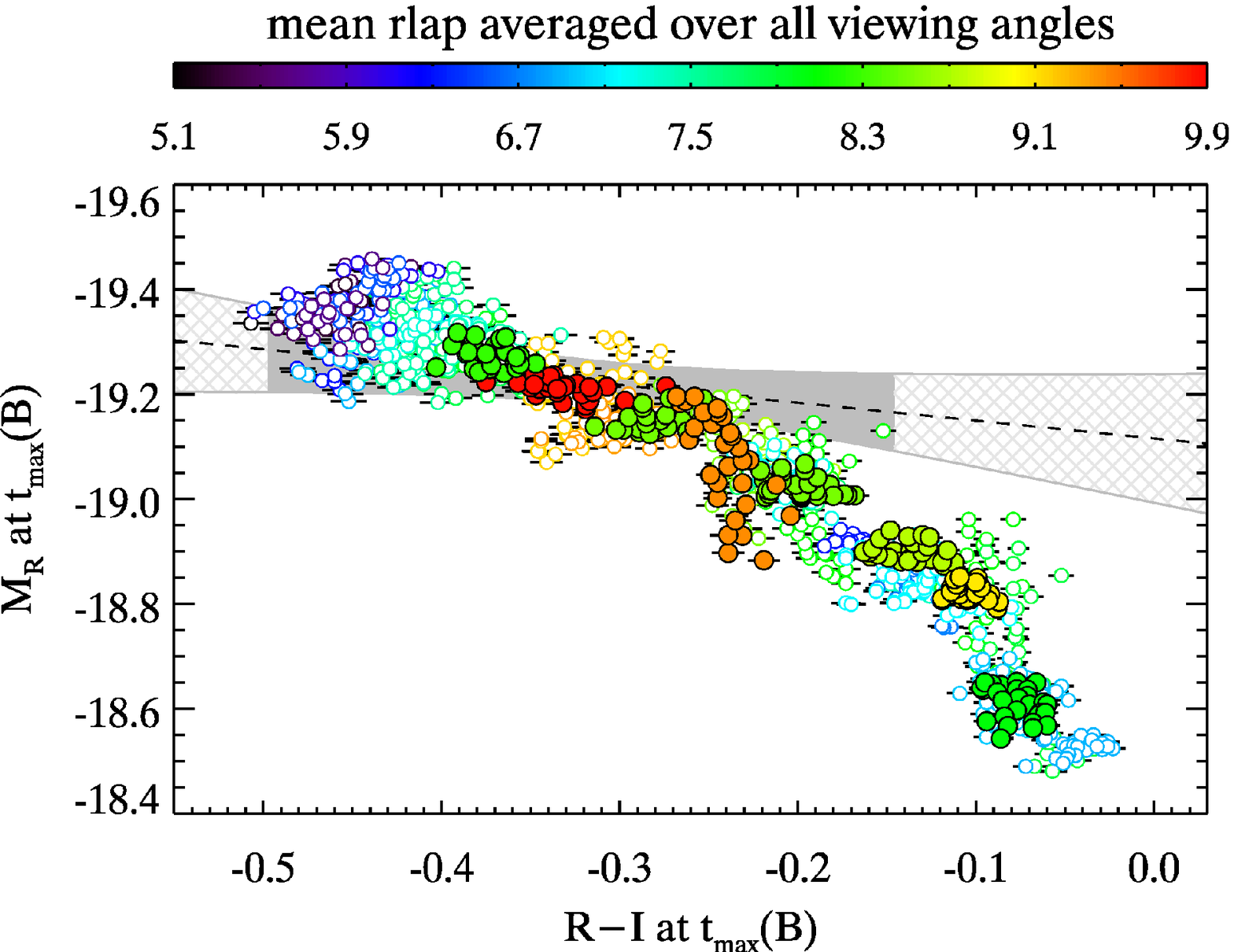}\hspace{.5cm}
\includegraphics[width=7cm]{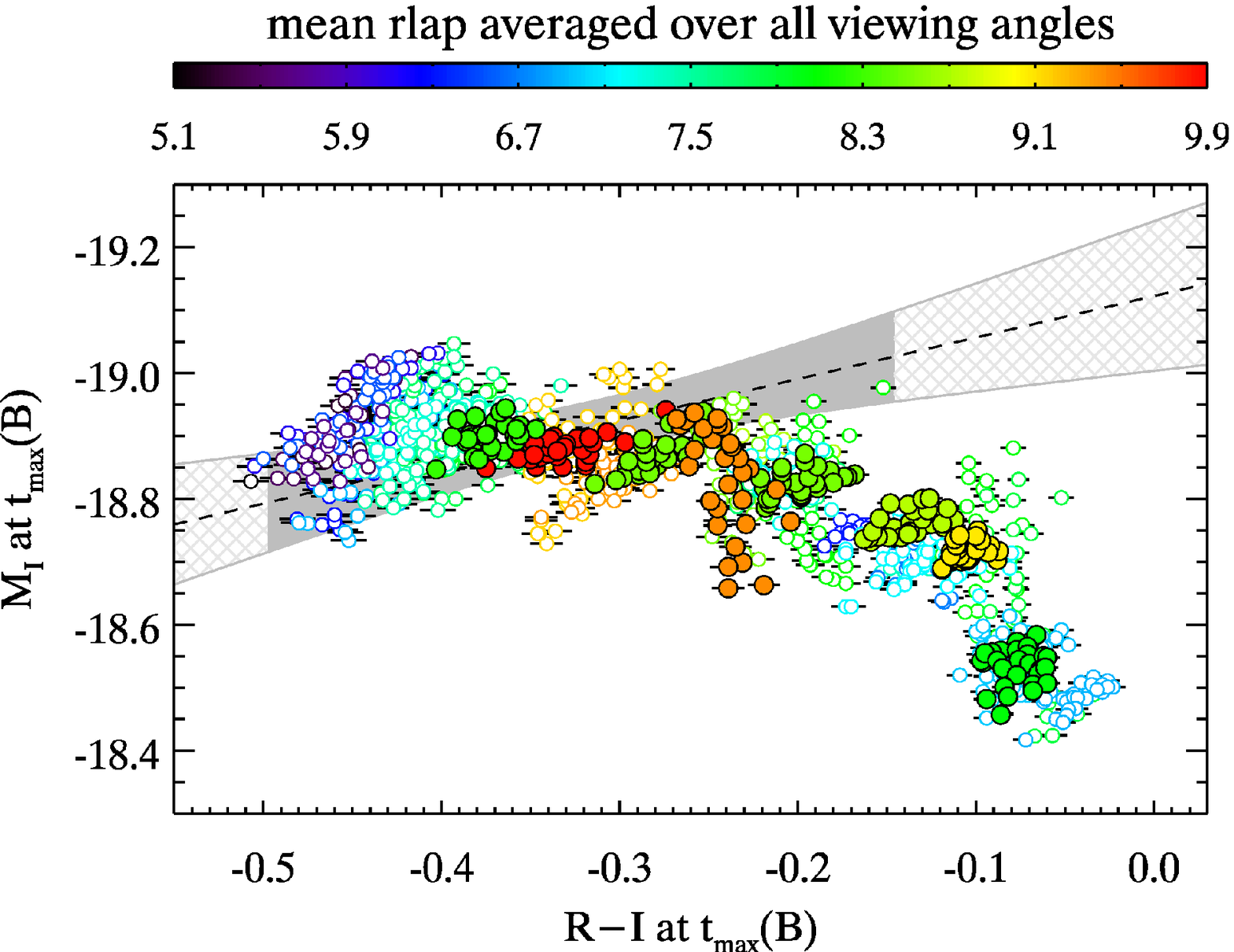}
\caption{\label{fig:bmaxcolors}
Intrinsic magnitudes {\it vs.} intrinsic colours at $B$-band
maximum. The colour-coding is the same as in
Fig.~\ref{fig:trdiff}. The grey contour corresponds to the 95\% 
confidence level of the observed relation inferred from BayeSN fits
to actual data ({\it dashed line}). The grey shaded area
highlights the colour range in the data used to determine the
observed relation, while the grey hatched area corresponds to
extrapolations of this relation.
}
\end{figure*}

More luminous \sneia\ are also intrinsically bluer (this is known as
the ``brighter-bluer'' relation, e.g. \citealt{Tripp:1998}). Its
degeneracy with extinction by dust (\sneia\ obscured by dust will
appear redder) is exploited by some light-curve fitters to describe
the effects of extinction and intrinsic colour variations using a
single parameter (e.g., SALT2: \citealt{SALT2}). 
The brighter-bluer relation (i.e. $M_{\rm max}(B)$ {\it vs}. intrinsic
$B-V$ at $B$-band maximum) derived 
from BayeSN fits to actual data is shown in the
upper-left panel of Fig.~\ref{fig:bmaxcolors} ({\it dashed line}). The
2D models of KRW09 also follow a 
brighter-bluer relation, but it is significantly steeper than
observed. This difference is not surprising given the lack of
overlap in $B-V$ colour between the models and the data
(Fig.~\ref{fig:colorcomp}).

The agreement between models and data is significantly better when
considering $M_{\rm max}(B)$ {\it vs.} $B-R$ at $B$-band maximum
(Fig.~\ref{fig:bmaxcolors}, {\it upper-right panel}), which forms an
extremely tight relation. Small deviations are apparent for
$B-R\gtrsim0$\,mag, where the models predict peak colours that
are much redder than inferred from observations
(Fig.~\ref{fig:colorcomp}). 
Of all the selected models, model 
DD2D\_asym\_01\_dc3 ({\it filled orange circles}) displays the largest
variation of $B-R$ colour (the same is true to a lesser extent when
considering the $B-V$ colour), illustrating the greater change of the
overall SED with viewing angle stemming from the asymmetric \nifs\
distribution (see Fig.~\ref{fig:dens}).

The lower panels of Fig.~\ref{fig:bmaxcolors}
show $M_R$ and $M_I$ {\it vs.} intrinsic $R-I$ colour (all quantities
are at $B$-band maximum). While $M_R$ has almost no relation to $R-I$
colour (the slope is consistent with zero at $\sim1\sigma$), $M_I$ {\it vs.}
$R-I$ appears to define a brighter-{\it redder}
relation\footnote{While this is not the subject of this paper, such a
relation would help break the degeneracy between extinction by dust
and intrinsic colour variations.}. In both cases, models that
predict redder $R-I$ colours than inferred from observations lie off
the relation.


\subsection{Colour evolution}\label{sect:colevol}

The early-time evolution of \snia\ colours results from a complex
interplay between a gradual cooling of the ejecta and changes
in ionization stages of several species, leading to order-of-magnitude
changes in the opacity at selected frequencies. In particular, the
{\sc iii}$\rightarrow${\sc ii} recombination of iron-group
elements is responsible for most of the
redistribution of flux from blue to red wavelengths and its timescale
controls the rate of colour evolution in
\sneia\ \citep[see][]{Kasen/Woosley:2007}.

\begin{figure*}
\centering
\resizebox{.8\textwidth}{!}{\includegraphics{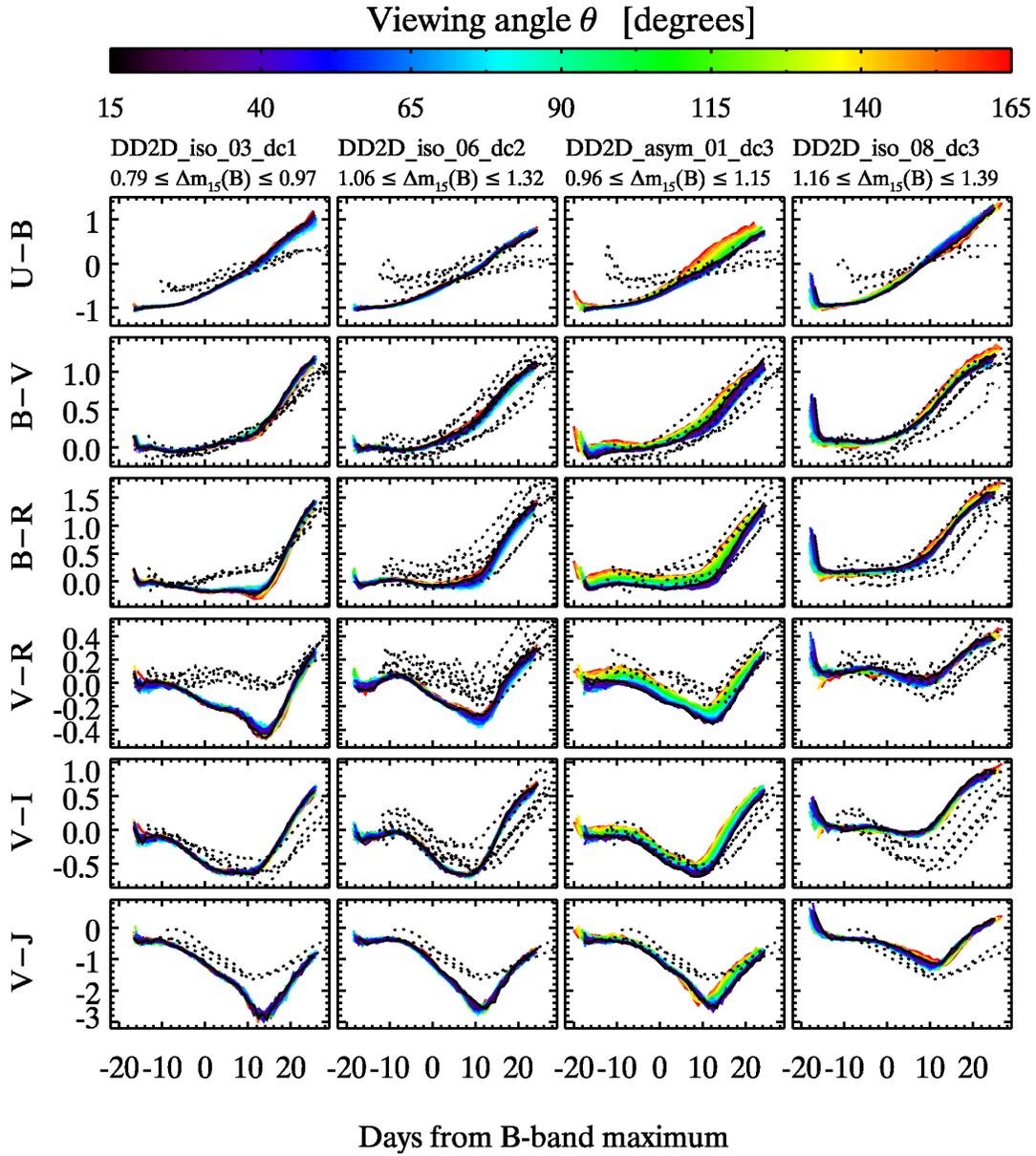}}
\caption{\label{fig:colevol}
Time evolution of intrinsic colours measured at
$B$-band maximum for four of our selected models, colour-coded
according to the viewing angle $\theta$. The dotted 
lines for $U-B$, $B-V$, $B-R$, $V-R$, and $V-I$ correspond to
dereddened colour curves from observed \sneia\ with low visual
extinction ($A_V<0.25$\,mag as inferred from BayeSN fits) in similar \dmft\
ranges. The dotted lines for $V-J$ correspond to templates for slow-
($0.8\lesssim \dmft \lesssim 1.0$) and mid-range ($1.0\lesssim \dmft
\lesssim 1.4$) decliners, as well as for SN~2001el
($\dmft\approx1.1$), from {\protect\cite{Krisciunas/etal:2004b}}. 
}
\end{figure*}

Fig.~\ref{fig:colevol} shows the time evolution of $U-B$, $B-V$, $B-R$,
$V-R$, $V-I$ and $V-J$ colours for our subset of
selected models (we have chosen one representative model for the
DD2D\_iso\_06 series, namely DD2D\_iso\_06\_dc2), colour-coded
according to the viewing angle $\theta$. Overplotted are observed
(dereddened) colour curves from \sneia\ in similar \dmft\
ranges, except for the $V-J$ colour curves where we show the
\cite{Krisciunas/etal:2004b} templates for slow- ($0.8\lesssim \dmft
\lesssim 1.0$) and mid-range ($1.0\lesssim \dmft \lesssim 1.4$)
decliners, as well as for SN~2001el ($\dmft\approx1.1$). To minimize
the uncertainties associated with extinction corrections, we only
include \sneia\ with a visual extinction $A_V<0.25$\,mag, as inferred
from BayeSN fits.

The $B-V$ colour evolution of the models is largely consistent with
that seen in the data, with small systematic offsets to redder colours
for model DD2D\_iso\_08\_dc3, which lies off the width-luminosity
relation. The agreement between the models and the data for the $B-R$
colour evolution is also satisfactory, apart for model
DD2D\_iso\_03\_dc1, whose $B-R$ colour becomes progressively bluer
with age until $\sim10$\,d past $B$-band maximum while the data become
redder over the same time interval. The disagreement is exacerbated
when considering $V-R$ for this model, but the $V-I$ colour evolution
is in good agreement (as is true for the other models of
Fig.~\ref{fig:colevol}). Interestingly, DD2D\_iso\_03\_dc1 is our best
overall model based on cross-correlations with observed spectra using
SNID (see \S~\ref{sect:rank}). Since SNID compares the {\it
  relative} shapes and strengths of spectral features, the overall SED
was not taken into account when ranking these models, which
allows for possibly large offsets in broadband colours between our
selected models and the data. Based
on the $V-R$ colour evolution, one would rank model DD2D\_iso\_08\_dc3
highest of the four models shown in Fig.~\ref{fig:colevol}, where in
fact it is the only model that does {\it not} lie on the observed
width-luminosity relation. As noted earlier, model
DD2D\_asym\_01\_dc3 displays the largest colour variation at any given
time.

The colour curves for $U-B$ and $V-J$, however, reveal the largest
inconsistencies between the models and the data. All the models have a
systematically bluer $U-B$ colour at early times (up until $\sim10$\,d
past $B$-band maximum), and increasingly redder $U-B$ colours at later
times. By 20\,d past $B$-band maximum, the models display a $U-B$
colour that is $\sim0.5$\,mag too red with respect to the data, a
difference that clearly cannot be explained by simple $U$-band
calibration uncertainties. 

The qualitative behaviour of the $V-J$ colour curves is
compatible with the data, but displays large offsets ($\gtrsim1$\,mag
for model DD2D\_iso\_03\_dc1) around $\sim10$\,d past $B$-band
maximum, which corresponds to the time of flux minimum between the
two $J$-band maxima (see Fig.~\ref{fig:lceg}). This is not the case
for model DD2D\_iso\_08\_dc3, which has {\it redder} $V-J$ colours at
any given time when compared to observations. The failure of the models
to properly reproduce the NIR flux in \sneia\ at all times could be due in
part to the atomic data used in the radiative transfer
calculations, as noted by \cite{Kasen:2006}. By comparing $J$-band
light curves computed with the 500,000 line Kurucz CD 23 and the 42
million line Kurucz CD 1, he showed that including more lines resulted
in $\sim1$\,mag more $J$-band flux between $-10$ and $+20$ days from
$B$-band maximum and a shallower flux minimum (his Fig.~5),
although it is not clear which lines contribute most to this
effect. Other approximations used in the radiative transfer
calculations (LTE, expansion opacity formalism, equivalent two-level
atom for flux redistribution etc.) could also explain part of these
discrepancies.

Fig.~\ref{fig:colevol} conveys the message that a direct comparison of
\snia\ models with observations must involve both photometric and
spectroscopic properties. Our highest-ranked model based on
cross-correlations with spectral templates using SNID is
DD2D\_iso\_03\_dc1, yet its colour curves present the largest level of
disagreement with observations. We also note that DD2D\_iso\_08\_dc3
fares best based on the evolution of its broadband colours, yet it
lies off the observed width-luminosity relation.


\section{Comparison of spectroscopic properties}\label{sect:spec}

It is difficult to resolve discrepancies between the models and data
based on photometric measurements alone, and we now turn to the
comparison of spectroscopic properties, first considering spectra at
maximum light (\S~\ref{sect:spectmax}), and then their evolution
with time (\S~\ref{sect:specevol}).

\subsection{Maximum-light spectra}\label{sect:spectmax}

\subsubsection{Spectroscopic diversity at a given \dmft}

\begin{figure*}
\centering
\resizebox{\textwidth}{!}{\includegraphics{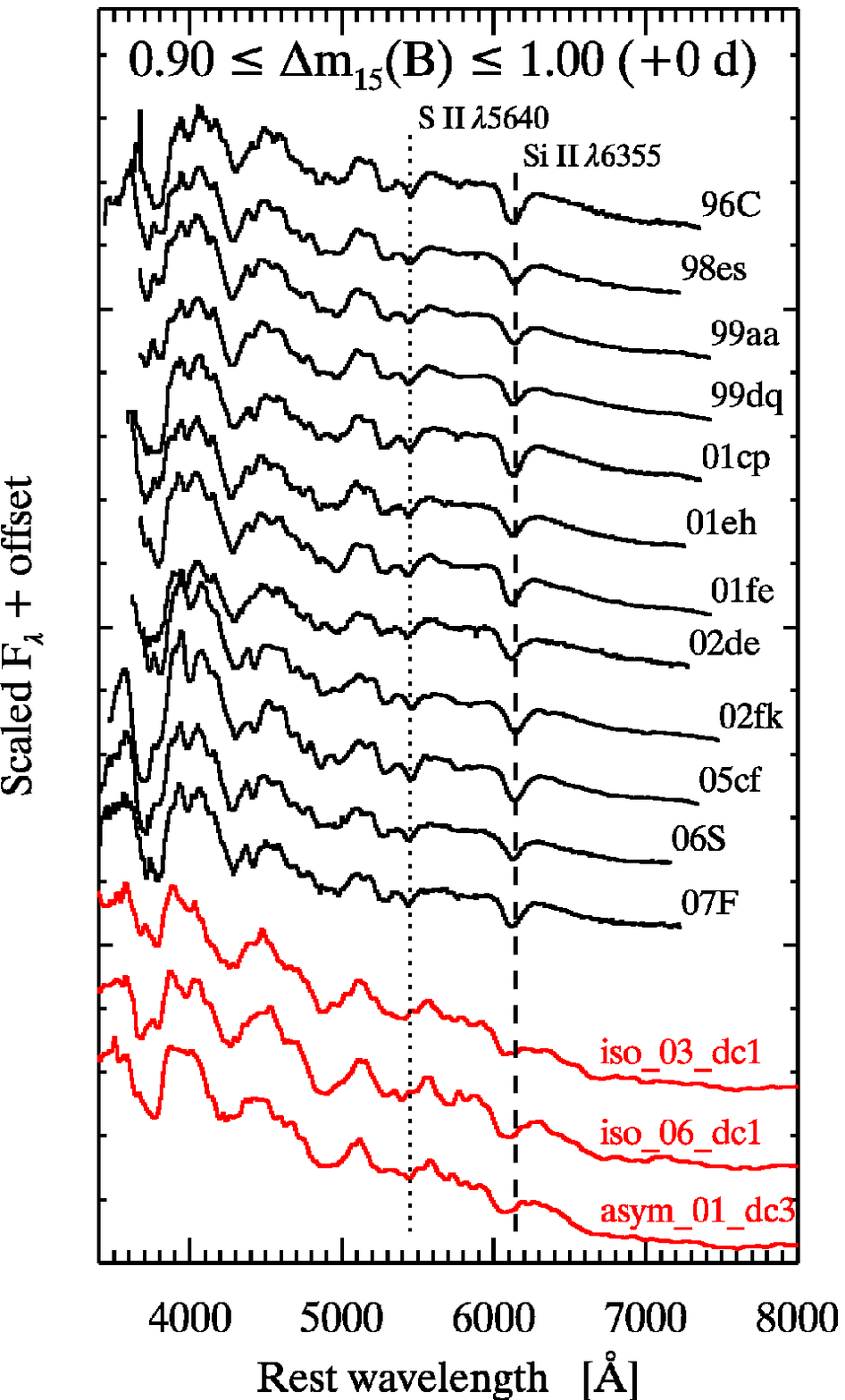}
\hspace{1.5cm}\includegraphics{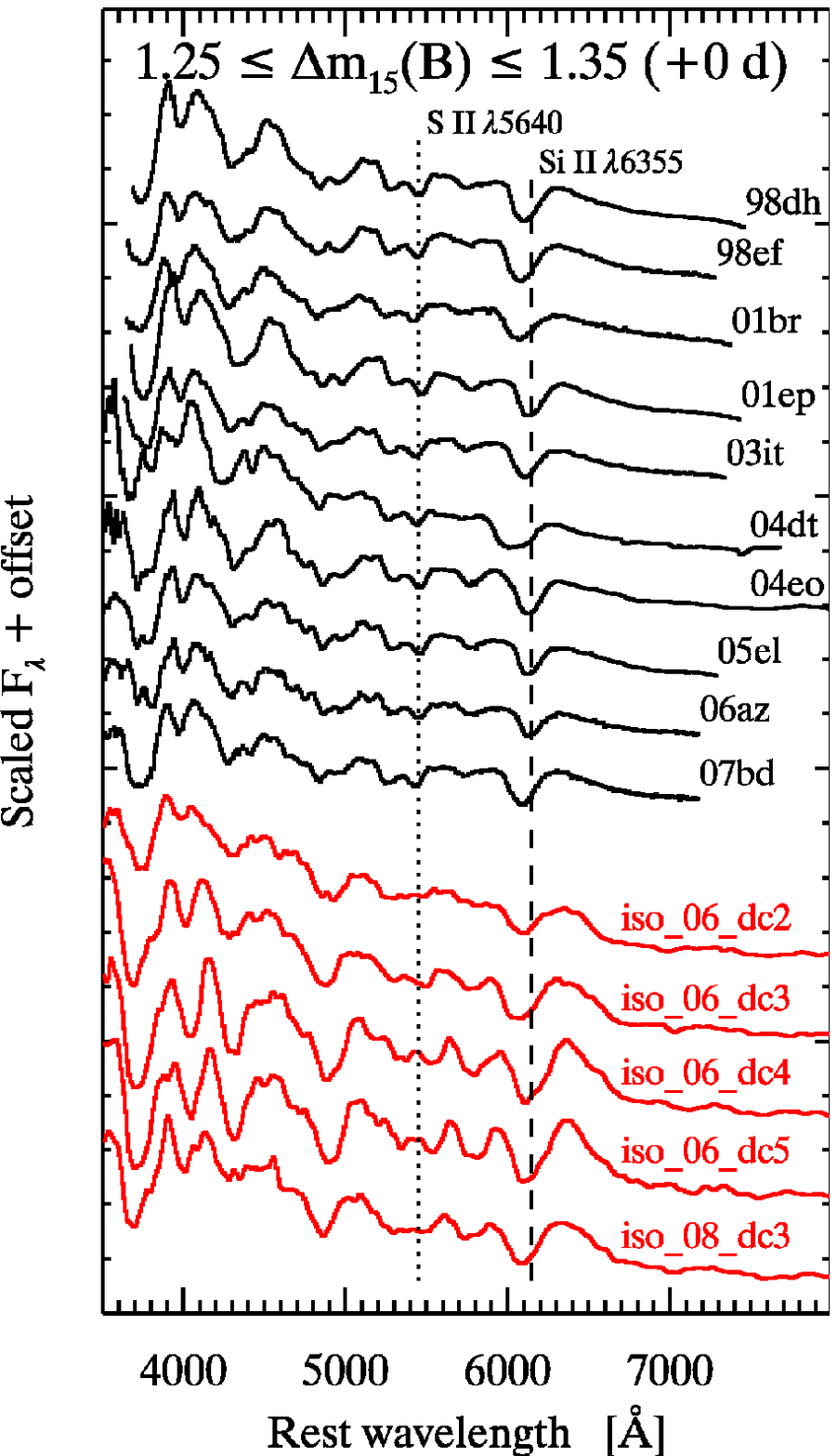}}
\caption{\label{fig:specvardm15_data}
Maximum-light spectra for $0.90\le\dmft\le1.00$ ({\it left})
and $1.25\le\dmft\le1.35$ ({\it right}) compared to models in the same
\dmft\ range. For sake of clarity, we only show spectra
corresponding to our subset of selected models, and select
the viewing angle corresponding to the decline rate closest to the
middle of the \dmft\ range considered. The vertical lines
  correspond to S\two\,\l5640 ({\it dotted}) and Si\two\,\l6355 ({\it
    dashed}) blueshifted by 10000\,\kms.
}
\end{figure*}

We would like to compare the general appearance of the model spectra
in specific \dmft\ ranges. To do this we select two narrow \dmft\
ranges ($\pm0.05$ around a central value) that include at least one
viewing angle for each of the eight models in our subset. Based on
Fig.~\ref{fig:philrel}, we see that three of our subset of selected
models overlap at $\dmft\approx0.95$ (DD2D\_iso\_03\_dc1,
DD2D\_iso\_06\_dc1, and DD2D\_asym\_01\_dc3), while the remaining five
overlap at $\dmft\approx1.3$ (DD2D\_iso\_06\_dc2 through dc5 and
DD2D\_iso\_08\_dc3). We thus define our \dmft\ ranges of interest as
0.90--1.00 and 1.25--1.35. Figure~\ref{fig:specvardm15_data} shows
observed maximum-light spectra in both \dmft\ ranges ({\it black})
compared to our subset of selected models ({\it red}), for which we only
show spectra for one viewing angle (the one that yields a \dmft\
closest to the middle of the \dmft\ range considered). The observed
spectra all have an estimated visual extinction $A_V<0.25$\,mag
and have been dereddened.

For $0.90\le\dmft\le1.00$, the observed sample consists of both
``normal'' and 1991T-like \sneia\ (of which SN~1998es, SN~1999aa,
SN~1999dq; all from \citealt{Matheson/etal:2008}). The latter are
characterized by hot ejecta, with evidence for lines
of doubly-ionized iron-group elements (Fe\three/Co\three) in place of lines of
Fe\two/Co\two\ in maximum-light spectra of ``normal'' \sneia, and
which explains part of the observed 
variation around $\sim4300$\,\AA. Redward of this, the spectra are
remarkably similar to one another, albeit with variations in the
strength of the absorption component of the prominent Si\two\,\l6355
line (1991T-like \sneia\ have shallower Si\two\ absorptions). The model
spectra in the same \dmft\ range are also similar to one another, but
they have a much bluer (hotter) SED than the data (as noted based on
the comparison of $U-B$ colours in \S~\ref{sect:col}), and shallower
Si\two\,\l6355 absorptions. This line is also broader and its
absorption component more blueshifted than observed in the data, both
evidence for larger ejecta velocities. Several small-scale features
present in the data are thus ``washed out'' in the models, as the
higher expansion velocities increase the line overlap. This is clearly
seen in the S\two\,\l\l5454,5640 doublet, which appears as two distinct
absorption components in all the observations but which is barely
resolved in the models (and not resolved at all for model
DD2D\_iso\_03\_dc1).

For $1.25\le\dmft\le1.35$, the observed sample consists almost
exclusively of ``normal'' \sneia, apart from SN~2004dt which displays
larger absorption velocities (clearly visible in the Si\two\,\l6355
line) and was found by \cite{WangL/etal:2006} to have the highest
degree of polarization ever measured in a \snia. SN~2004eo was labeled
``transitional'' by \cite{Pastorello/etal:2007b} due to its
intermediate properties between normal and underluminous \sneia. The
model spectra in the same \dmft\ range show a large degree of
heterogeneity, both in the slope of the overall SED and in the
relative shapes and strengths of spectral features. Model
DD2D\_iso\_06\_dc2 is too blue compared to observations, and displays
relatively shallow absorption features; conversely, models
DD2D\_iso\_06\_dc4 and dc5 have a significantly redder SED with broader
and deeper absorption features than in the observations, as clearly
seen from the three prominent Si\two\ absorption features at
$\sim4000$\,\AA, $\sim5800$\,\AA, and $\sim6100$\,\AA. Model
DD2D\_iso\_08\_dc3 fares better both in terms of SED slope and relative
strengths of spectral features, albeit with less prominent absorption
around $\sim4300$\,\AA\ and lack of small-scale structure around
$\sim4800$\,\AA\ (both features are predominantly due to lines of
Fe\two/Co\two, with important contributions from Mg\two\ and Ti\two). As was
the case for the model spectra with smaller \dmft\ 
(Fig.~\ref{fig:specvardm15_data}, {\it left panel}), the 
S\two\,\l\l5454,5640 doublet is not resolved in this model (the same is
also true for DD2D\_iso\_06\_dc2).

\begin{figure*}
\centering
\resizebox{0.8\textwidth}{!}{
\includegraphics{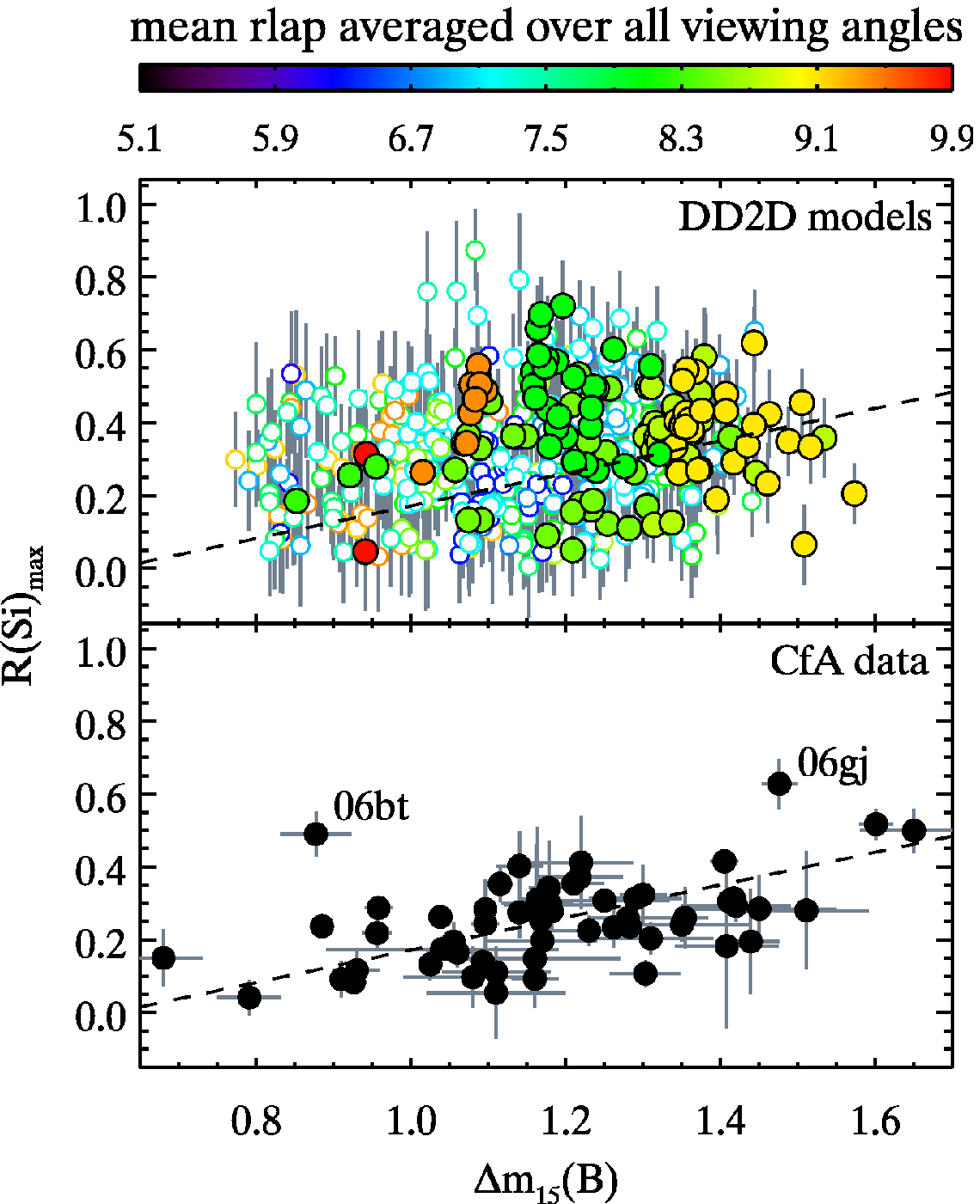}\hspace{2cm}
\includegraphics{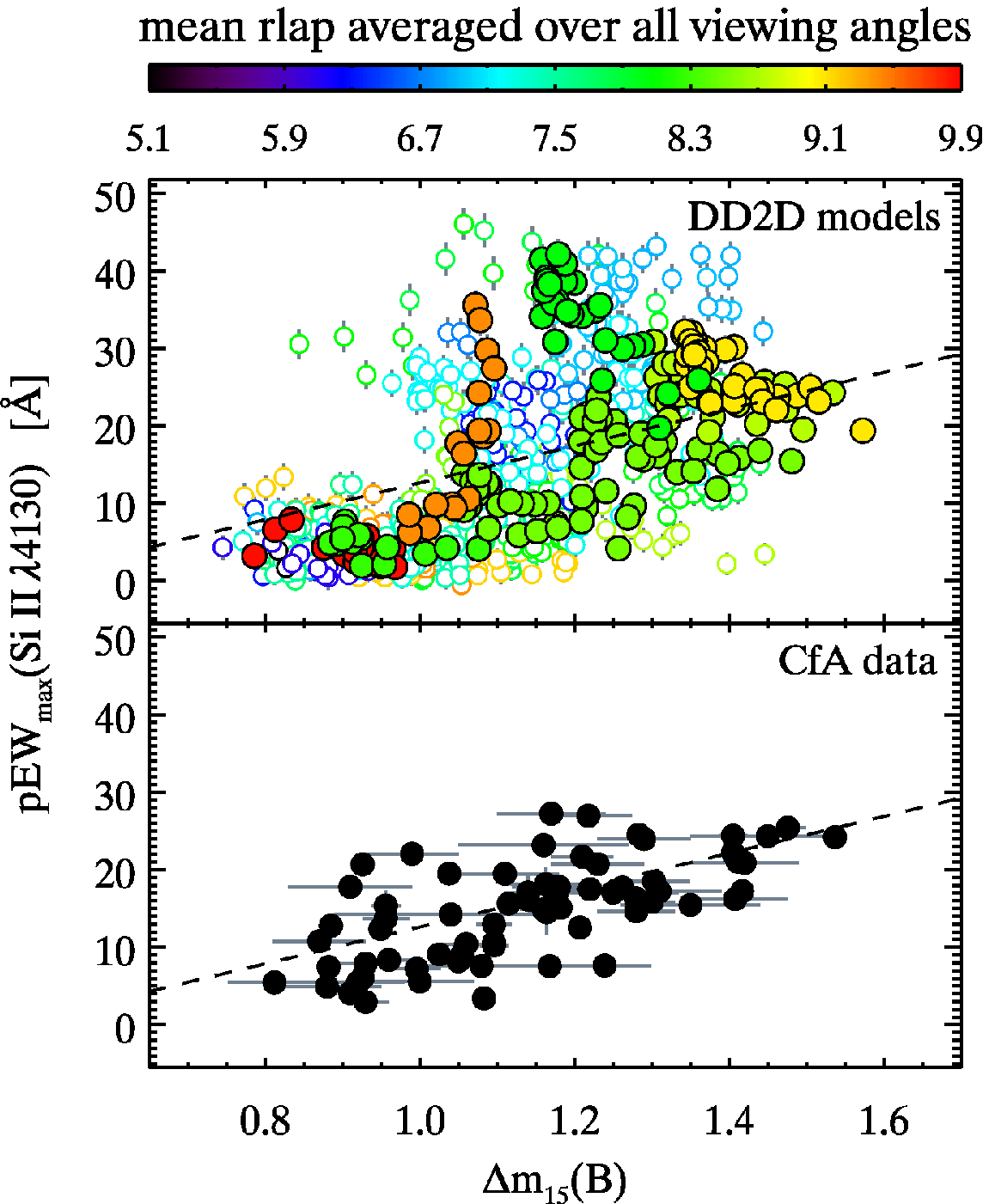}
}
\caption{\label{fig:rsi}
{\it Left:}
Spectroscopic indicator \rsi\ {\it vs.} \dmft, for the 2D
delayed-detonation models of KRW09 ({\it top}) and for \snia\ data
from the CfA SN Program ({\it bottom}). The colour-coding is the same
as in the right panel of Fig.~\ref{fig:philrel}. The dashed line is a
linear fit to the 
CfA data, and is overplotted in the upper panel. The points corresponding
to the outliers SN~2006bt and SN~2006gj are labeled accordingly.
{\it Right:} pseudo-EW of the Si\two\,\l4130 line {\it vs.} \dmft.
}
\end{figure*}

The partial failure of the models to reproduce the relative shapes and
strength of the Si\two\ lines impacts their ability to reproduce the
observed correlation of several spectroscopic indicators with
\dmft\ \citep[see, e.g.,][]{Blondin/Mandel/Kirshner:2011}. 
Fig.~\ref{fig:rsi} shows two 
such indicators, namely the \rsi\ ratio of \cite{Nugent/etal:1995}
(defined as the ratio of the relative absorption depth of the
Si\two\,\l5972 line to that of Si\two\,\l6355) and the {\it pseudo}
equivalent width (pEW) of Si\two\,\l4130 
\citep[see][]{Arsenijevic/etal:2008,Walker/etal:2010,Blondin/Mandel/Kirshner:2011,Chotard/etal:2011}. 
There is a clear correlation
between both indicators and \dmft\ in our spectroscopic data
set ({\it lower panels}), albeit with two significant outliers for
\rsi\ (one of which SN~2006bt, whose peculiar nature has been discussed at
length by \citealt{SN2006bt}). A linear fit to the data is shown in
both cases ({\it dashed line}) and overplotted on measurements from
the models ({\it upper panels}). There is no clear correlation between
\rsi\ and \dmft\ in the models, but the measurements have a large
associated error due to Monte Carlo noise affecting the weaker
Si\two\,\l5972 line. As for Si\two\,\l4130, there is  a general
trend of increasing pEW with \dmft, but with a large scatter for
$\dmft\gtrsim1.1$ caused by some model spectra with unusually broad and
deep absorptions.

\subsubsection{Composite spectra at a given \dmft}

\begin{figure*}
\centering
\resizebox{\textwidth}{!}{\includegraphics{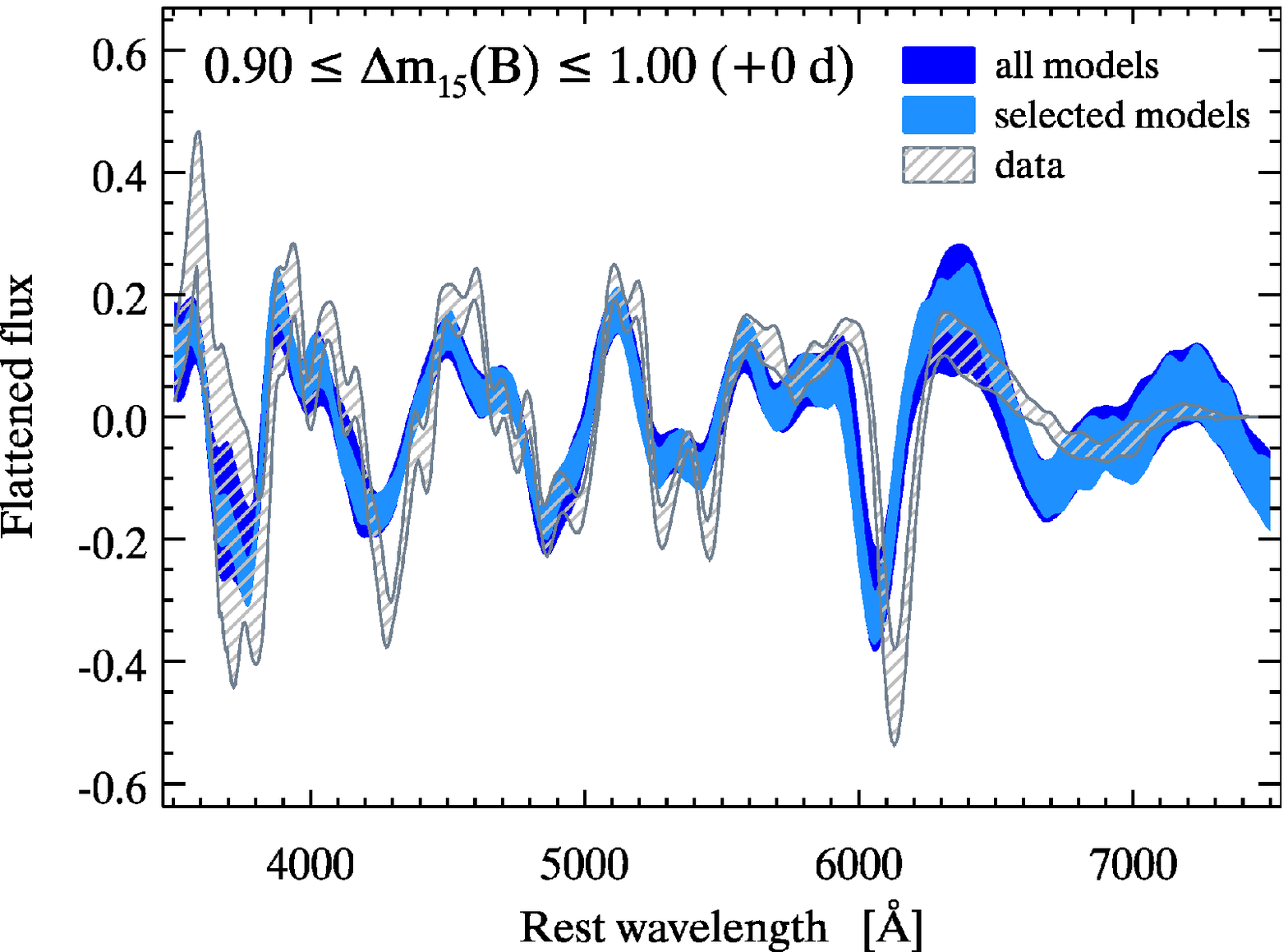}
\hspace{1.5cm}\includegraphics{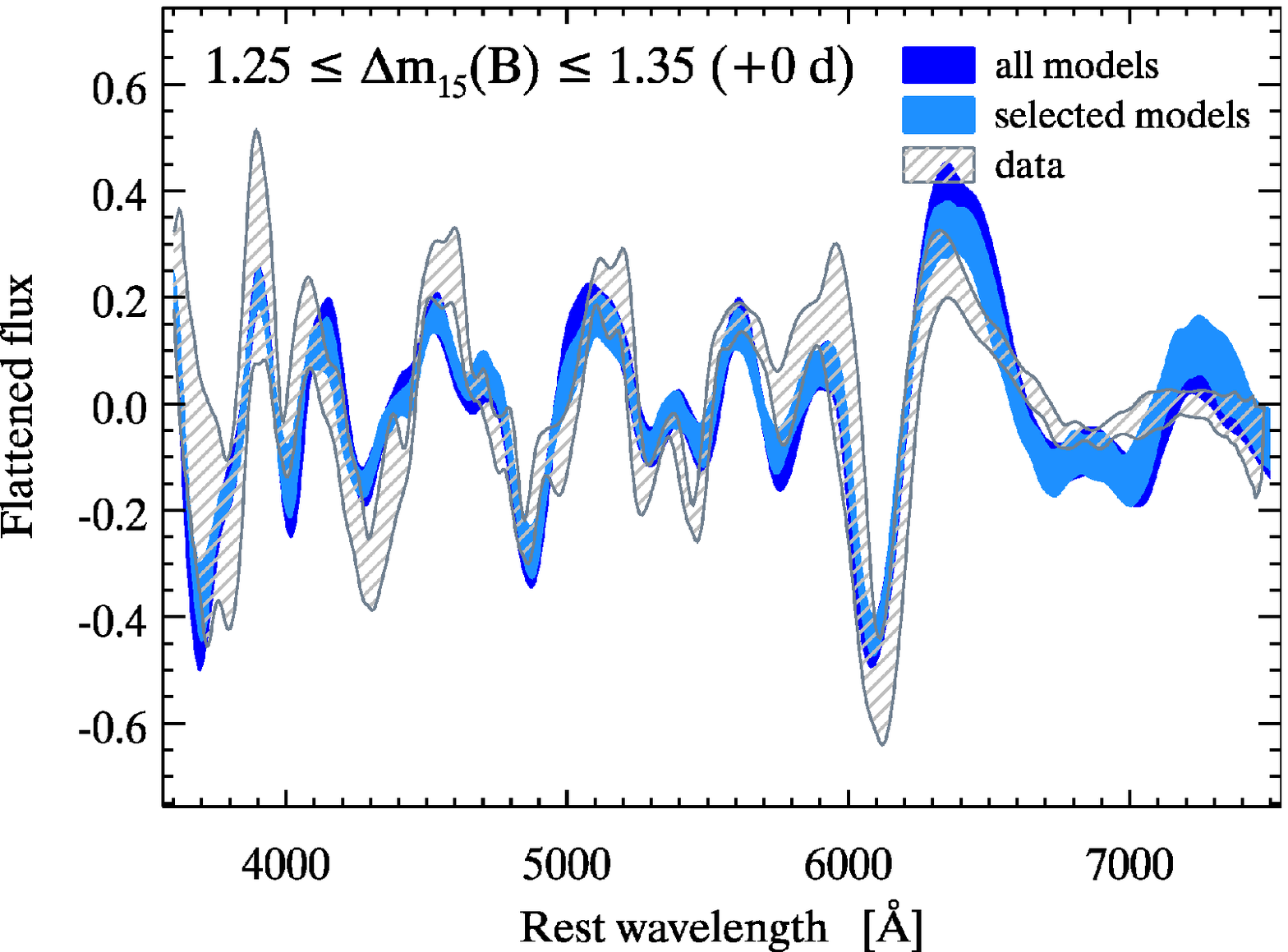}}
\caption{\label{fig:specvardm15}
Composite maximum-light spectra for $0.90\le\dmft\le1.00$ ({\it left})
and $1.25\le\dmft\le1.35$ ({\it right}). The shaded bands correspond
to the standard deviation about the mean maximum-light spectrum. We
show the flattened composite spectra for all DD2D models ({\it filled
  dark blue}), our subset of selected models ({\it filled light blue})
and as observed based on the sample of spectra shown in
Fig.~\ref{fig:specvardm15_data} ({\it hatched grey}).
}
\end{figure*}

To better study overall systematic differences between the models and
observations, we generate composite spectra using the same
pre-processing as done by SNID: the individual spectra are
``flattened'' through division by a {\it pseudo} continuum. We then
compute the mean flux in each wavelength
bin, as well as the standard deviation from the mean. The result is a
composite spectrum with error bands, which we show in
Fig.~\ref{fig:specvardm15}. Note that the 
composite spectra based on observations ({\it hatched grey}) only uses
the sample of spectra shown in Fig.~\ref{fig:specvardm15_data}, while
those based on the models comprise all viewing angles which yield a
\dmft\ in the appropriate range. The composite spectra for
all (selected) models thus consists of 234 (50) and 150 (41) individual
spectra for $0.90\le\dmft\le1.00$ and $1.25\le\dmft\le1.35$,
respectively.

The composite spectra for $0.90\le\dmft\le1.00$ confirm the shallower
absorptions and larger blueshifts of spectral lines in the models. The
location of 
maximum absorption in the Si\two\,\l6355 line is typically
$\sim100$\,\AA\ too blue in the models, corresponding to
a $\sim5000$\,\kms\ difference in absorption velocity. The small-scale
structure in the two iron-group-dominated absorption features
($\sim4300$\,\AA\ and $\sim4800$\,\AA) present in the data is not seen
in the models, as mentioned earlier. The emission component of the
Si\two\,\l6355 line appears to be relatively stronger in the models,
but this is an artefact of the division by a pseudo continuum. The
flux level in the models drops significantly redward of
$\sim6500$\,\AA\ compared to the data, artificially enhancing the
relative strength of the Si\two\,\l6355 emission profile after
division by this pseudo continuum.

The composite spectra for $1.25\le\dmft\le1.35$ do not seem to support
the apparent heterogeneity of the model spectra in
Fig.~\ref{fig:specvardm15_data}. This is not surprising since
Fig.~\ref{fig:specvardm15_data} only shows 5 of the 150 (41) spectra used to
generate the composite spectrum for all (selected) models. It is also
possible that the division by a pseudo continuum attenuates some of
these differences. Nonetheless, the composite spectra for this \dmft\
range do reveal some interesting properties of the models. The
absorption velocity offset is less pronounced in the Si\two\
lines. The location of maximum absorption in the Si\two\,\l4130 line
even appears to be at larger wavelengths in the models, but this could
result from line overlap. The two iron-group-dominated features at
$\sim4300$\,\AA\ and $\sim4800$\,\AA\ again appear to be smoothed by a
larger expansion velocity field. The S\two\ doublet is again shallower
in the models, although better resolved than for
$0.90\le\dmft\le1.00$. Both absorption components appear to be
slightly offset to redder wavelengths when compared to the data,
although the difference is typically $\lesssim25$\,\AA, or
$\lesssim1500$\,\kms. The emission 
component of the Si\two\,\l6355 line is again artificially enhanced
via the division by a pseudo continuum, although part of the
difference is real and clearly seen in the model spectra shown in
Fig.~\ref{fig:specvardm15_data} (in particular for models
DD2D\_iso\_06\_dc4 and dc5). We will explore some of these systematic
differences in more detail by focusing on this line in the
next section.

\subsubsection{Si\two\,\l6355 absorption velocity}

\begin{figure}
\centering
\resizebox{0.475\textwidth}{!}{
\includegraphics{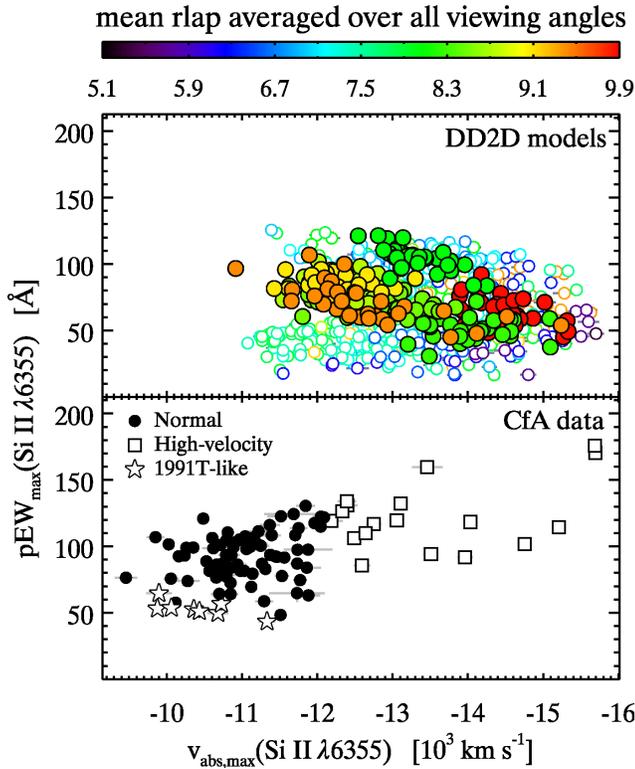}
}
\caption{\label{fig:specclass}
Pseudo-equivalent width of the Si\two\,\l6355 line {\it
  vs.} its absorption velocity at maximum light. 
The colour-coding is the same as in Fig.~\ref{fig:trdiff}. 
The symbols for the data points correspond to the spectroscopic subclasses
defined by {\protect\cite{WangX/etal:2009b}}.
}
\end{figure}

\cite{WangX/etal:2009b} introduced a classification scheme for \sneia\
based on the Si\two\,\l6355 absorption velocity at maximum light. By
considering deviations from an empirical mean trend, they classified
their \snia\ sample into ``normal'' and ``high-velocity'' subclasses, with
separate subclasses for 1991T- and 1991bg-like \sneia. Fig.~\ref{fig:specclass} ({\it
lower panel}) shows the relation between the pseudo equivalent
width of the Si\two\,\l6355 line with its absorption velocity
(i.e. the velocity at maximum absorption, $v_{\rm abs}$) within
3\,d from maximum light in our spectroscopic sample from the 
CfA SN Program \citep[see][their Fig.~2]{WangX/etal:2009b}. Only
\sneia\ with $\dmft\lesssim1.6$ are included, to match the \dmft\
range of the models. We added a random component drawn from a
$500$\,\kms-wide uniform distribution to account for velocity
measurement errors resulting from the 10\,\AA\ binning in the
synthetic spectra.
We have also run one model (DD2D\_iso\_06\_dc2) with a larger
  number of photon packets ($10^9$ instead of $10^8$) and find the
  measurements to be consistent with the lower-S/N model run.
Almost all the models have Si\two\,\l6355
absorption velocities consistent with the ``high-velocity'' subclass,
regardless of the viewing angle, in agreement with the large
blueshifts discussed in the previous section. Several models have
significantly lower pEW than in the data (at a given $v_{\rm abs}$),
again consistent with the shallower absorptions seen in some of the
individual spectra (in particular, those in the $0.90\le\dmft\le1.00$
range).

\begin{figure}
\centering
\resizebox{0.475\textwidth}{!}{
\includegraphics{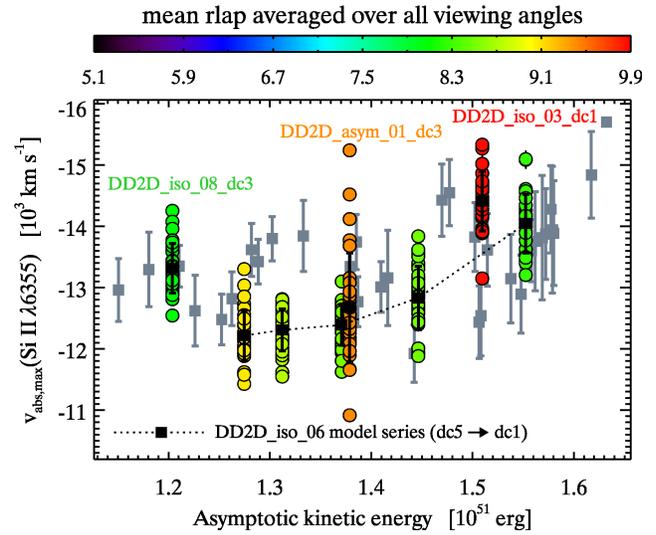}
}
\caption{\label{fig:vabsekin}
Absorption velocity of the Si\two\,\l6355 line at maximum light
{\it vs.} asymptotic kinetic energy. The colour-coding is the same
as in Fig.~\ref{fig:trdiff}. For our subset of selected models
({\it filled circles}), we show the complete set of 30 measurements
(one per viewing angle). For all other models, only the mean 
$v_{\rm abs}$ and its standard deviation ({\it grey squares}) are
plotted. Last, the dotted line corresponds to the relation of  $v_{\rm 
  abs,max}$(Si\two\,\l6355) with kinetic energy for the
DD2D\_iso\_06 model series.
}
\end{figure}

These measurements reveal further properties of the models
themselves. Model DD2D\_asym\_01\_dc3 ({\it filled orange circles})
displays the largest variation 
in Si\two\,\l6355 absorption velocity ($-11000\gtrsim v_{\rm
  abs}\gtrsim -15000$\,\kms), the larger blueshifts corresponding
to the largest extent of the \nifs\ distribution.
For models with a symmetric distribution of \nifs, one expects
the expansion rate to scale approximately with
$\sqrt{E_{\rm kin}/M}$. Fig.~\ref{fig:specclass}  shows that
one of our selected models with the lowest kinetic energy
(DD2D\_iso\_06\_dc5, {\it filled yellow circles}; $E_{\rm
  kin}\approx1.3\times10^{51}$\,erg, see Table~\ref{tab:modelinfo}) has
lower $|v_{\rm abs}|$ than one of our selected models with the highest
kinetic energy (DD2D\_iso\_03\_dc1, {\it filled red circles}; $E_{\rm
  kin}\approx1.5\times10^{51}$\,erg). To see whether this holds for the
other 2D models of KRW09, we show the relation of $v_{\rm
  abs,max}$(Si\two\,\l6355) {\it vs.} kinetic energy in
Fig.~\ref{fig:vabsekin}. 

We do not see a general trend of increasing $|v_{\rm abs}|$ with
kinetic energy, although the models with the highest mean absorption
velocity all have $E_{\rm kin}>1.45\times10^{51}$\,erg. The lack of a
general trend may seem contrary to expectations, but it reflects
the impact of the different 
distributions of ignition points in each model series on the
distribution of intermediate-mass elements in the ejecta. Within a
given model series (i.e. for a given ignition setup), the mean
$|v_{\rm abs}|$  increases with the kinetic energy of the
explosion, which is directly related to the criterion for
deflagration-to-detonation transition ($E_{\rm kin}$ decreases
steadily from dc5 to dc1 within a model series; see
Table~\ref{tab:modelinfo}). We illustrate this for the DD2D\_iso\_06
model series in Fig.~\ref{fig:vabsekin} ({\it dotted line}).

\begin{figure}
\centering
\resizebox{0.475\textwidth}{!}{
\includegraphics{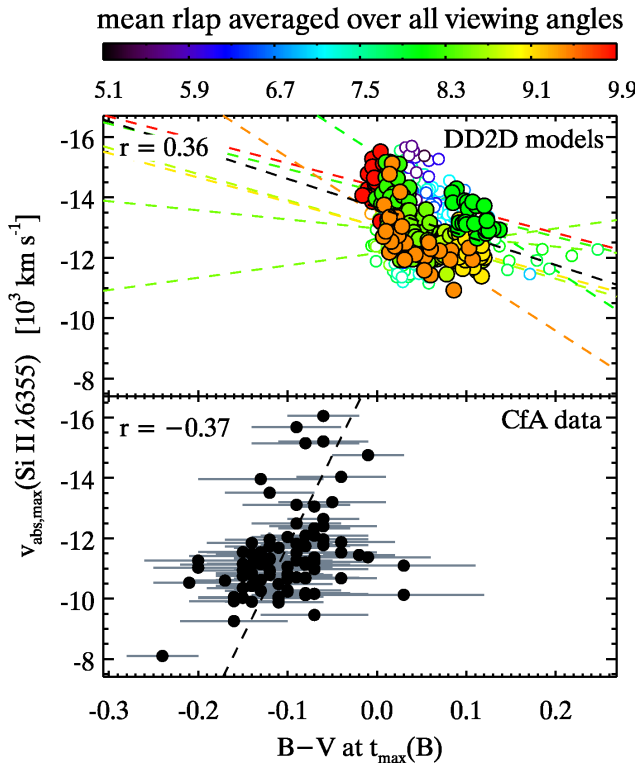}
}
\caption{\label{fig:vabsbmv}
Absorption velocity of the the Si\two\,\l6355 line at maximum light
{\it vs.} intrinsic $B-V$ colour at $B$-band maximum, in the 2D models
of KRW09 ({\it upper panel}) and in data from the CfA SN Program ({\it
 lower panel}). The colour-coding is the same
as in Fig.~\ref{fig:trdiff}. In both cases, we show a linear fit to
the entire sample ({\it dashed black line}) and report the Pearson
correlation coefficient ($r$) of $v_{\rm abs,max}$ with intrinsic
$B-V$ colour.
We also show the linear relations obtained from fits to
individual models ({\it dashed colour lines}).
}
\end{figure}

In a recent paper, \cite{Foley/Kasen:2011} noted a correlation
between the Si\two\,\l6355 absorption velocity at maximum light with
$B^{\rm max}-V^{\rm max}$ pseudo-colour\footnote{difference
  between the $B$ magnitude at $B$-band maximum and the $V$  
magnitude at $V$-band maximum.} at the same age, redder
\sneia\ having higher $|v_{\rm abs}|$. The lower panel of
Fig.~\ref{fig:vabsbmv} shows the relation of $v_{\rm
  abs,max}$(Si\two\,\l6355) with intrinsic $B-V$ colour at $B$-band
maximum (as opposed to $B^{\rm max}-V^{\rm max}$ pseudo-colour) in
data from the CfA SN Program. Although the correlation between
both quantities is rather weak ($r=-0.37$; but note the large error on
intrinsic $B-V$ colour), the slope of the relation is in
agreement with that found by \cite{Foley/Kasen:2011}.
For the 2D models of KRW09 (Fig.~\ref{fig:vabsbmv}, {\it upper
panel}), the general trend is in the opposite direction ({\it dashed
black line}). Within each model, the
trend of absorption velocity with colour generally also has the
opposite trend as that seen in the data.

\cite{Foley/Kasen:2011} provide a simple explanation for the observed
correlation by invoking the greater line opacity in the $B$-band for
\sneia\ with high-velocity ejecta, both through increased line overlap
and a greater impact of Fe\two/Co\two\ lines in the outer cooler regions of the
ejecta where higher-velocity spectral features are formed. Based on
radiative transfer calculations by \cite{Kasen/Plewa:2007}, they show
that the detonating failed deflagration (DFD) 
model of \cite{Plewa:2007} is consistent with the association of
redder $B-V$ colour with larger Si\two\,\l6355 blueshifts, when viewed
from different viewing angles. As mentioned earlier, the 2D models of
KRW09 also display variations of the Si\two\,\l6355 absorption
velocity with viewing angle, with associated variations in $B-V$ colour, but
the correlation in most cases is in the opposite direction as that
seen in the DFD model, and in all cases much weaker than shown by
\cite{Foley/Kasen:2011} (their Fig.~8). Since the radiative transfer
code used by KRW09 and \cite{Kasen/Plewa:2007} is the same, this
discrepancy must be traced back to differences 
in the explosion models (whether intrinsic or resulting from different
physical approximations and numerical treatment thereof) or
nucleosynthetic post-processing (\citealt{Kasen/Plewa:2007} used an
approximate alpha-network scheme which for instance does not include
sodium, while KRW09 used a more elaborate nuclear network to compute
detailed abundances for a representative model, and interpolated the
output for all other models). Analysis of more DFD models would be
needed to check whether the correlation between Si\two\,\l6355
absorption velocity and maximum-light $B-V$ colour is a generic
feature of this type of explosion. This is apparently not the case for
delayed-detonation models of \sneia, although Fig.~\ref{fig:vabsbmv}
once again shows that the 2D models of KRW09 lie in a different region
of parameter space than the data (larger Si\two\,\l6355 blueshifts and
redder $B-V$ colours).

One possible explanation resides in the different distributions
of iron-group and intermediate-mass elements in both DFD and
delayed-detonation models. In the DFD model, the deflagration is
ignited on one side of the ejecta, but detonates on the opposite side,
causing both IGE and IME to be ejected at high velocities on the
ignition side. In delayed-detonation models, the IGE and IME are
preferentially ejected in opposite directions (see
Fig.~\ref{fig:dens}, {\it bottom row}). Since the distribution of
IGE sets the $B-V$ colour to a large extent \cite[see,
  e.g.][]{Kasen/Woosley:2007}, one might expect the two detonation
models to show opposite trends for $|v_{\rm abs}|$ and $B-V$.

\subsection{Spectral evolution}\label{sect:specevol}

The comparison of maximum-light spectral properties in the previous
section has several practical advantages: more data are available at
this age and they are of better quality (the use of a Monte Carlo
radiative transfer code means this is also true of the model spectra,
which consist of a higher number of photon packets at maximum
light). Any thorough validation of a \snia\ model should, 
however, include a comparison of the time evolution of its spectra
with observations. As the supernova expands, the layers where
the spectrum is formed recede to deeper regions of the ejecta, where
the composition is different and the 
expansion velocity smaller. By measuring various parameters of
individual spectral features and their evolution with time, one has a
complete census of a model's failures and successes: a poor model
might reproduce certain spectral features at certain times, but a good
model should reproduce {\it all} features at {\it all} times.

\subsubsection{Overall evolution}

\begin{figure}
\centering
\resizebox{.465\textwidth}{!}{\includegraphics{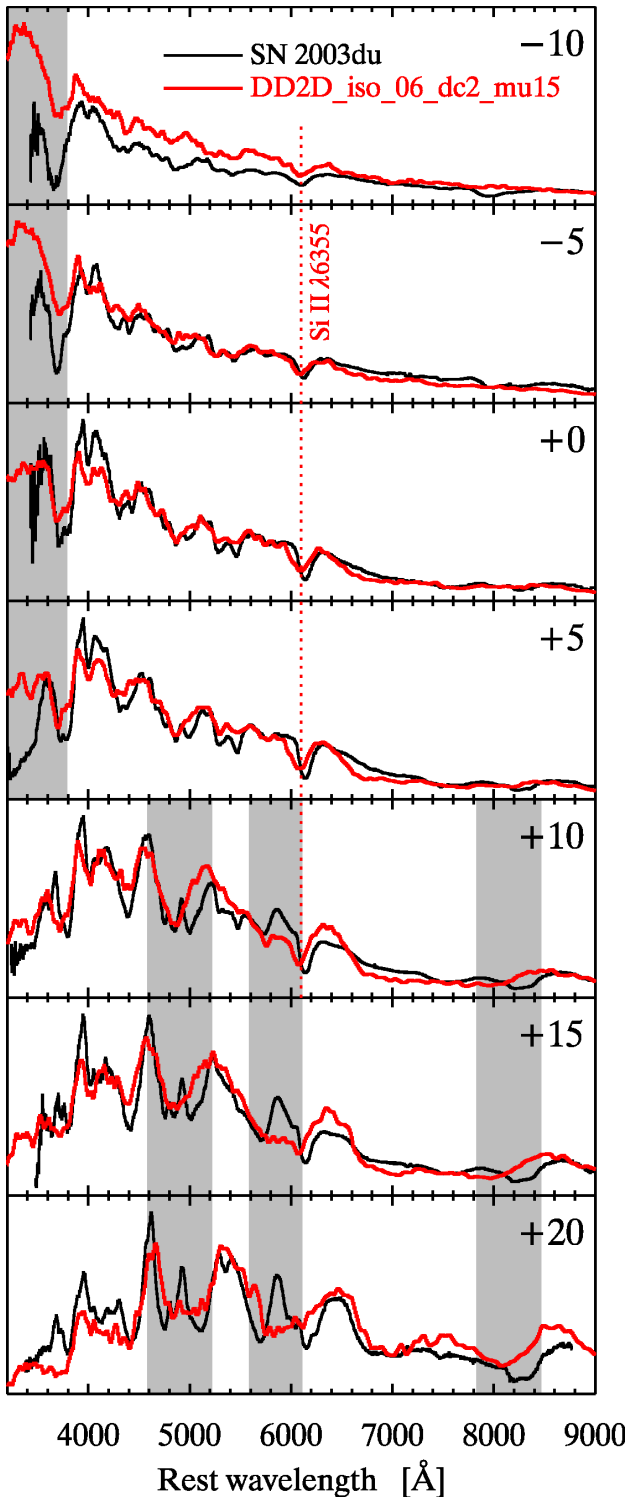}}
\caption{\label{fig:speceg}
Spectroscopic evolution of model DD2D\_iso\_06\_dc2 viewed along
$\theta=88^{\circ}$ ({\it red}), compared to spectra of SN~2003du
({\it black};
\citealt{Anupama/Sahu/Jose:2005,Stanishev/etal:2007}).
The synthetic and observed spectra have been scaled to match their
respective rest-frame absolute $B$-band magnitude (see Fig.~\ref{fig:lceg}).
The age of each
spectrum is indicated in days from $B$-band maximum in the top-right
corner of each plot. The grey shaded regions highlight the model
$U$-band excess (between $-10$\,d and +5\,d), and several
disagreements around 5000\,\AA, 5800\,\AA, and 8200\,\AA\
(respectively attributed to Fe\two/Co\two, Na\one, and Ca\two) from +10\,d
onwards. The 
vertical red dotted line indicates the wavelength location of maximum
absorption in Si\two\,\l6355 in the models, which remains almost
constant between $-10$\,d and +10\,d.
}
\end{figure}

In Fig.~\ref{fig:speceg} we show the spectral evolution of model
DD2D\_iso\_06\_dc2 viewed along $\theta=88^{\circ}$ (see
Fig.~\ref{fig:lceg} for the corresponding $UBVRIJHK_s$ light curves),
in five-day increments between $-10$ and +20\,d from $B$-band maximum
(i.e. between $\sim10$\,d and $\sim40$\,d past explosion),
compared to observations of SN~2003du. Both models and observations
correspond to a \snia\ with $\dmft\approx1$.

At $-10$\,d, the model spectrum displays a much bluer (hotter)
  SED with weaker features than SN~2003du. The absorption due 
to Ca\two\,H\&K ($\sim3700$\,\AA) is much weaker, and that due to the
Ca\two\ IR triplet ($\sim8000$\,\AA) is non-existent, both clear signs
of over-ionization in the ejecta. 
The weak  Ca\two\,H\&K line certainly contributes to the excess of $U$-band flux
inferred from comparisons of $U-B$ colour curves
(Fig.~\ref{fig:colevol}), but most of the discrepancy appears to
originate blueward of this, in a region dominated by absorption
by iron-group elements ({\it grey highlighted region}). This is
most clearly seen in the +5\,d spectrum of SN~2003du, which unlike the
earlier ones extends blueward of 3600\,\AA. As noted earlier based on
the $U-B$ colour curves, this excess of $U$-band flux remains
important up until +5\,d, at which point the absorption due to
Ca\two\,H\&K has significantly strengthened in the models though
remains weaker than in 
SN~2003du. The S\two\ doublet ($\sim5400$\,\AA), well fit at $-5$\,d
is too weak in the models between +0\,d and +10\,d. The same is true
of the two iron-group dominated absorption complexes at
$\sim4300$\,\AA\ and $\sim4800$\,\AA, which also reveal some of the
model shortcomings from maximum light onwards. The velocity offset in
the Si\two\,\l6355 absorption (broader yet shallower in the models;
{\it red dotted line}) also becomes apparent at maximum
light. Interestingly, the Si\two\,\l6355 absorption velocity appears
to remain almost constant between $-10$\,d and +10\,d while it
decreases steadily in SN~2003du.

From +10\,d onwards, we highlight several discrepancies in other
wavelength regions. The model appears to lack an absorption around
$\sim5000$\,\AA, which in SN~2003du and other \sneia\ is part of a
complex Fe\two/Co\two\ absorption feature. The pseudo equivalent width of
this feature at +10\,d is systematically $\sim50$\,\AA\ smaller in all
the models with respect to observations. It is difficult to
disentangle the effects of temperature (if too hot, as clearly the
case before maximum light, this would delay the {\sc
  iii}$\rightarrow${\sc ii} recombination timescale of iron-group
elements), the impact of the LTE approximation (non-LTE
effects are particularly important in treating the Fe\two/Co\two\ opacities;
see \citealt{Baron/etal:1996}), and the high ejecta velocities (as
clearly seen from the large blueshift of the Ca\two\ IR triplet
absorption)-- which would enhance line overlap and smear out
small-scale absorption features. 

Another striking feature is the lack of emission in the model 
around 5800\,\AA. This feature is commonly attributed to the
Na\one\,D doublet.
The absence of a strong 5800\,\AA\ feature is a generic feature of
these models, and not just particular to model DD2D\_iso\_06\_dc2.
The radiative transfer calculations we have at our disposal do not extend
beyond +25\,d or so past $B$-band maximum, so we cannot say whether
this line appears at later ages once the ejecta have cooled
down, although KRW09 present 
synthetic spectra of model DD2D\_iso\_06\_dc2 at +31\,d that do not
show evidence for this feature (their Fig.~2). Artificially removing
this line in the spectra of SN~2003du can lead to $R$-band magnitude 
differences of up to $\sim0.2$\,mag, and could explain part of the
discrepancy in $B-R$ and $V-R$ colours between models and observations
past maximum light (see Fig.~\ref{fig:colevol}).

Non-LTE calculations by \cite{Baron/etal:2006} fail to reproduce
this feature (their Fig.~4), so the LTE approximation used by
KRW09 is not necessarily the cause of this discrepancy.
If this line is  Na\one\,D, the LTE approximation should be
able to partly reproduce this strong resonance line if sodium is
abundant in its neutral stage ($X_{\rm Na}=10^{-5}$ throughout the
ejecta in all the models of KRW09, roughly corresponding to one third
of the solar value; e.g., \citealt{Asplund/etal:2009}). The absence of
this line would then be related to an ionization problem (Na\one\ has
a low ionization potential of $\sim5.1$\,eV), which would
likely persist until later times due to nonthermal ionization in the
ever-thinning ejecta \citep[e.g.,][]{Kozma/etal:2005}. Another
possibility is that this feature is not due to Na\one, since
its presence is difficult to reconcile with the presence of singly-
and doubly-ionized species, but we do not explore this possibility
further at this stage.

\subsubsection{Absorption velocities}

Studying the blueshifts of individual absorptions and their evolution
with time enables one to indirectly probe the dynamics of the
expansion. As a general rule, the blueshift decreases with time (as
does $|v_{\rm abs}|$), since the spectra form in deeper layers of the
ejecta (i.e. at lower mass coordinate) as the SN expands. Various
inhomogeneities and radiative transfer effects can, however, affect
this simple picture. For example, \cite{Quimby/etal:2007a} invoke a
shell-like density structure (formed within the context of a pulsating
delayed detonation or WD-WD merger scenario) to explain the $\sim10$\,d-long
plateau in absorption velocity of the Si\two\,\l6355 line in
SN~2005hj. As for radiative transfer effects,
\cite{Dessart/etal:2011} recently showed how taking into account
time-dependent terms in the gas energy and statistical equilibrium
equations can lead to an {\it increase} of the He\one\,\l10830
blueshift with time (contrary to all other He\one\ lines) in the
context of Type Ib/c supernovae.

In what follows we explore the evolution of the absorption velocities
in two strong lines in \snia\ spectra: Ca\two\,\l3945
(i.e. Ca\two\,H\&K) and the characteristic Si\two\,\l6355. The Ca\two\
line profile spans a wavelength region where the dominant source
of opacity is due to bound-bound transitions \citep[see,
  e.g.][]{Pinto/Eastman:2000b}, and line overlap is an issue \citep[see
  also][]{Blondin/etal:2006}. Moreover, observations of the Ca\two\ IR
triplet in early-time spectra of \sneia\ have revealed the presence of
high-velocity absorption components \citep[see][]{Mazzali/etal:2005}. In
what follows, and in the presence of Ca\two\,\l3945 profiles with
multiple absorptions, we report the velocity associated with the
{\it deepest} absorption. The Si\two\,\l6355 line is less affected by line
overlap (at least until +10\,d or so past $B$-band maximum), although
some \sneia\ appear to have a high-velocity component associated with
this line at early times (e.g., SN~2005cf;
\citealt{Garavini/etal:2007b,WangX/etal:2009a}).

\begin{figure*}
\centering
\resizebox{\textwidth}{!}{
\includegraphics{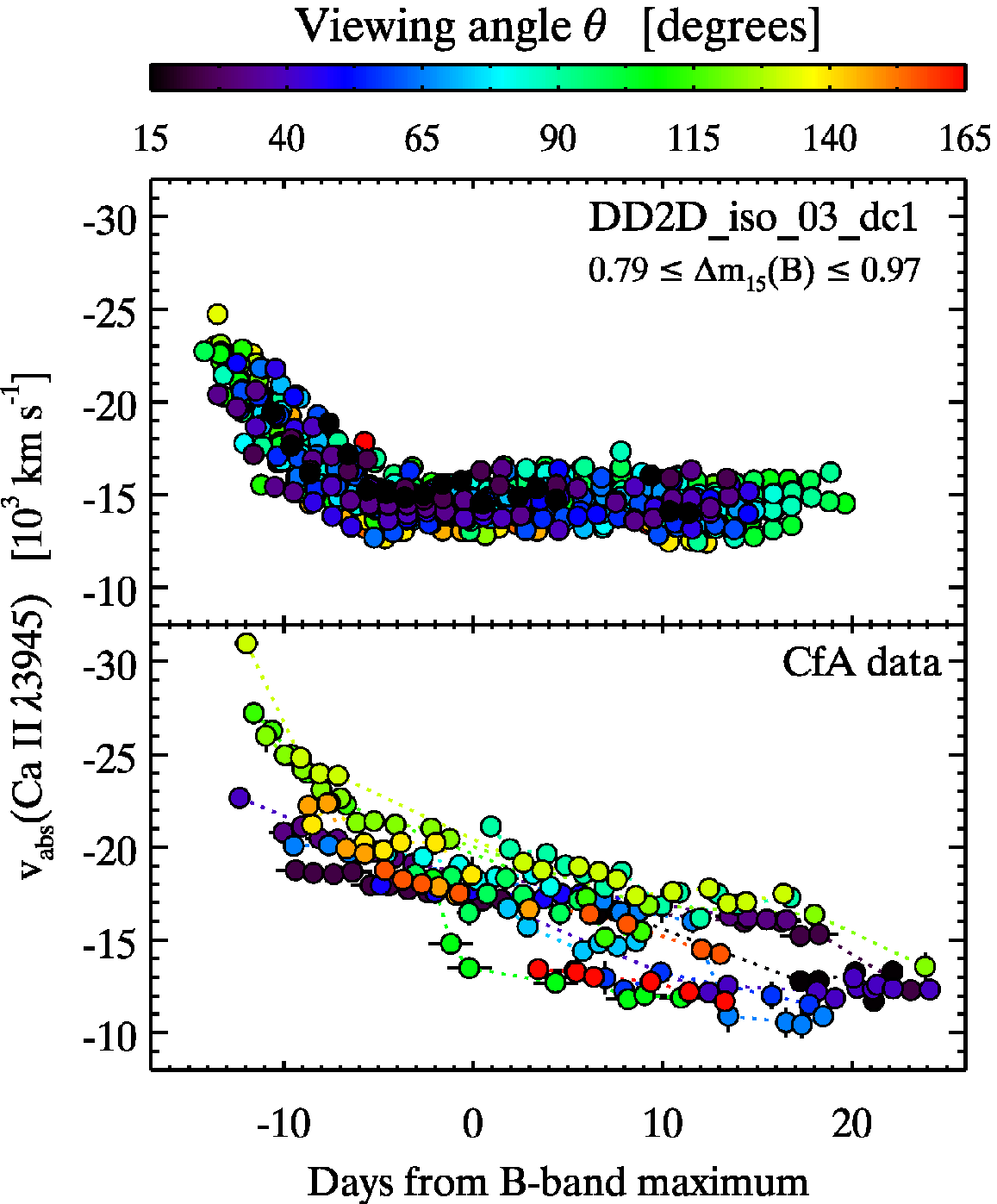}
\includegraphics{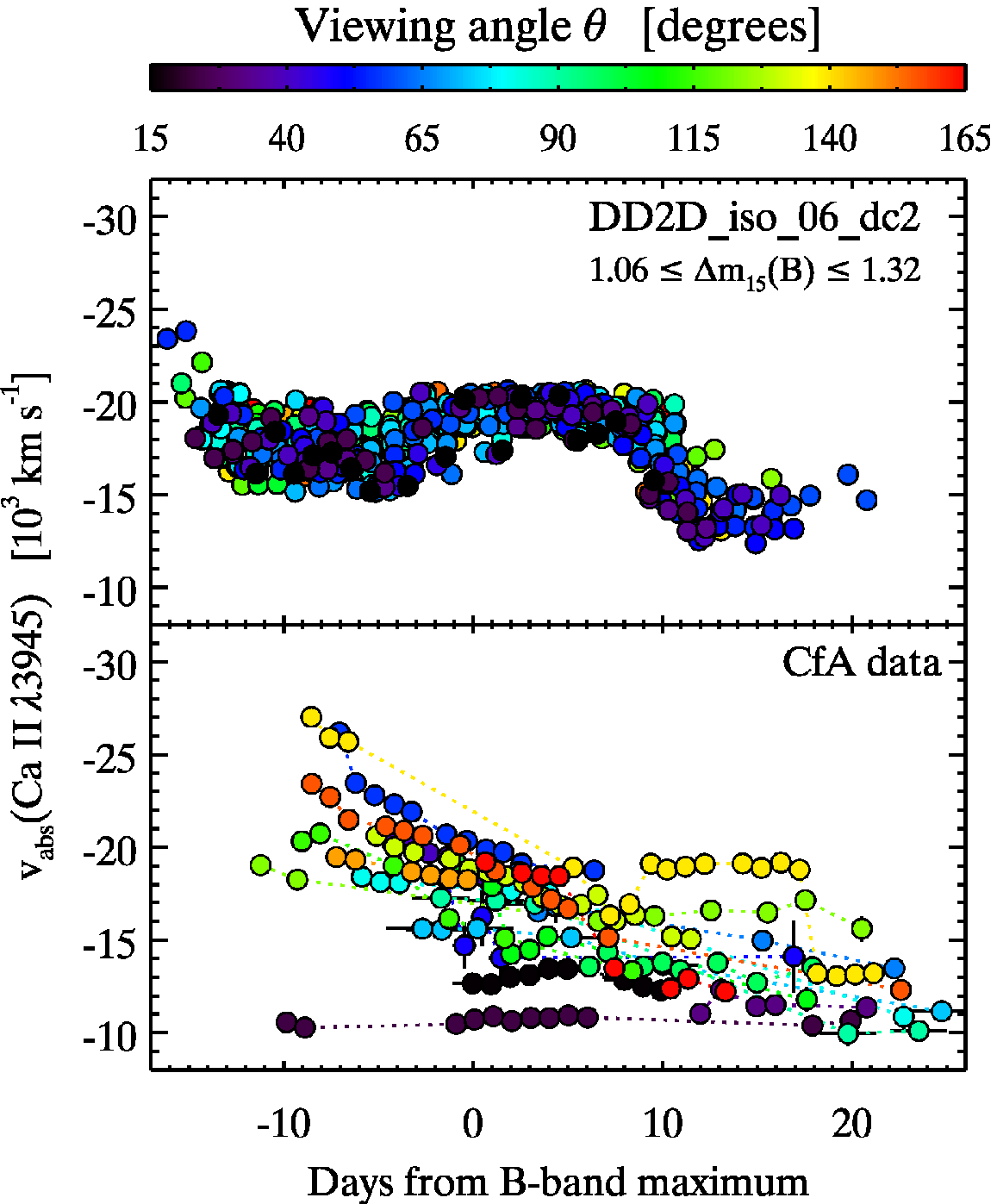}
\includegraphics{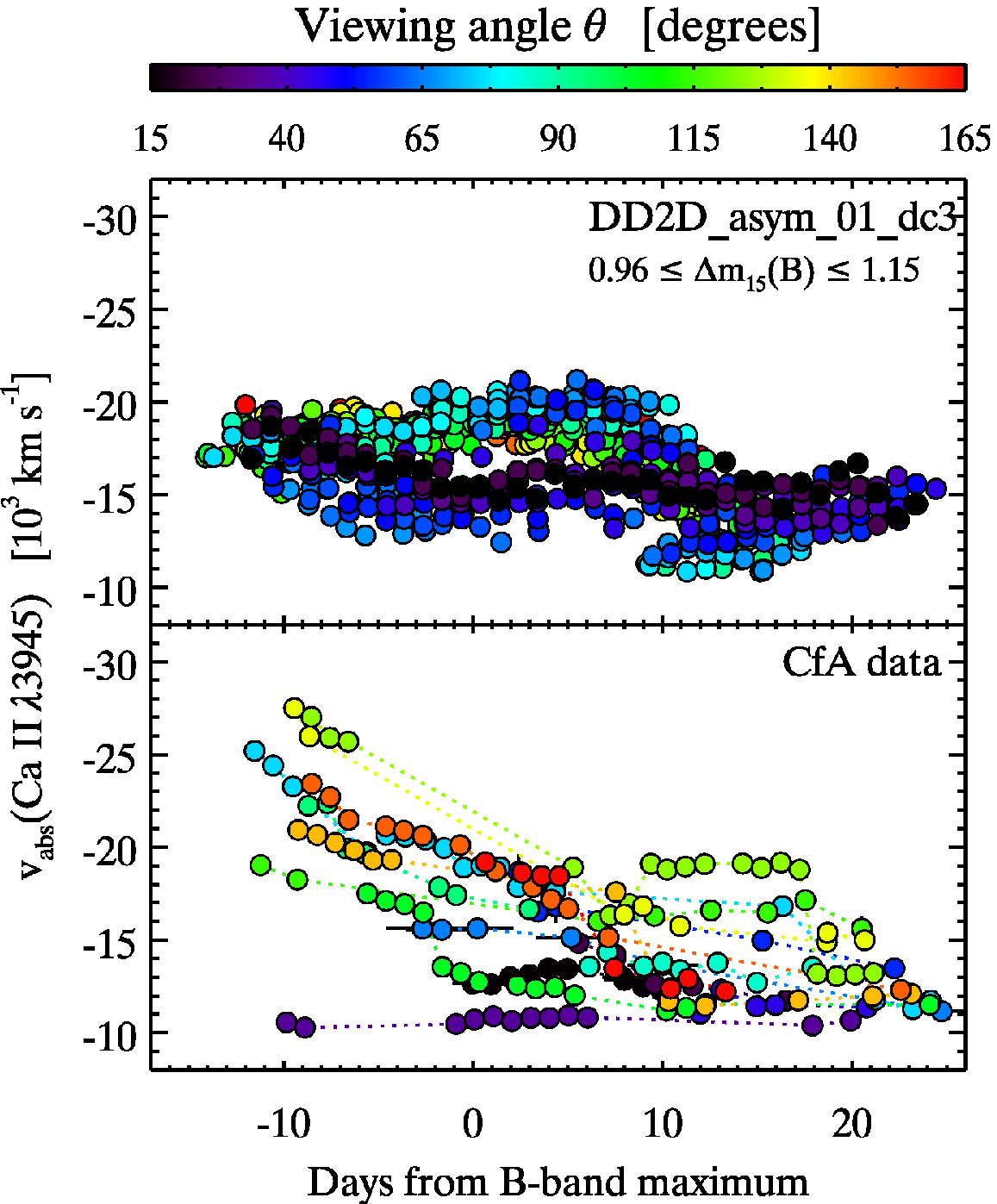}
\includegraphics{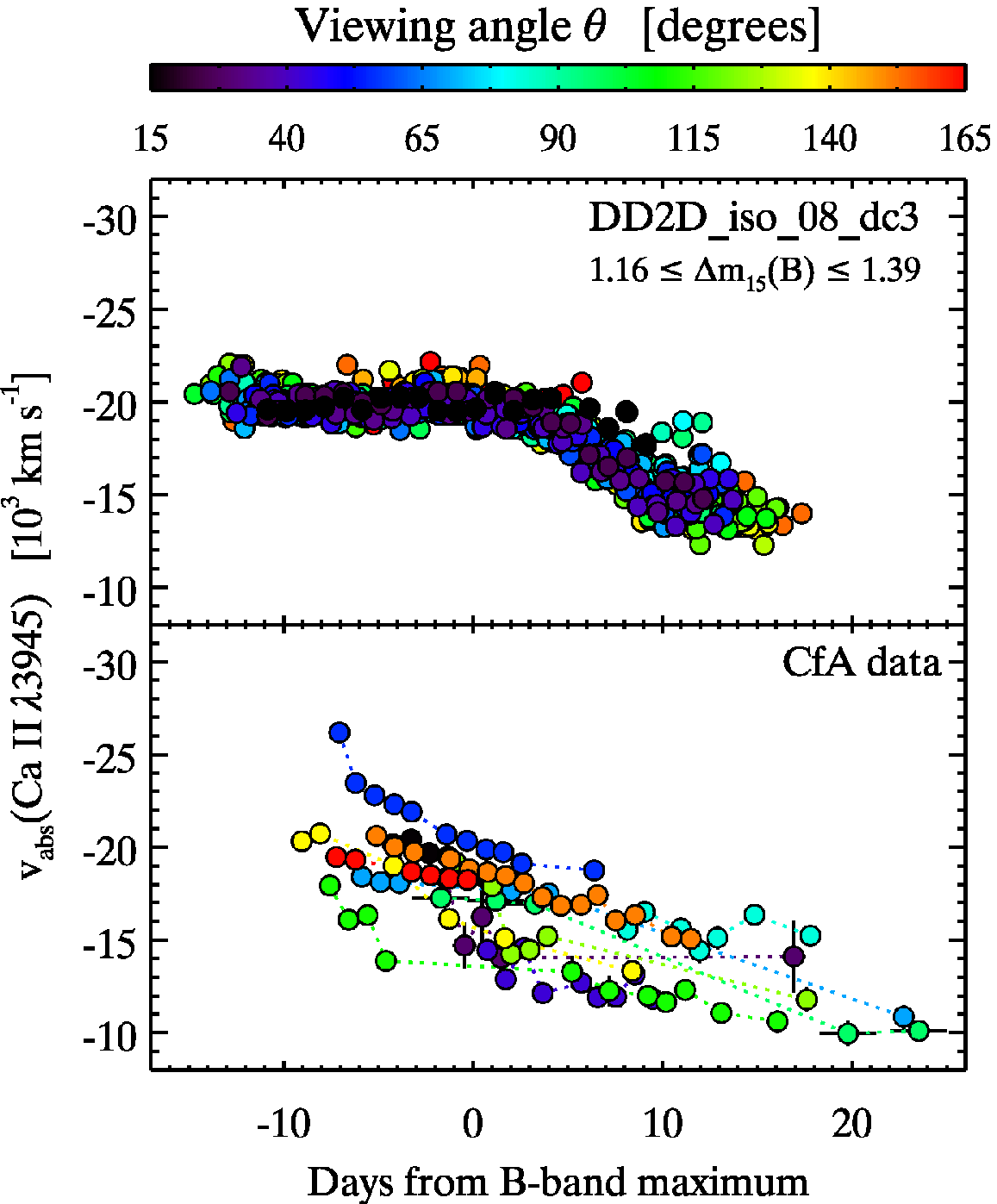}
}

\vspace{1cm}
\resizebox{\textwidth}{!}{
\includegraphics{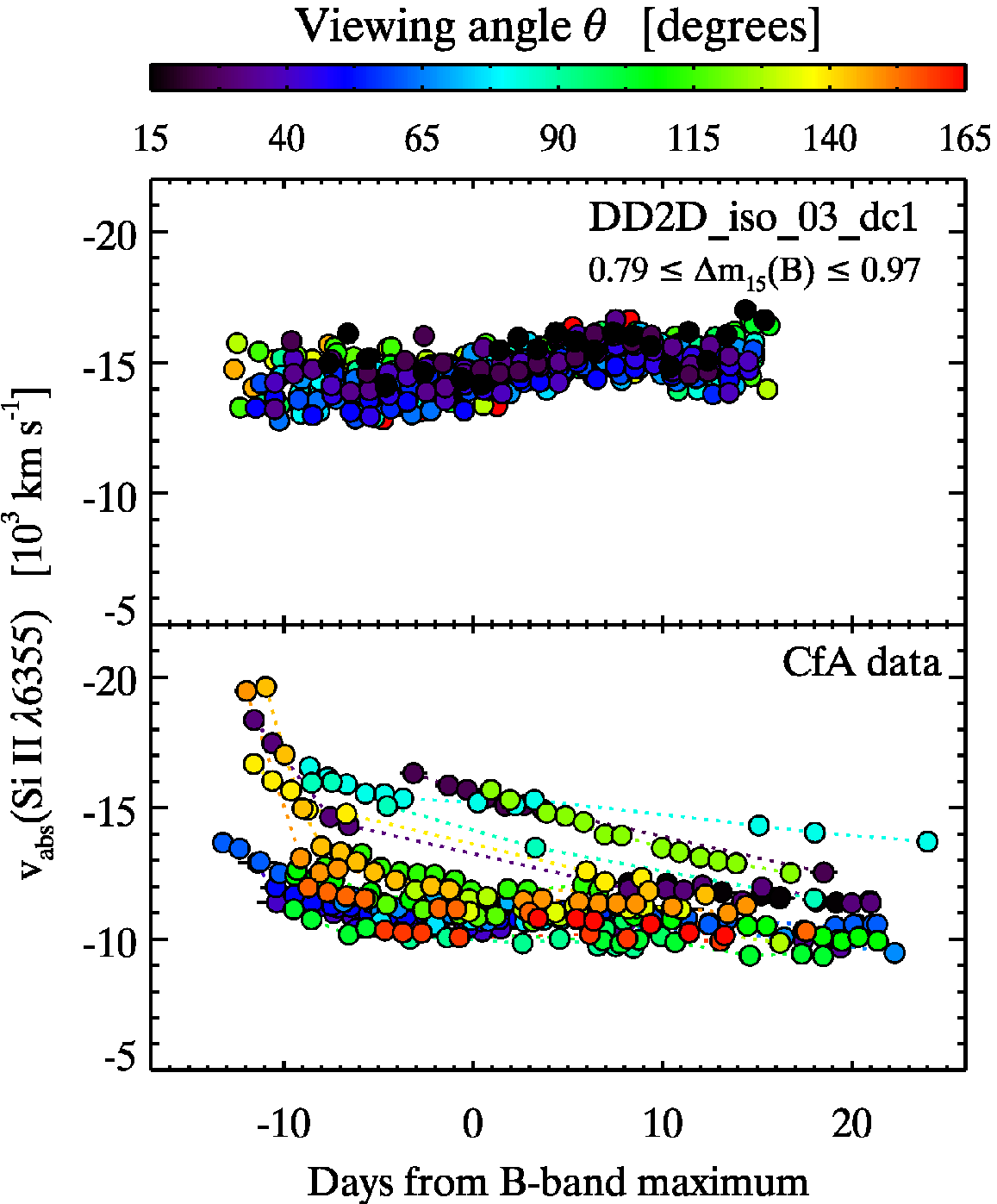}
\includegraphics{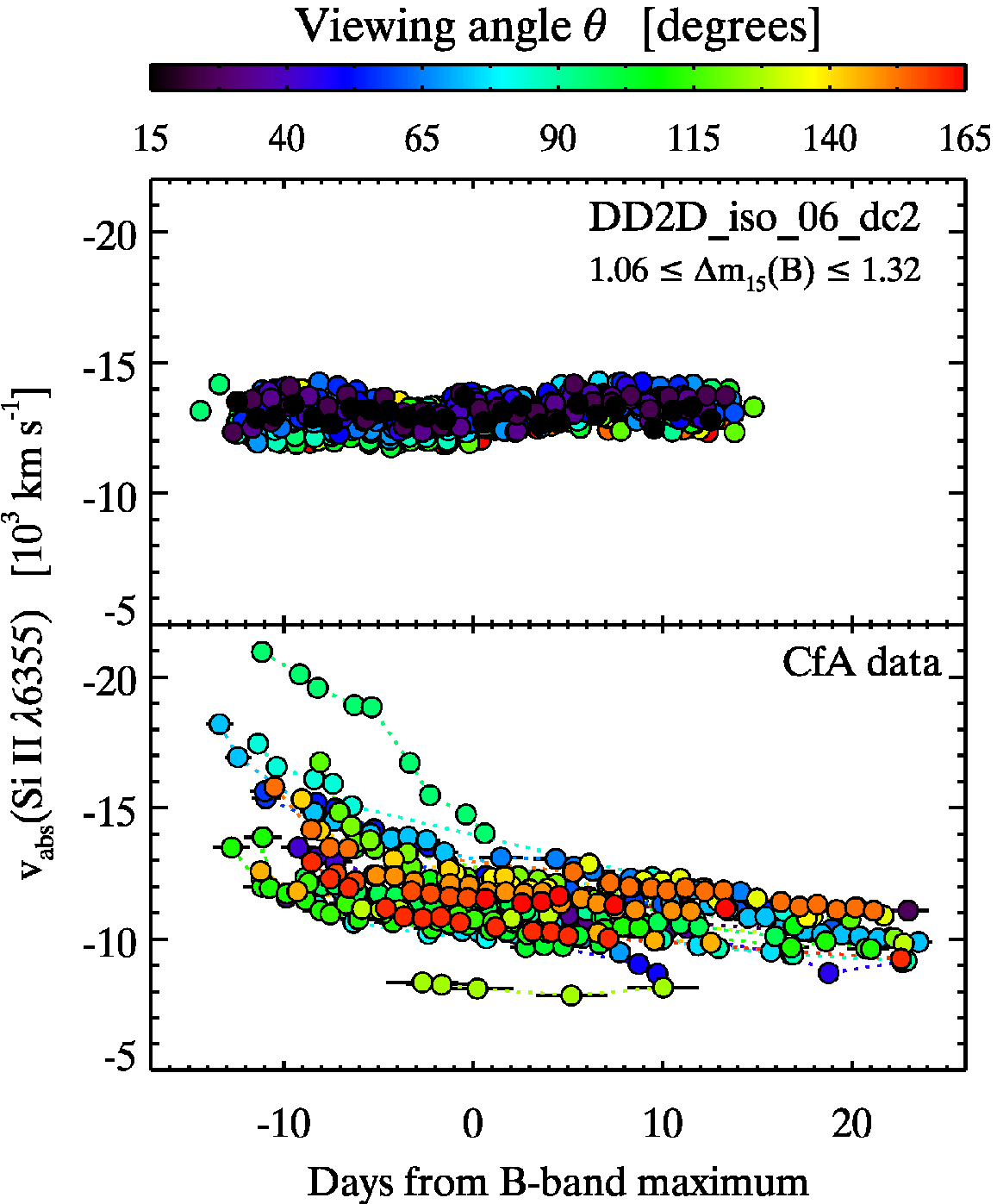}
\includegraphics{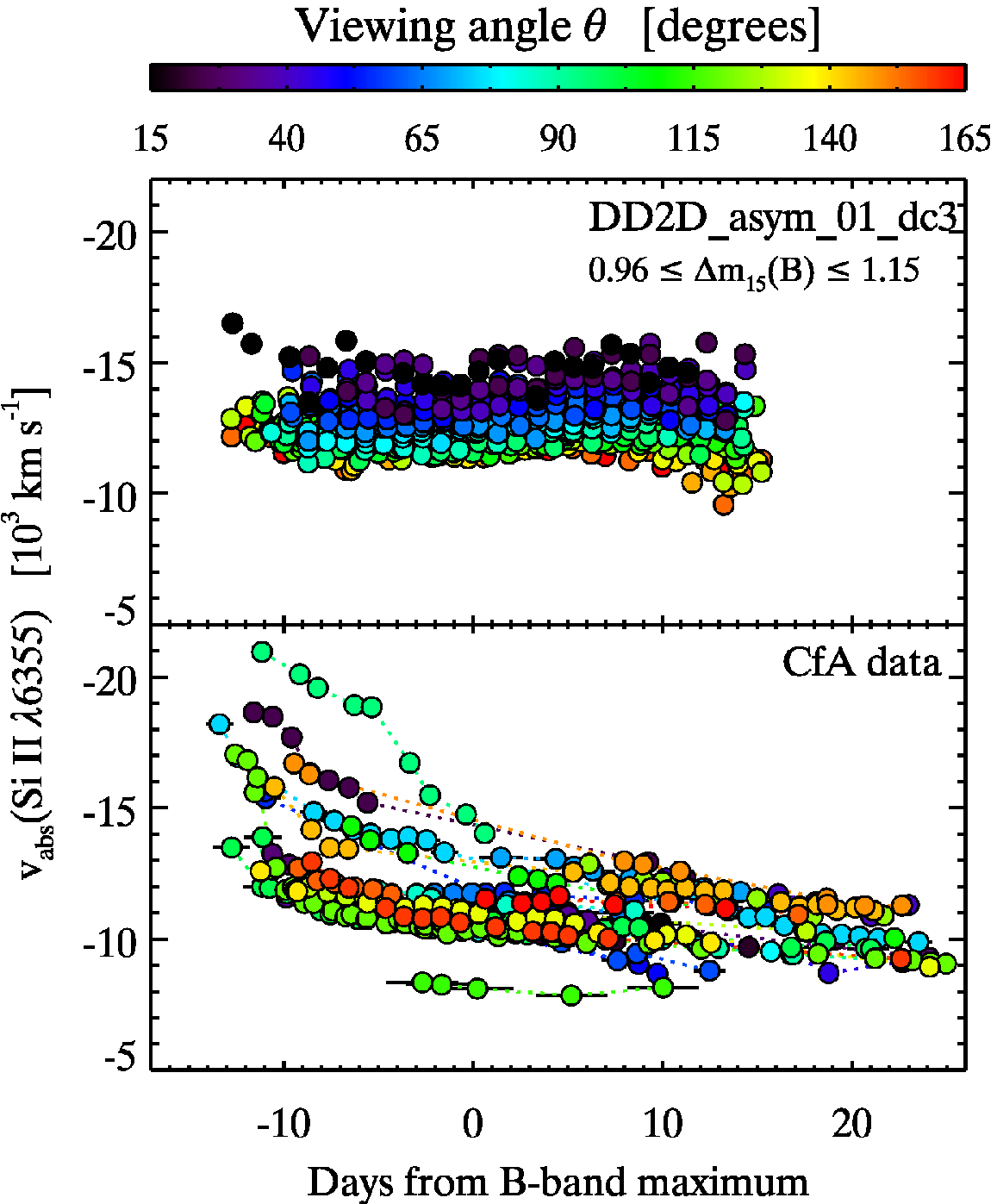}
\includegraphics{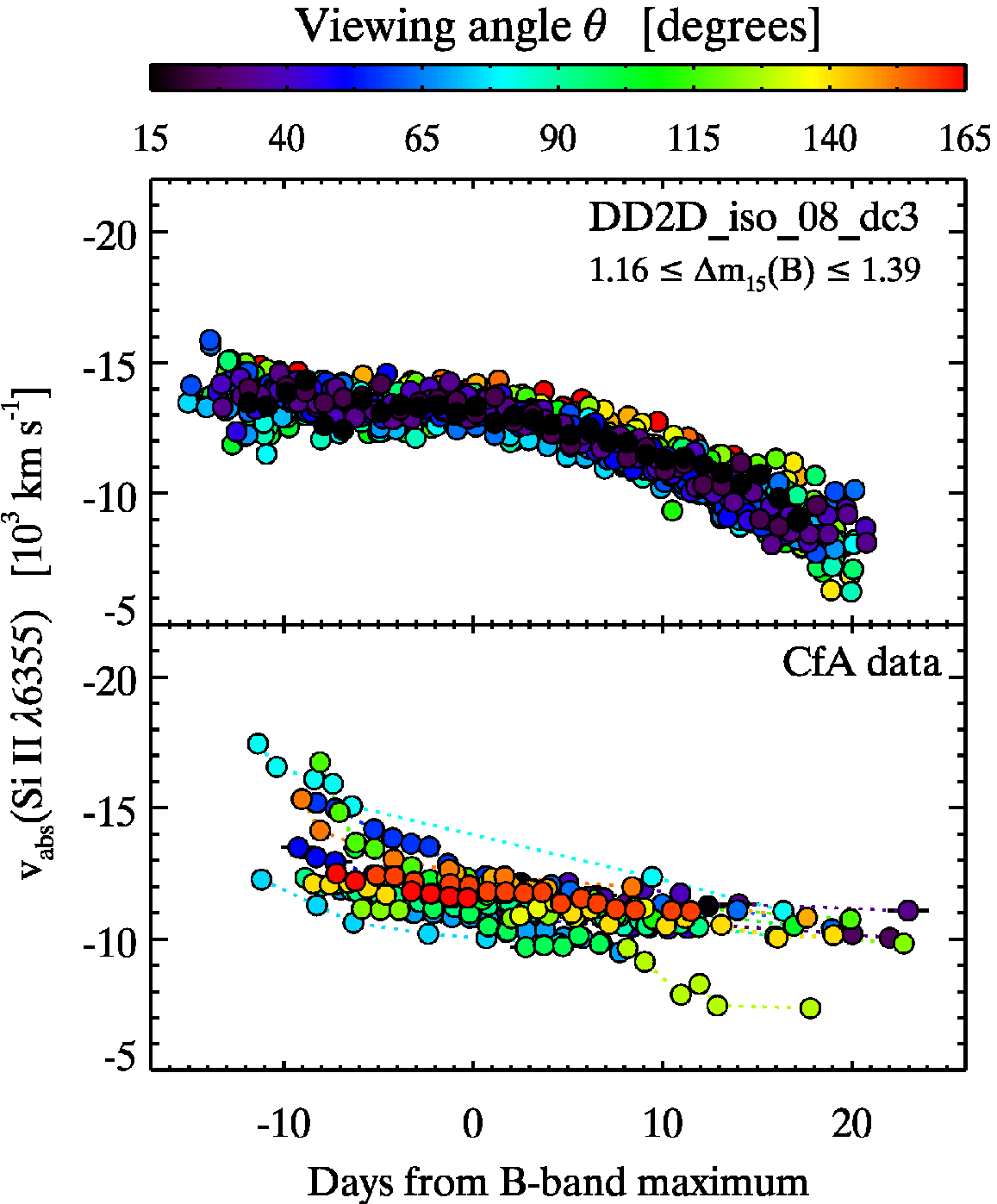}
}
\caption{\label{fig:vabs}
{\it Top row:}
Evolution of the Ca\two\,\l3945 absorption velocity with time for four
of our selected models ({\it upper panels}), colour-coded according to
the viewing angle $\theta$. The lower panels shows similar curves
for \sneia\ in the same \dmft\ range from our spectroscopic
sample. The colour-coding in this case denotes individual
\sneia, whose measurements are connected by dotted lines.
{\it Bottom row:} same as above for the Si\two\,\l6355 line.
}
\end{figure*}

In the top row of Fig.~\ref{fig:vabs} we show the evolution of the
Ca\two\,\l3945 absorption velocity, for four of our selected models
(where we have again chosen model DD2D\_iso\_06\_dc2 as representative
of the DD2D\_iso\_06 model series), colour-coded according to viewing
angle. The lower panel in each case shows similar measurements on
\sneia\ in the same \dmft\ range from our spectroscopic sample, 
where the colour-coding is used to distinguish individual supernovae. The
models exhibit a strong diversity in absorption velocity
evolution. Model DD2D\_iso\_03\_dc1 has a steeply decreasing
$|v_{\rm abs}|$ between $-15$\,d and $-5$\,d from $B$-band maximum
(from 25000\,\kms\ to 15000\,\kms), and little to no variation
at later times. Such a variation is also apparent in some observed
\sneia, although others display a steadier decrease in $|v_{\rm abs}|$
beyond +10\,d. Model DD2D\_iso\_06\_dc1 ({\it not shown};
$0.85\le\dmft\le1.08$) has a similar behaviour. The other models of
the DD2D\_iso\_06 series, however, all display little variation in the
Ca\two\ absorption velocity between $-15$\,d and nearly +10\,d, with a
sharp $\sim5000$\,\kms\ drop in $|v_{\rm abs}|$ around +10\,d,
followed by a second phase of almost no variation. 
The discontinuous jump in Ca\two\ absorption velocity could be in
part related to a measurement artefact due to overlap and changes in
relative strength between Si\two\,\l3858 and Ca\two\,\l3945.
The same overall pattern is visible in model DD2D\_iso\_08\_dc3, but
it is not present in the data. 
Interestingly, model
DD2D\_asym\_01\_dc3 appears to display both types of behaviour,
depending on the viewing angle.

As noted earlier, the Si\two\,\l6355 line is far less prone to
measurement uncertainties, and is a better probe of the ejecta
dynamics than Ca\two\,\l3945. The bottom row of Fig.~\ref{fig:vabs}
shows the evolution of the Si\two\,\l6355 absorption velocity in the
same four models, again compared to observations in the same \dmft\
range. Apart from model DD2D\_iso\_08\_dc3, which displays little
variation in $|v_{\rm abs}|$ between $-15$\,d and maximum light, and a
steady decrease thereafter, the other three models show almost no
variation for a given viewing angle (as expected, the largest scatter
at any given age occurs in model DD2D\_asym\_01\_dc3). Models
DD2D\_iso\_03\_dc1 and DD2D\_iso\_06\_dc2 even display a slight {\it
  increase} of $|v_{\rm abs}|$ with time. For most models we were
unable to reliably measure the location of maximum absorption in
Si\two\,\l6355 past +15\,d or so. 

Also noticeable are the apparent lack of correlation between the
Si\two\ and the Ca\two\ absorption velocities, as well as the absence of high
$|v_{\rm abs}|$ measurements before $-10$\,d. Several \sneia\ from our
spectroscopic sample display $\sim20000$\,\kms\ blueshifts at early
times, where the models seem constrained to $|v_{\rm
  abs}|\lesssim15000$\,\kms. Thus, while all the models are consistent
with the ``high-velocity'' subclass of \cite{WangX/etal:2009b} based
on the Si\two\,\l6355 absorption velocity at maximum light (see
Fig.~\ref{fig:specclass}), they would be considered ``normal'' when
considering the same measurement at $-10$\,d.

\subsubsection{Physical interpretation and constraints on models}\label{sect:interpretation}

\begin{figure}
\centering
\resizebox{.475\textwidth}{!}{\includegraphics{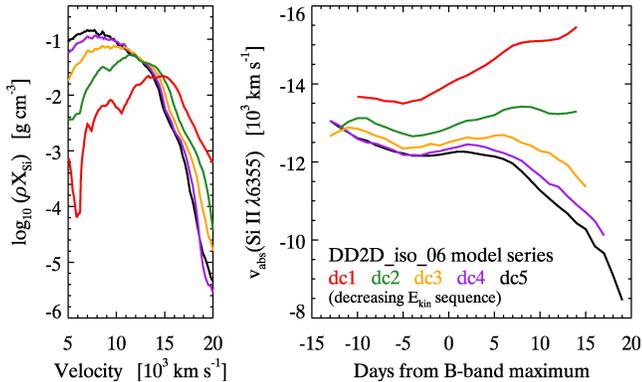}}
\caption{\label{fig:vabsgrad_rhosi}
{\it Right:} Angle-averaged silicon mass density ($=\rho X_{\rm Si}$) profiles at
$t=100$\,s past explosion for the DD2D\_iso\_06 model series (dc1
through dc5).
{\it Left:} Mean absorption velocity curves for Si\two\,\l6355 in the
same model series.
}
\end{figure}

Within the DD2D\_iso\_06 model series, model dc1 shows a
steady {\it increase} of $|v_{\rm abs}$(Si\two\,\l6355)$|$ with
time, model dc2 shows almost no variation, and models
dc3 through dc5 display a pre-maximum plateau followed by a
progressively steeper post-maximum decline
(Fig.~\ref{fig:vabsgrad_rhosi}, {\it right}). The post-maximum
velocity gradient is in the range 50--150\,\kms\,day$^{-1}$ for 
these latter models, which represents a sizeable fraction of the
range of velocity gradients observed by \cite{Benetti/etal:2005}.
The dc1 through dc5 sequence corresponds to decreasing
asymptotic kinetic energy within a model series (see
Table~\ref{tab:modelinfo}), which directly affects the amount and
radial distribution of intermediate-mass elements. 

The left panel of
Fig.~\ref{fig:vabsgrad_rhosi} shows the angle-averaged silicon mass
density profiles ($=\rho X_{\rm Si}$) for the dc1--dc5 models of the
DD2D\_iso\_06 model series. Models with higher kinetic energy have a
Si density distribution which peaks at higher velocities ($\sim
14000$\,\kms\ for model dc1 cf. $\sim7000$\,\kms\ for model dc5),
resulting in a higher $|v_{\rm abs}|$ at any given age (see also
Fig.~\ref{fig:vabsekin}). These models also synthesize more \nifs\ at
the expense of intermediate-mass elements, hence both the height and
width of the Si density distribution decrease with increasing $E_{\rm
  kin}$. This conditions the strength of the Si\two\,\l6355 line and
its velocity evolution. In models with higher kinetic energy, the
lines are expected to be weaker, owing to the lower peak Si density,
and to leave a more transient imprint on the spectra: in model dc1,
the Si\two\,\l6355 absorption profile is identifiable out to
$\sim15$\,d, while it is clearly visible out to $\sim20$\,d in model
dc5. The narrower Si distributions (with less Si at low velocities)
result in a more modest change of the absorption velocity past maximum
light: the Si\two\ velocity gradient is almost flat in model dc2, and
progressively steeper for models dc3 through dc5. The apparent
increase in $|v_{\rm abs}|$ with time for model dc1 is likely due to
the weakness of the Si\two\ line in that model, whose absorption
profile is biased by neighbouring lines.

The absence of high-velocity ($|v_{\rm abs}| \gtrsim 15000$\,\kms)
measurements in the models at early times is a consequence of the drop
in the silicon mass fraction in the outer regions of the ejecta
($v\gtrsim15000$\,\kms). To reproduce the
Si\two\,\l6355 absorption profile in SN~2002bo, a \snia\ with a large
absorption velocity at early times ($v_{\rm abs}\approx -18000$\,\kms\
at $-13$\,d), \cite{Stehle/etal:2005} infer a homogeneous $X_{\rm
  Si}\approx0.3$ for $v\gtrsim11000$\,\kms. In model
DD2D\_iso\_06\_dc2, $X_{\rm Si}$ drops by two orders of magnitude
(from $\sim0.3$ to $\sim0.001$) between $\sim16000$ and
$\sim21000$\,\kms. The lack of a significant fraction of Si at high
velocities in the models of KRW09 naturally accounts for the failure
of the models to reproduce the high $|v_{\rm abs}|$ measurements
observed in some \sneia, but we note that the nucleosynthetic yields
in these low-density regions are subject to a large uncertainty. 

The absence of $|v_{\rm abs}|>25000$\,\kms\ measurements for
Ca\two\,\l3945 is a consequence of the velocity cutoff used
when remapping the hydrodynamical output on the 2D cylindrical grid
for the radiative transfer calculations (see \S~\ref{sect:model}). We
do note, however, that a low mass fraction $X_{\rm
  Ca}\approx10^{-5}$ is sufficient to yield a Ca\two\ absorption at
the cutoff velocity, illustrating the strength of this doublet.

The distribution of intermediate-mass elements typically extends
out to $\sim20000$\,\kms\ in the models of KRW09, regardless of the
initial distribution of ignition points. However, where almost all
models with an isotropic distribution of ignition points have
IME down to the centre of the ejecta ($X_{\rm IME}>10^{-3}$ at
$v\approx0$\,\kms), most of the models with an anisotropic
  distribution (DD2D\_asym)
have no IME below $\sim10000$\,\kms, and hence display narrow
5000-10000\,\kms\ ``shells'' of IME in their ejecta. These IME shells
are a direct consequence of the weakness of the deflagration due
  to its one-sided ignition, resulting in a small pre-expansion and an
IME synthesis in a thin outer layer of the ejecta,
and are clearly incompatible with the
large $>10000$\,\kms\ widths of absorption profiles observed in \snia\
spectra. Not surprisingly, 5 of the 6 rejected models (see
Table~\ref{tab:modelinfo}) have narrow $\lesssim5000$\,\kms\ IME
shells in their ejecta.

\section{Discussion: are asymmetric ignitions ruled out by
  observations?}\label{sect:disc}

\begin{figure}
\centering
\resizebox{.475\textwidth}{!}{
\includegraphics[width=5.5cm]{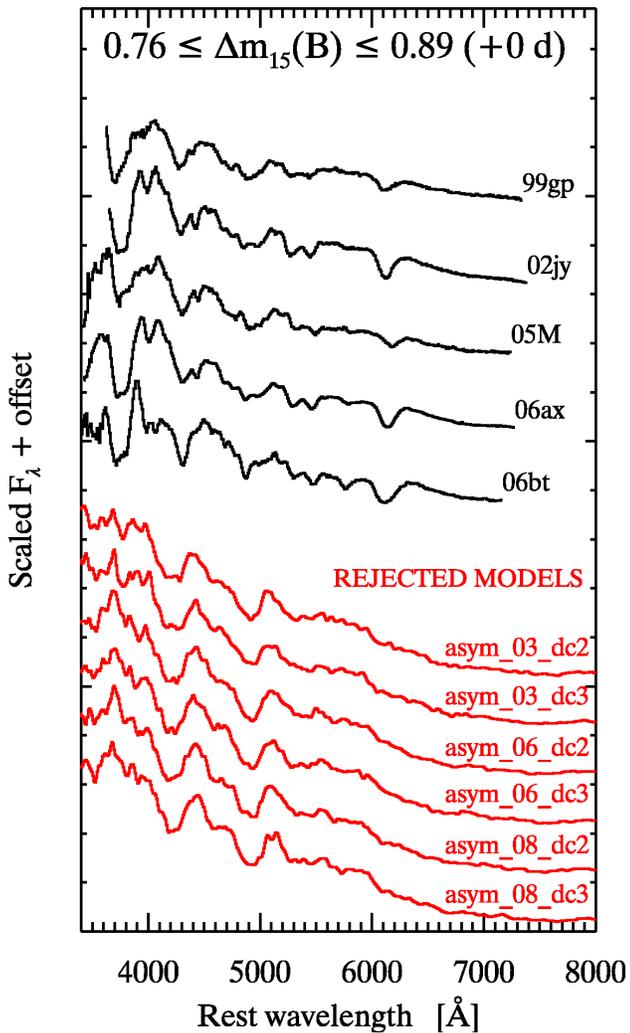}
}
\caption{\label{fig:rejected}
Comparison of observed maximum-light spectra ({\it black}) with the
subset of rejected models viewed along $\theta=40^\circ$, all in the
range $0.76\le\dmft\le0.89$. SN~1999gp and SN~2005M are both
  1991T-like \sneia.
}
\end{figure}

In section~\ref{sect:rank} we rejected six models that do
not yield satisfactory matches to observed \snia\ spectra for some viewing
angles. 
Fig.~\ref{fig:rejected} shows spectra for these models viewed along
$\theta=40^\circ$ at maximum light, compared
with observed maximum-light spectra in the same \dmft\ range.
The mismatch with observations is clear (and these objects would
{\it not} be classified as \sneia!\footnote{nor do they yield good
matches to SN spectra of other types.}, but the synthetic spectra
share some common properties: all have a very blue SED characteristic
of hot ionized ejecta. Moreover, they show
almost no sign of lines from intermediate-mass elements (Ca, S, Si)
characteristic of \snia\ spectra, and instead are dominated by
iron-peak elements. This agrees fully with the output from the
hydrodynamical simulations: these models have the highest kinetic
energy ($E_{\rm kin}\gtrsim1.6\times 10^{51}$\,erg) and synthesize
the most \nifs\ ($M(\nifs)\gtrsim1$\,\msun), at the expense of
intermediate-mass elements ($M({\rm IME})\lesssim0.15$\,\msun). The
association of anisotropic distributions of ignition points with
greater amounts of 
synthesized \nifs\ holds for the other DD2D\_asym models of KRW09. The
deflagration burns less material in these models, and by the time 
the detonation is triggered, the WD will have expanded less and the
burning will occur at higher densities, increasing the fraction of
material burnt to nuclear statistical equilibrium, and resulting in a
globally symmetric ejecta. A direct
observational consequence is that these explosions will tend to
yield low \dmft\ values. This is  seen in this model series,
where the DD2D\_asym models have \dmft\ values in a narrow range
($\dmft\approx0.8$), whereas the DD2D\_iso models are evenly
distributed across $0.9\lesssim\dmft\lesssim1.4$.

These rejected models may cast some doubts on the validity of the
assumption of an anisotropic distribution of ignition points in
  \snia\ deflagrations. This 
is in apparent contradiction with the recent study by
\cite{Maeda/etal:2010c}, who associate the observed diversity in the
spectroscopic evolution of normal \sneia\ with viewing angle effects
in off-center delayed-detonation models. 
By using nebular lines from stable iron-group elements as probes of
the distribution of deflagration ashes
\citep[see][]{Maeda/etal:2010a}, they showed that \sneia\ displaying
low velocity gradients in the Si\two\,\l6355 line correspond to
off-center explosions viewed from the ignition (deflagration) side,
while SN with high velocity gradients resulted from the same explosion
viewed from the opposite direction.

In the present study, we see a variation of the Si\two\ velocity
gradient in explosions with an isotropic distribution of
  ignition points in which the criterion
for deflagration-to-detonation transition is varied. This criterion primarily
affects the asymptotic kinetic energy of the explosion and results
in a change in the distribution and abundance of silicon in the ejecta
(see Fig.~\ref{fig:vabsgrad_rhosi}). While this does not invalidate
the interpretation of \cite{Maeda/etal:2010c}, it nonetheless shows
that multiple parameters can affect the steepness of the Si\two\
velocity gradient.

We also note that the most DD2D\_asym models of KRW09 yield good
matches to observations, such as DD2D\_asym\_01\_dc3 which is part of
our subset of selected models. Moreover, not all DD2D\_asym models
with high explosion energies yield poor matches to observed
spectra. Models DD2D\_asym\_07\_dc2 and dc3 have similar properties as
the rejected models of Fig.~\ref{fig:rejected}, yet their
maximum-light spectra are compatible with observed spectra of
high-luminosity 1991T-like \sneia\ 
(see SN~1999gp and SN~2005M in Fig.~\ref{fig:rejected}), and with the
peculiar SN~2000cx \citep{Li/etal:2001b}, both characterized by hot,
ionized, and energetic ejecta. Asymmetric ignitions are thus not
ruled out as a whole in the present study, but observations suggest
that some anisotropic configurations do not occur in Nature.

While the models with an asymmetric distribution of \nifs\
result in a larger variation of photometric and spectroscopic properties with
viewing angle, we were unable to find {\it specific} measurements
which correlate with the degree of asymmetry in the explosion. Ejecta
asymmetries are bound to leave an imprint on the spectral line-profile
morphology (see \citealt{Dessart/Hillier:2011} in the context of Type
II supernovae), but such signatures would probably be drowned in Monte
Carlo noise in the synthetic spectra studied here. Spectropolarimetric
observations of \sneia\ still provide the most direct (yet
observationally expensive) probe of ejecta asymmetries (see
\citealt{WangL/Wheeler:2008} for a review), while nebular line
profiles offer a means to assess the geometry of the explosion by
probing the innermost regions of the ejecta. The limited number of
photon packets used in the radiative transfer simulations and the
assumption of LTE prevents us to investigate spectropolarimetric
signatures or spectra during the nebular phase in the present study.


\section{Conclusions}\label{sect:ccl}

We have presented a detailed comparison of a recent survey of 2D
delayed-detonation explosion models by KRW09 with observations of Type
Ia supernovae. We apply standard methods used by \snia\
observers to compare the model light curves and spectra with empirical
templates. This represents a significant step forward in the
realism of the models.
Running several light-curve fitters (MLCS2k2, SALT2,
SNooPy) on synthetic $(U)BVRI$ light curves, we find some tension
between the light-curve shape of the models and actual data, 
the models having longer rise times.
Based on cross-correlations with a library of \snia\ spectra, we
quantified the overall resemblance of individual models to observed 
\sneia, and found that the best models/viewing angles
lied systematically on the observed width-luminosity relation.

Comparison of several photometric properties of the models (rise
times, maximum-light colours and their evolution with time) shows a
broad agreement with observations, but reveals some problems with flux
redistribution from 
the near-UV to the near-IR bands, a key mechanism needed to explain
both the width-luminosity relation and the secondary maxima in the NIR
light curves, and mediated by the {\sc iii}$\rightarrow${\sc ii}
recombination timescale of iron-group elements
\citep{Kasen:2006,Kasen/Woosley:2007}. Subsequent
investigation of spectra for a subset of selected models confirmed the
excess of $U$-band flux in the models at early times, likely caused by
a hot ionized ejecta and subsequent lack of absorption by
Fe\two/Co\two. Interestingly, one of our selected models (DD2D\_iso\_08\_dc3)
that shows the best overall agreement in optical and NIR colour
evolution with observations lies off the width-luminosity relation
(i.e. the colours match, but not the luminosity). This reveals one
limitation of our approach, which relies on maximum-light
spectra to rank the different models, whereas a combination of
photometric and spectroscopic properties is needed for a proper
evaluation.

Comparison of maximum-light spectra show the models have
systematically large absorption velocities (most visible in the
Si\two\,\l6355 line), affecting the relative shapes and strengths of
spectral features and smoothing out small substructures observed in
iron-dominated absorption complexes at $\sim4300$\,\AA\ and
$\sim4800$\,\AA. Consequently, correlations between several
spectroscopic indicators and \dmft\ decline rate have a much larger
scatter in the models. The relation found by \cite{Foley/Kasen:2011}
between Si\two\,\l6355 absorption velocity and intrinsic $B-V$ colour
(redder \sneia\ having larger $|v_{\rm abs}|$) is also not reproduced
in the models, most showing a weak correlation in the opposite
direction. However, the observed correlation is weak and 
subject to a large uncertainty given the errors on intrinsic $B-V$
colour inferred from the data.
Nonetheless, we identify a trend of larger absorption blueshifts for
higher kinetic energy for models in which only the criterion for
deflagration-to-detonation transition is varied.

The overall evolution of the model spectra compares well with
observations, as illustrated by the comparison of synthetic spectra
for model DD2D\_iso\_06\_dc2 between $-10$\,d and +20\,d from
$B$-band maximum with observed spectra of SN~2003du, but several
discrepancies characteristic of most models are apparent. The
synthetic spectra are too blue (i.e. too hot) at early times, and the
ionization that ensues affects their subsequent evolution. Most
notably, the models appear to lack an absorption feature around
$\sim5000$\,\AA\ (attributed to Fe\two/Co\two), and fail to reproduce the strong
emission feature at $\sim5800$\,\AA\ (attributed to Na\one\,D) from
+10\,d onwards. Non-LTE
calculations of \snia\ spectra also fail to reproduce this line
\citep{Baron/etal:2006}.

The evolution of the Ca\two\,\l3945 absorption velocity with time
exhibits a strong diversity in the models 
which contrasts with the steady and smooth decrease
seen in the data. Model DD2D\_asym\_01\_dc3 has an asymmetric
\nifs\ distribution and shows all types of behaviour depending on the
viewing angle, due to the varying radial Ca distribution with different
  lines of sight.
The evolution of the
Si\two\,\l6355 velocity shows little variation before maximum light,
while the post-maximum evolution (the velocity gradient) appears
conditioned by the kinetic energy of the explosion, affecting the
abundance and radial distribution of silicon. While this does not
contradict the recent findings of \cite{Maeda/etal:2010c}, who
associate the observed diversity in velocity gradients with viewing
angle effects in off-center explosions, it shows that the
interpretation of such gradients depends on more than a single
parameter of the explosion.

We reject six models of KRW09 with highly asymmetric ignition
conditions and are characterized by large amounts ($\gtrsim
1$\,\msun) of \nifs. We do not reject off-center delayed-detonation
models for \sneia\ as a whole, but note the extreme sensitivity of the amount and
distribution of burning products in the deflagration phase to the
initial distribution of ignition points \citep[see
  also][]{Livne/Asida/Hoeflich:2005}.

Throughout this paper we have focused on discrepancies between the
models and observations more than we have highlighted their mutual
agreement, but this merely results from the unprecedented level of
detail of our study. Such detail is necessary to use the
predictive power of the models to provide a physical basis to some
observed trends, as well as use the data to impose meaningful
constraints on the models. The 2D delayed-detonation models of KRW09
have a degree of fidelity which makes them amenable to the same
analysis we use on observations of \sneia, but they still
require some adjustments to accurately match the data. 
The ability to reproduce the bolometric/multi-band light curves
and the width-luminosity relation is a necessary but not a sufficient
condition for a model to be considered a valid approximation of real
\sneia.
Further insights from three-dimensional hydrodynamical simulations, more
accurate nucleosynthetic post-processing, and full non-LTE radiative
transfer calculations are all part of the solution.
We are confident that a detailed comparison of light curves and
spectra from grids of models using the framework developed in this
paper will lead to a better understanding of \snia\ explosion
mechanisms.


\section*{Acknowledgments}

SB acknowledges useful discussions with Luc Dessart, Ryan Foley,
Alexei Khokhlov and Masaomi Tanaka. 
This research has been supported by the DOE
SciDAC Program (DE-FC02-06ER41438).  Computing time was
provided by ORNL through an INCITE award and by NERSC.
The work of FKR is supported by the Deutsche Forschungsgemeinschaft
via the Emmy Noether programme (RO 3676/1-1).
Support for supernova research at Harvard University, including the
CfA Supernova Archive, is provided in part by NSF grant AST
09-07903.


\bibliographystyle{mn2e}
\bibliography{modelcomp}


\begin{table*}
\caption{Asymptotic kinetic energies, abundances, peak bolometric
  luminosities and decline-rate ranges for the 44 2D
  delayed-detonation models of KRW09.}\label{tab:modelinfo} 
\begin{tabular}{lccccccccl}
\hline\hline
Model & $E_{\rm kin}$   & $M(\nifs)$   & $M$(stable IGE) & $M$(IME)     & $M$(O)       & $M$(C)       & $L_{\rm bol,peak}$            & \dmft\ & Notes \\
      & ($10^{51}$ erg) & (M$_{\sun}$) & (M$_{\sun}$) & (M$_{\sun}$) & (M$_{\sun}$) & (M$_{\sun}$) & ($10^{43}$ erg s$^{-1}$) &        &       \\
\hline
DD2D\_iso\_01\_dc2   & 1.548 & 0.942 & 0.199 & 0.215 & 0.044 & 0.007 & 1.921--2.133 & 0.91--1.03 &              \\
DD2D\_iso\_01\_dc3   & 1.506 & 0.889 & 0.194 & 0.253 & 0.061 & 0.011 & 1.818--2.026 & 0.99--1.08 &              \\
DD2D\_iso\_01\_dc4   & 1.463 & 0.832 & 0.193 & 0.289 & 0.078 & 0.016 & $\cdots$     & $\cdots$   & not included \\
DD2D\_iso\_01\_dc5   & 1.442 & 0.809 & 0.192 & 0.302 & 0.085 & 0.019 & 1.614--1.812 & 1.07--1.17 &              \\
DD2D\_iso\_02\_dc2   & 1.569 & 0.901 & 0.256 & 0.210 & 0.034 & 0.007 & 1.863--2.172 & 0.88--1.16 &              \\
DD2D\_iso\_02\_dc3   & 1.515 & 0.796 & 0.253 & 0.298 & 0.051 & 0.010 & 1.563--1.915 & 0.97--1.25 &              \\
DD2D\_iso\_02\_dc5   & 1.410 & 0.612 & 0.241 & 0.456 & 0.082 & 0.016 & 1.281--1.430 & 1.15--1.48 &              \\
DD2D\_iso\_03\_dc1   & 1.509 & 0.799 & 0.274 & 0.280 & 0.045 & 0.009 & 1.763--1.850 & 0.79--0.97 & subset       \\
DD2D\_iso\_03\_dc2   & 1.385 & 0.584 & 0.263 & 0.448 & 0.097 & 0.016 & 1.277--1.357 & 0.90--1.16 &              \\
DD2D\_iso\_03\_dc3   & 1.288 & 0.441 & 0.257 & 0.548 & 0.138 & 0.028 & 0.925--1.017 & 1.08--1.33 &              \\
DD2D\_iso\_04\_dc1   & 1.502 & 0.774 & 0.275 & 0.298 & 0.051 & 0.011 & 1.499--1.938 & 0.77--0.99 &              \\
DD2D\_iso\_04\_dc2   & 1.379 & 0.562 & 0.264 & 0.470 & 0.095 & 0.017 & 0.983--1.519 & 0.85--1.19 &              \\
DD2D\_iso\_04\_dc3   & 1.263 & 0.399 & 0.255 & 0.583 & 0.145 & 0.027 & 0.633--1.228 & 1.06--1.38 &              \\
DD2D\_iso\_04\_dc4   & 1.226 & 0.369 & 0.252 & 0.593 & 0.162 & 0.033 & 0.581--1.147 & 0.99--1.44 &              \\
DD2D\_iso\_05\_dc1   & 1.470 & 0.718 & 0.260 & 0.357 & 0.060 & 0.011 & 1.560--1.594 & 0.87--1.11 &              \\
DD2D\_iso\_05\_dc2   & 1.333 & 0.496 & 0.249 & 0.530 & 0.114 & 0.019 & 1.111--1.131 & 1.06--1.22 &              \\
DD2D\_iso\_05\_dc3   & 1.210 & 0.330 & 0.237 & 0.648 & 0.160 & 0.033 & 0.754--0.826 & 1.21--1.44 &              \\
DD2D\_iso\_05\_dc4   & 1.151 & 0.293 & 0.232 & 0.648 & 0.189 & 0.046 & 0.691--0.725 & 1.23--1.44 &              \\
DD2D\_iso\_06\_dc1   & 1.552 & 0.882 & 0.256 & 0.227 & 0.036 & 0.009 & 1.859--1.959 & 0.85--1.08 & subset       \\
DD2D\_iso\_06\_dc2   & 1.446 & 0.698 & 0.247 & 0.378 & 0.073 & 0.012 & 1.443--1.564 & 1.06--1.32 & subset       \\
DD2D\_iso\_06\_dc3   & 1.371 & 0.567 & 0.240 & 0.485 & 0.099 & 0.019 & 1.151--1.288 & 1.21--1.48 & subset       \\
DD2D\_iso\_06\_dc4   & 1.312 & 0.472 & 0.235 & 0.561 & 0.117 & 0.025 & 0.972--1.064 & 1.30--1.53 & subset       \\
DD2D\_iso\_06\_dc5   & 1.275 & 0.421 & 0.231 & 0.599 & 0.129 & 0.029 & 0.896--0.959 & 1.34--1.57 & subset       \\
DD2D\_iso\_07\_dc2   & 1.282 & 0.444 & 0.246 & 0.556 & 0.139 & 0.023 & 0.867--1.079 & 1.04--1.24 &              \\
DD2D\_iso\_07\_dc3   & 1.181 & 0.346 & 0.238 & 0.603 & 0.183 & 0.039 & 0.640--0.861 & 1.18--1.30 &              \\
DD2D\_iso\_08\_dc1   & 1.477 & 0.719 & 0.275 & 0.341 & 0.060 & 0.012 & 1.534--1.767 & 0.80--1.07 &              \\
DD2D\_iso\_08\_dc2   & 1.302 & 0.447 & 0.259 & 0.549 & 0.130 & 0.023 & 0.930--1.072 & 1.00--1.38 &              \\
DD2D\_iso\_08\_dc3   & 1.204 & 0.329 & 0.250 & 0.624 & 0.169 & 0.035 & 0.695--0.785 & 1.16--1.39 & subset       \\
DD2D\_asym\_01\_dc2  & 1.416 & 0.677 & 0.224 & 0.410 & 0.081 & 0.015 & 1.095--1.811 & 0.93--1.09 &              \\
DD2D\_asym\_01\_dc3  & 1.379 & 0.644 & 0.221 & 0.419 & 0.103 & 0.021 & 0.997--1.696 & 0.96--1.15 & subset       \\
DD2D\_asym\_02\_dc2  & 1.387 & 0.614 & 0.229 & 0.454 & 0.096 & 0.016 & 1.234--1.451 & 0.96--1.23 &              \\
DD2D\_asym\_02\_dc3  & 1.252 & 0.458 & 0.216 & 0.558 & 0.147 & 0.027 & 0.886--1.139 & 1.18--1.38 &              \\
DD2D\_asym\_03\_dc2  & 1.580 & 0.982 & 0.234 & 0.157 & 0.028 & 0.006 & 1.895--2.460 & 0.79--0.88 & rejected     \\
DD2D\_asym\_03\_dc3  & 1.572 & 0.965 & 0.233 & 0.171 & 0.031 & 0.007 & 1.834--2.426 & 0.79--0.88 & rejected     \\
DD2D\_asym\_04\_dc2  & 1.577 & 0.974 & 0.201 & 0.190 & 0.036 & 0.007 & 1.722--2.262 & 0.80--0.91 &              \\
DD2D\_asym\_04\_dc3  & 1.562 & 0.955 & 0.198 & 0.204 & 0.042 & 0.009 & 1.666--2.243 & 0.81--0.96 &              \\
DD2D\_asym\_05\_dc2  & 1.538 & 0.939 & 0.208 & 0.211 & 0.042 & 0.007 & 1.979--2.054 & 0.77--1.00 &              \\
DD2D\_asym\_05\_dc3  & 1.508 & 0.883 & 0.205 & 0.253 & 0.057 & 0.011 & 1.876--1.981 & 0.82--1.06 &              \\
DD2D\_asym\_06\_dc2  & 1.628 & 1.082 & 0.221 & 0.087 & 0.014 & 0.003 & 2.175--2.562 & 0.78--0.88 & rejected     \\
DD2D\_asym\_06\_dc3  & 1.622 & 1.068 & 0.213 & 0.104 & 0.019 & 0.004 & 2.144--2.537 & 0.78--0.89 & rejected     \\
DD2D\_asym\_07\_dc2  & 1.579 & 1.053 & 0.175 & 0.144 & 0.030 & 0.005 & 2.192--2.362 & 0.74--0.94 &              \\
DD2D\_asym\_07\_dc3  & 1.558 & 1.034 & 0.170 & 0.158 & 0.038 & 0.008 & 2.164--2.332 & 0.76--0.95 &              \\
DD2D\_asym\_08\_dc2  & 1.632 & 1.103 & 0.206 & 0.081 & 0.014 & 0.003 & 2.113--2.586 & 0.76--0.87 & rejected     \\
DD2D\_asym\_08\_dc3  & 1.617 & 1.081 & 0.189 & 0.113 & 0.020 & 0.005 & 2.043--2.578 & 0.77--0.89 & rejected     \\
\hline
\end{tabular}

\end{table*}

\label{lastpage}

\end{document}